\documentclass[10pt,prd,nofootinbib,preprint]{revtex4}
\usepackage[a4paper, total={7in, 10.5in}]{geometry}
\usepackage[utf8]{inputenc}
\usepackage{bm}
\usepackage{color}
\usepackage{xcolor}
\usepackage{float}
\usepackage{booktabs}
\usepackage{multirow}
\usepackage{amsmath}
\usepackage{graphicx}
\usepackage{esint}
\usepackage{epsfig}
\usepackage{caption}
\usepackage{subcaption}
\usepackage{hyperref}
\usepackage{cancel}
\makeatletter

\DeclareFontEncoding{LGR}{}{}

\ProvideTextCommand{\~}{LGR}[1]{\char126#1}

\newcommand{\lyxmathsym}[1]{\ifmmode\begingroup\def\b@ld{bold}
  \text{\ifx\math@version\b@ld\bfseries\fi#1}\endgroup\else#1\fi}

\providecommand{\tabularnewline}{\\}

\@ifundefined{textcolor}{}
{%
 \definecolor{BLACK}{gray}{0}
 \definecolor{WHITE}{gray}{1}
 \definecolor{RED}{rgb}{1,0,0}
 \definecolor{GREEN}{rgb}{0,1,0}
 \definecolor{BLUE}{rgb}{0,0,1}
 \definecolor{CYAN}{cmyk}{1,0,0,0}
 \definecolor{MAGENTA}{cmyk}{0,1,0,0}
 \definecolor{YELLOW}{cmyk}{0,0,1,0}
}

\usepackage{xcolor}
\usepackage{epsfig}\usepackage{slashed}\usepackage{appendix}
\usepackage{tabularx}
\usepackage{pdfpages}
\usepackage{epstopdf}
\usepackage{multirow}
\usepackage{siunitx}
\usepackage{soul}

\usepackage[T1]{fontenc}
\begin{document}
\begin{center}
{\bf\Large\boldmath
The Angular Observables of 
$\Lambda_b \to \Lambda_c(\to \Lambda^0 \pi^+) \, 
\tau^-(\to \pi^- \nu_\tau)\, \bar{\nu}_\tau$ 
within the Paradigm of FCCC Anomalies
}\\[5mm]
\par\end{center}
\begin{center}
\setlength{\baselineskip}{0.2in} {Muhammad Arslan$^{a,}$\footnote{arslan.hep@gmail.com (corresponding author)},
Ishtiaq Ahmed$^{b,}$\footnote{ishtiaqmusab@gmail.com} and Muhammad Jamil Aslam $^{a,}$\footnote{jamil@qau.edu.pk}
}\\[5mm] $^{a}$~\textit{Department of Physics, Quaid-i-Azam
University, Islamabad 45320, Pakistan.}\\
 $^{b}$~\textit{National Center for Physics, Islamabad 44000, Pakistan.}\\[5mm]
\par\end{center}

\begin{abstract}
We present a global analysis of the current $B$-meson flavor anomalies and extend it to the baryonic sector through the decay $\Lambda_b^0 \to \Lambda_c^+(\to \Lambda^0 \pi^+) \tau^-(\to \pi^- \nu_\tau)\bar{\nu}_\tau$. The lepton flavor universality ratios $R_{\tau/(\mu,e)}(D^{(*)})$, measured by BaBar, Belle, and LHCb, exhibit a combined $3.8\sigma$ deviation from Standard Model (SM) predictions. Using the latest HFLAV averages and imposing $B_c$-lifetime constraints on the branching ratio, $\mathcal{B}(B_c \to \tau \nu) < 60\%, 30\%, 10\%$, we perform a global fit to the anomaly data and propagate the preferred new physics (NP) solutions to the full cascade decay $\Lambda_b^0 \to \Lambda_c^+(\to \Lambda^0 \pi^+) \tau^-(\to \pi^- \nu_\tau)\bar{\nu}_\tau$. The mixed vector-scalar scenario $(C_{V_L},C_{S_R})$ emerges as the most favored NP solution, yielding the largest pull from the SM while remaining insensitive to branching-ratio constraints. The single-operator $C_{V_L}$ case identified as the next most competitive scenario. We study the impact of NP vector, scalar and tensor operators on a complete set of angular observables on the five-fold $\Lambda_b$ decay using Lattice-QCD form factors and find that the scenarios $(\Re[C_{S_L}=4C_T],\Im[C_{S_L}=4C_T])$ and $(C_{S_L},C_{S_R})$ generate the largest deviations from the SM predictions. In particular, the observables $\mathcal{K}_{1c}$, $\mathcal{K}_{2ss}$, $\mathcal{K}_{2cc}$, and $\mathcal{K}_{4s}$ show the highest sensitivity to NP effects. The correlation analysis reveals distinctive NP patterns: the $(\Re[C_{S_L}=4C_T],\Im[C_{S_L}=4C_T])$ scenario exhibits inverse correlations among $\mathcal{K}_{1c}$ and $\mathcal{K}_{2ss,2cc,4s}$ and direct correlations between $\mathcal{K}_{2ss}$ and $\mathcal{K}_{2cc,4s}$, pointing to destructive helicity interference and a possible CP-violating phase, while the $(C_{S_L},C_{S_R})$ scenario displays complementary behavior consistent with CP-conserving dynamics. These results establish baryonic semileptonic decays as a powerful and independent probe of the $R_{\tau/(\mu,e)}(D^{(*)})$ anomalies, with future measurements providing critical tests of the underlying NP structure.\\[5mm]
\noindent\textit{Keywords:} $B$-physics anomalies, Lepton Flavor Universality Violation, Semileptonic baryon decays, New Physics, Angular observables

\end{abstract}

\maketitle

\section{Introduction}\label{sec1}
The semileptonic and leptonic decays of hadrons containing a $b$ quark play a central role in the determination of the elements of the Cabibbo--Kobayashi--Maskawa (CKM) matrix. In this context, studies of flavor-changing charged-current (FCCC) and flavor-changing neutral-current (FCNC) transitions of $b-$hadrons are of particular importance. Among these, FCNC transitions are especially sensitive to effects of physics beyond the Standard Model (SM), as they are forbidden at tree level in the SM and occur only through loop- or box-level diagrams, where contributions from heavy virtual particles may enter.

Despite the analysis of Run-3 data from the Large Hadron Collider (LHC), collected at a center-of-mass energy of $13.6~\text{TeV}$ with an integrated luminosity of $39.7~\text{fb}^{-1}$, no direct evidence for new particles has been observed so far. Nevertheless, indirect indications of physics beyond the SM arise from phenomena such as neutrino oscillations and persistent anomalies in flavor observables, which suggest the existence of new physics (NP) \cite{Arbey:2021gdg,Gonzalez-Garcia:2022pbf}. Under these circumstances, FCNC decays provide a powerful probe for testing the SM and searching for NP effects. In particular, the exclusive decay modes
$B^{\pm}\!\to K^{(*)\pm}\ell^{+}\ell^{-}$,
$B^{0}\!\to K^{0}\ell^{+}\ell^{-}$, and
$B_s^{0}\!\to \phi\,\ell^{+}\ell^{-}$,
with $\ell=e,\mu$, have been extensively studied experimentally \cite{LHCb:2014cxe,LHCb:2014auh,LHCb:2015wdu,LHCb:2015svh,LHCb:2016ykl,LHCb:2021zwz,LHCb:2021xxq}.
Of particular interest are measurements of lepton-flavor universality violation (LFUV) in $B\to K^{(*)}\ell^{+}\ell^{-}$ decays \cite{LHCb:2022vje,LHCb:2022qnv,Smith:2024xgo,CMS:2024syx}, for which the dependence on CKM matrix elements and the associated hadronic form-factor uncertainties largely cancel. These observables have therefore been widely analyzed within various NP scenarios; see, \textit{e.g.}, Refs.~\cite{Celis:2017doq,Buttazzo:2017ixm,Aebischer:2019mlg,Alasfar:2020mne,Isidori:2021tzd,Ciuchini:2022wbq}. Recent experimental results indicate that these measurements are consistent with the SM predictions within approximately $0.2\sigma$ \cite{Hiller:2003js,Bordone:2016gaq,Mishra:2020orb,Isidori:2020acz,Bernlochner:2021vlv,Fischer:2021sqw,London:2021lfn,Crivellin:2021sff,Crivellin:2022qcj}.

Unlike FCNC transitions, the signatures of NP in semileptonic FCCC transitions $b\rightarrow c\ell\nu_{\ell}$ ($\ell=e,\mu,\tau$) are not diminished, where the measured values of the lepton flavor universality (LFU) ratio in $B\to D^{(*)}$ decays, \textit{i.e.}, $R_{\tau/\mu,e}\left(D^{(*)}\right)\equiv\mathcal{B}\left(B\rightarrow D^{(*)}\tau^-\overline{\nu}_{\tau}\right)/\mathcal{B}\left(B\rightarrow D^{(*)}\ell^-\bar{\nu}_{\ell}\right)$, with $\ell = e,\mu$, by BABAR \cite{BaBar:2012obs,BaBar:2013mob}, Belle \cite{Belle:2015qfa,Belle:2019rba,Belle:leptonphoton}
and LHCb \cite{LHCb:2015gmp,LHCb:2017smo,LHCb:2017rln,LHCb:2023zxo,LHCb:2023uiv}
deviate from their SM predictions. Recently, the
 Heavy Flavor Averaging Group (HFLAV), took averages of almost ten years data of all these experiments and showed $3.8\sigma$ combined deviation from the SM results  \cite{MILC:2015uhg, Na:2015kha, Aoki:2016frl, Fajfer:2012vx, Bigi:2016mdz, Bernlochner:2017jka, Bigi:2017jbd, Jaiswal:2017rve, Gambino:2019sif, Bordone:2019vic, Martinelli:2021onb, HFLAV:2024link, HFLAV:2025link}. Their most up-to-date averaged results reported by HFLAV group in 2025 \cite{HFLAV:2025link}, and the corresponding SM predictions are:
 \begin{eqnarray}
 R_{\tau/\mu,e}\left(D\right) &=& 0.347\pm0.025\;, \quad\quad R_{\tau/\mu,e}\left(D^*\right) = 0.288\pm 0.012\;, \label{HFLAV-RDDs}\\
R^{\text{SM}}_{\tau/\mu,e}\left(D\right)&=& 0.296\pm 0.004\;,\quad\quad R^{\text{SM}}_{\tau/\mu,e}\left(D^*\right) = 0.254\pm 0.005\;. \label{SM-RDDs} 
\end{eqnarray}
Governed by the same quark level FCCC transitions, the LFU ratio in $R_{\tau/\mu}^{LHCb}\left(J/\psi\right)$ and $R_{\tau/\mu}^{CMS2023, 2024}\left(J/\psi\right)$ are measured by the the LHCb \cite{LHCb:2017vlu} and CMS \cite{RJSi:CMS2023, RJSi:CMS2024} collaborations in $B_c \to J/\psi \ell \nu_\ell$ decays. Also, the $R_{\tau/\ell}\left(\Lambda_{c}\right)$ is measured by LHCb collaboration \cite{LHCb:2022piu} in $\Lambda_b \to \Lambda_c \tau^- \bar{\nu}_{\tau}$ decay and the naive average of $ R_{\tau/\mu}\left(J/\psi\right)$ and corresponding result for $R_{\tau/\ell}\left(\Lambda_{c}\right)$ are
\begin{equation}
 R_{\tau/\mu}\left(J/\psi\right) = 0.61\pm 0.18\;,\quad\quad  R_{\tau/\ell}\left(\Lambda_{c}\right) =  0.242 \pm 0.076\;.\label{Exp-RJpsiLC}
\end{equation}
These experimental results differ from the corresponding SM predictions
\begin{eqnarray}
 R^{\text{SM}}_{\tau/\mu}\left(J/\psi\right) &=&   0.258 \pm 0.038,\; \cite{Watanabe:2017mip, Harrison:2020nrv} \label{SN-RJPsi}\\
R^{\text{SM}}_{\tau/\ell}\left(\Lambda_c\right) &=& 0.324\pm0.004,\; \cite{Detmold:2015aaa, Bernlochner:2018kxh} \label{Exp-Lambdac}
\end{eqnarray}
by $1.9\sigma$ and $1.1\sigma$, respectively. In the case $R_{\tau/\mu}\left(J/\psi\right)$, the only shortcoming is the uncertainty in the measurement of the lifetime of $B_c$ meson because the pure leptonic decay $B_c \to \tau \nu_\tau$ is not measured yet \cite{Celis:2016azn, Alonso:2016oyd}. To take care of it, an upper limit of
$60\%, 30\%$, and $10\%$ on its branching ratio is imposed in the literature \cite{Gershtein:1994jw,Bigi:1995fs,Beneke:1996xe,Chang:2000ac,Kiselev:2000pp,Akeroyd:2017mhr}. 

In addition to the deviations observed in the LFU ratios, polarization observables in
$B \to D^{*}\tau\nu_{\tau}$ decays also provide important tests of the SM.
In particular, the longitudinal polarization asymmetry of the $\tau^{-}$ lepton,
$P_{\tau}(D^{*})$, and the longitudinal polarization fraction of the $D^{*}$ meson,
$F_{L}(D^{*})$, measured by the Belle collaboration
\cite{Belle:2017ilt,Belle:2016dyj,Belle:2019ewo},
exhibit deviations at the level of approximately $1.5\sigma$ from the corresponding SM
predictions \cite{Alok:2016qyh,Iguro:2020cpg}.
For $F_{L}(D^{*})$, the LHCb collaboration has reported results obtained by combining the
Run~1 data set with a subset of Run~2 data, covering the full kinematic range in $q^{2}$
\cite{LHCb:2023ssl,Chen:2024zot}.
Combining the Belle and LHCb measurements yields \cite{Iguro:2024hyk}
\begin{equation}
F_{L}(D^{*}) = 0.49 \pm 0.05\;, \label{FLDs}
\end{equation}
which is consistent with the SM prediction within $1\sigma$.

Motivated by these anomalies, numerous studies have explored possible NP
interpretations; see, \textit{e.g.},
Refs.~\cite{Azizi:2018axf,Azizi:2019aaf,Blanke:2018yud,Blanke:2019qrx,Huang:2018nnq,
Alok:2019uqc,Sahoo:2019hbu,Shi:2019gxi,Bardhan:2019ljo,Fedele:2022iib,
Asadi:2019xrc,Murgui:2019czp,Mandal:2020htr,Cheung:2020sbq,Colangelo:2020vhu,
Arslan:2023wgk,Yasmeen:2024cki, Huang:2025kof, Tang:2022nqm}.
In many of these analyses, dimension-six operators involving only left-handed (LH)
neutrinos in $b \to c \tau \bar{\nu}_{\tau}$ transitions were considered.
More general scenarios, including right-handed (RH) neutrinos and/or RH quark currents
within the model-independent weak effective Hamiltonian (WEH), have also been investigated;
see, \textit{e.g.},
Refs.~\cite{Greljo:2018ogz,Azatov:2018kzb,Heeck:2018ntp,Babu:2018vrl,He:2017bft,
Gomez:2019xfw,Alguero:2020ukk,Dutta:2013qaa,Dutta:2017xmj,Dutta:2017wpq,
Dutta:2018jxz}.
By performing global fits to the available $b \to c \tau \bar{\nu}_{\tau}$ data,
constraints on the NP Wilson coefficients (WCs) $C_{O_i}$, with
$i = V_{L,R},\, S_{L,R},\, T$, were obtained in
Refs.~\cite{Freytsis:2015qca,Alok:2017qsi,Arslan:2025zph}.
Following updated data releases from Belle and LHCb, revised constraints based on the
2024 HFLAV averages \cite{HFLAV:2024link} and their implications for various FCCC decays
were presented in Ref.~\cite{Arslan:2025zph}.

In light of the evolving experimental situation, particularly the updated 2025 HFLAV
averages for $R_{\tau/\mu,e}(D)$, $R_{\tau/\mu,e}(D^{*})$, as well as the updated measurements of the observables
$F_{L}(D^{*})$, $P_{\tau}(D^{*})$, $R_{\tau/\ell}(\Lambda_{c})$, and
$R_{\tau/\mu}(J/\psi)$ (see
Eqs.~\eqref{Exp-RJpsiLC}–\eqref{FLDs}) \cite{HFLAV:2025link},
it is timely to revisit and update the theoretical analyses.
As a first step, following
Refs.~\cite{Alok:2017qsi,Arslan:2025zph}, we perform an updated global fit  within the framework of the
model-independent WEH, assuming left-handed neutrinos and real NP WCs,
and incorporating the latest constraints on
$R_{\tau/\mu,e}(D)$, $R_{\tau/\mu,e}(D^{*})$, $F_{L}(D^{*})$, and $P_{\tau}(D^{*})$.

The baryonic decay $\Lambda^0_b \to \Lambda^+_c \tau^- \bar{\nu}_{\tau}$ is currently consistent
with SM expectations within experimental uncertainties.
Nevertheless, a potential window for NP effects remains, particularly toward the lower
end of the experimentally allowed ranges.
Unlike the mesonic transitions $B \to D$ and $B \to D^{*}$, both the initial and final
states in the baryonic decay, $\Lambda_b$ and $\Lambda_c$, carry spin-$\tfrac{1}{2}$.
As a result, all effective operators contribute to the decay
$\Lambda_b^0 \to \Lambda_c^+ \tau^- \bar{\nu}_{\tau}$.
These baryonic transitions exhibit kinematic structures and form-factor dependencies
distinct from their mesonic counterparts, thereby providing complementary sensitivity to
the underlying $b \to c \tau \bar{\nu}_{\tau}$ dynamics.
The primary limitation arises from the comparatively less precise knowledge of the
$\Lambda_b \to \Lambda_c$ form factors relative to those in $B \to (D,D^{*})$ decays.

The decay $\Lambda^0_b \to \Lambda_c \tau^- \bar{\nu}_{\tau}$ has been studied extensively
within the SM and in a variety of NP scenarios, together with dedicated calculations of
the relevant form factors; see, \textit{e.g.},
Refs.~\cite{Detmold:2015aaa,Azizi:2018axf,Bernlochner:2018kxh,Gutsche:2015rrt,
Gutsche:2015mxa,Shivashankara:2015cta,Dutta:2015ueb,Faustov:2016pal,
Li:2016pdv,Celis:2016azn,Datta:2017aue}.
An angular analysis of the cascade decay
$\Lambda_b^0 \to \Lambda_c^+(\to \Lambda^0 \pi^+) \tau^- \bar{\nu}_{\tau}$
requires knowledge of the polar and azimuthal angles of the $\tau^-$ lepton
\cite{Gutsche:2015mxa,Shivashankara:2015cta,Dutta:2015ueb,Faustov:2016pal,
Li:2016pdv,Celis:2016azn,Datta:2017aue}.
Since the $\tau^-$ is accompanied by an undetected neutrino, these angles cannot be
reconstructed precisely, rendering the corresponding angular distributions inaccessible
experimentally \cite{Bhattacharya:2020lfm}.
A viable alternative is to consider the subsequent decay
$\tau^- \to \pi^- \nu_{\tau}$, leading to the full cascade process $
\Lambda_b^0 \to \Lambda_c^+(\to \Lambda^0 \pi^+)\,
\tau^-(\to \pi^- \nu_{\tau})\,\bar{\nu}_{\tau}
$.
This channel has been analyzed assuming an unpolarized $\Lambda_b$ initial state and
including all possible Lorentz structures of the NP effective operators in \cite{Hu:2020axt}.
In this case, the three-momenta of the final-state particles
$\Lambda^0$, $\pi^+$, and $\pi^-$ are experimentally accessible.
After integrating over the relevant kinematic and angular variables, several observables,
such as the $\Lambda_c$ polarization $P_{\Lambda_c}(q^2)$ and the forward--backward
asymmetry of the $\pi^-$, $A_{FB}(q^2)$, can be extracted.

In this work, we employ the latest form factors for the
$\Lambda_b \to \Lambda_c$ transition calculated using lattice QCD
\cite{Bernlochner:2018kxh} and perform a detailed analysis of the angular
structure of the five-fold differential decay
$\Lambda_b^0 \to \Lambda_c^+(\to \Lambda^0 \pi^+)\,
\tau^-(\to \pi^- \nu_\tau)\bar{\nu}_\tau$ within the framework of the
model-independent WEH, assuming left-handed neutrinos.
In particular, we study the full set of angular coefficients
$\mathcal{K}_i$, which are linearly related to the helicity amplitudes
and therefore provide a cleaner separation of possible new-physics (NP)
effects compared to inclusive observables such as the LFU ratio
$R_{\tau/\ell}(\Lambda_{c})$.
To quantify the impact of NP in this decay, we use the allowed NP
parameter space obtained from constraints on
$R_{\tau/\mu,e}(D)$, $R_{\tau/\mu,e}(D^{*})$, $F_{L}(D^{*})$, and $P_{\tau}(D^{*})$ using latest HFLAV averages
\cite{HFLAV:2025link} and compare our results with the corresponding
SM predictions and available experimental measurements.
We find that NP scenarios capable of explaining the anomalies observed in
$\bar{B} \to D^{(*)}\tau^- \bar{\nu}_\tau$ decays can also induce sizable
effects in several observables of the baryonic decay
$\Lambda_b^0 \to \Lambda_c^+(\to \Lambda^0 \pi^+)\,
\tau^-(\to \pi^- \nu_\tau)\bar{\nu}_\tau$.

The main benchmarks of the present analysis are summarized as follows:
\begin{itemize}
    \item We incorporate the HFLAV Spring~2025 averages for
    $R_{\tau/\mu,e}(D^{(*)})$, which indicate an overall discrepancy of
    approximately $3.8\sigma$ from the SM, with a correlation coefficient
    of $-0.39$.
    \item We include constraints from the branching ratio
    $\mathcal{B}(B_c \to \tau \bar{\nu}_\tau)$, imposing upper bounds of
    $60\%$, $30\%$, and $10\%$, together with current LHC collider limits
    based on $\tau^\pm\nu_\tau$ searches at an integrated luminosity of
    $139~\mathrm{fb}^{-1}$, evaluated at the scale $\mu = m_b$.
    \item In contrast to earlier studies favoring the
    $(C_{V_L},C_{S_R})$ and $(C_{V_L},C_{S_L}=-4C_T)$ solutions with high
    $p$-values (approximately $87\%$ and $85\%$) and a strong
    $\sim4.2\sigma$ pull from the SM, our updated analysis reveals a
    distinct NP preference driven by $(\Re[C_{S_L}=4C_T],\Im[C_{S_L}=4C_T])$
    interactions.
    \item We demonstrate that the angular coefficients
    $\mathcal{K}_{1c}$, $\mathcal{K}_{2ss}$, $\mathcal{K}_{2cc}$, and
    $\mathcal{K}_{4s}$ are the most sensitive to NP effects, providing a
    novel strategy to identify helicity-driven signatures of physics
    beyond the SM.
    \item We find that the observed direct--inverse correlation patterns
    between real $(C_{S_L},C_{S_R})$ and complex
    $(\Re[C_{S_L}=4C_T],\,\Im[C_{S_L}=4C_T])$ NP scenarios
    offer a distinctive diagnostic tool for disentangling different NP
    contributions.
    \item We highlight that baryonic angular correlations, in contrast to
    mesonic ones, exhibit rich interference patterns that allow for the
    isolation of both CP-conserving and CP-violating effects.
\end{itemize}

The paper is organized as follows.
In Sec.~\ref{sec2}, we introduce the WEH,
including the SM operators and possible NP vector, scalar, and tensor
interactions.
For completeness, analytic expressions for $R_{\tau/\mu,e}(D^{(*)})$, $P_\tau(D^{(*)})$, and $F_L(D^*)$,
expressed in terms
of the NP Wilson coefficients, are collected in
Appendix~\ref{AppendixA}.
In Sec.~\ref{subsec3}, we perform a global fit to the latest experimental
data to determine the allowed parameter space for NP Wilson
coefficients, taking into account constraints from
$\mathcal{B}(B_c \to \tau \bar{\nu}_\tau)$ and collider bounds.
Section~\ref{secIII} presents the transversality amplitudes and angular
coefficients for the decay
$\Lambda_b^0 \to \Lambda_c^+(\to \Lambda^0 \pi^+)\,
\tau^-(\to \pi^- \nu_\tau)\bar{\nu}_\tau$, together with the five-fold
differential decay distribution and related observables.
A phenomenological analysis based on lattice-QCD form factors is discussed in Sec.~\ref{Num-anlaysis}. The
correlations among the various observables are discussed in
Sec.~\ref{correlation}, and our conclusions are summarized in
Sec.~\ref{sec6}.
Additional technical details concerning
derivation of the five-fold $\Lambda_b^0 \to \Lambda_c^+(\to \Lambda^0 \pi^+)\,
\tau^-(\to \pi^- \nu_\tau)\bar{\nu}_\tau$ decay
observables in terms of helicity amplitudes are provided in Appendix \ref{obsRLc}.

\section{Theoretical Framework and Bounds on the NP parameter space}\label{sec2}
\subsection{Weak Effective Hamiltonian (WEH)}

We consider the dimension-six semileptonic operators contributing at tree level to the WEH for the transition
$b \to c \tau^- \bar{\nu}_\tau$. Matching these operators onto the Standard Model Effective Field Theory (SMEFT) yields the relations among the corresponding
WCs. The most general WEH for $b \to c \tau^- \bar{\nu}_\tau$, including all Lorentz-invariant structures, is given by
\cite{Freytsis:2015qca,Alok:2017qsi}
\begin{equation}
H_{\text{eff}}
=\frac{4G_{F}V_{cb}}{\sqrt{2}}
\left[
\left(C_{V_{L}}\right)_{\text{SM}}\mathcal{O}_{V_{L}}
+\frac{\sqrt{2}}{4G_{F}V_{cb}}\frac{1}{\Lambda^{2}}
\sum_{i} C_{i}\mathcal{O}_{i}
\right],
\label{weh}
\end{equation}
where $G_{F}$ is the Fermi constant, $V_{cb}$ the CKM matrix element, and
$P_{R,L}=(1\pm\gamma_{5})/2$ are the chiral projectors. The SM contribution is normalized to
$(C_{V_{L}})_{\text{SM}}=1$.
The NP operators $\mathcal{O}_{i}$ with
$i=V_{L},V_{R},S_{L},S_{R},T$ are \cite{Buchmuller:1985jz,Grzadkowski:2010es,Aebischer:2015fzz}
\begin{align}
\mathcal{O}_{V_{L}} &= (\bar{c}\gamma^{\mu}P_{L}b)(\bar{\tau}\gamma_{\mu}P_{L}\nu), \qquad
\mathcal{O}_{V_{R}} = (\bar{c}\gamma^{\mu}P_{R}b)(\bar{\tau}\gamma_{\mu}P_{L}\nu), \nonumber \\
\mathcal{O}_{S_{L}} &= (\bar{c}P_{L}b)(\bar{\tau}P_{L}\nu), \qquad\qquad
\mathcal{O}_{S_{R}} = (\bar{c}P_{R}b)(\bar{\tau}P_{L}\nu), \nonumber \\
\mathcal{O}_{T} &= (\bar{c}\sigma^{\mu\nu}P_{L}b)(\bar{\tau}\sigma_{\mu\nu}P_{L}\nu).
\label{eq2}
\end{align}

The relevant low-energy scale for $b \to c \tau \bar{\nu}$ transitions is
$\mu_{b}=m_{b}$. The WCs at $\mu_{b}$ are related to those at the NP scale
$\Lambda=2~\text{TeV}$ through renormalization-group evolution
\cite{Gonzalez-Alonso:2017iyc,Blanke:2018yud}:
\begin{align}
\widetilde{C}_{V_{L}}(m_{b}) &= 1.12\,C_{V_{L}}^{\text{eff}}(2~\text{TeV}), \qquad
\widetilde{C}_{V_{R}}(m_{b}) = 1.07\,C_{V_{R}}^{\text{eff}}(2~\text{TeV}), \nonumber \\
\widetilde{C}_{S_{R}}(m_{b}) &= 2\,C_{S_{R}}^{\text{eff}}(2~\text{TeV}), \nonumber \\
\begin{pmatrix}
\widetilde{C}_{S_{L}}(m_{b}) \\
\widetilde{C}_{T}(m_{b})
\end{pmatrix}
&=
\begin{pmatrix}
1.91 & -0.38 \\
0 & 0.89
\end{pmatrix}
\begin{pmatrix}
C_{S_{L}}^{\text{eff}}(2~\text{TeV}) \\
C_{T}^{\text{eff}}(2~\text{TeV})
\end{pmatrix}.
\label{eq3}
\end{align}
For the WEH in Eq.~(\ref{weh}), the physical observables considered in this work can be
expressed in terms of the NP WCs evaluated at $\mu_{b}=m_{b}$. Their explicit expressions are
available in Refs.~\cite{Watanabe:2017mip,Iguro:2018vqb,Asadi:2018wea,Asadi:2018sym,
Ligeti:2016npd,Robinson:2018gza,Gomez:2019xfw,Cardozo:2020uol,Fedele:2022iib,
Mandal:2020htr,Kamali:2018bdp,Iguro:2022yzr} and are summarized in
Appendix~\ref{AppendixA}.

\subsection{Analysis of the parameter space of NP WCs}\label{subsec3}

In this section, we explore the parameter space of NP WCs using the most recent HFLAV data on FCCC transitions \cite{HFLAV:2025link}. To this end, we employ the fitting strategy developed in Ref.~\cite{Blanke:2018yud} and implemented in Ref.~\cite{Arslan:2025zph}. The analysis includes NP WCs that can be either real or complex.

Our global fit incorporates four observables,
\[
N_{\text{obs}}=4:\qquad
R_{\tau/\mu,e}(D),\;
R_{\tau/{\mu,e}}(D^{*}),\;
P_{\tau}(D^{*}),\;
F_{L}(D^{*}),
\]
and is performed under two scenarios:  
(i) a one-dimensional (1D) fit, where only one NP WC is switched on at a time, and  
(ii) a two-dimensional (2D) fit, where two NP WCs are allowed to be nonzero simultaneously. Accordingly, the number of fit parameters is $N_{\text{par}}=1\,(2)$ for the 1D (2D) case, yielding the number of degrees of freedom
\[
N_{\text{dof}} = N_{\text{obs}} - N_{\text{par}} = 3\,(2).
\]

Within this framework, we determine the best-fit points (BFPs), their $1\sigma$ and $2\sigma$ ranges, the minimum $\chi^2$, the corresponding $p$-values, and the SM pull for all 1D scenarios. These results are summarized in Table~\ref{1d-table}. The effects
$\mathcal{B}(B_c^- \to \tau^- \bar{\nu}_\tau) < 10\%,\,30\%,\,60\%$
constraints are consistently included in the fits.

From Table~\ref{1d-table}, we observe that the fit strongly prefers a NP contribution in $C_{V_L}$, yielding an excellent $p$-value of about $93\%$, corresponding to an approximately $4\sigma$ pull from the SM. Pure scalar scenarios involving $C_{S_L}$ or $C_{S_R}$ are moderately allowed but provide a poorer description of the data. In contrast, the relation $C_{S_L}=4C_T$ is strongly disfavored, with a $p$-value of only $\sim 0.05\%$.

\begin{figure}[H]
\centering{} \subfloat[]{\includegraphics[width=6cm, height=5cm]{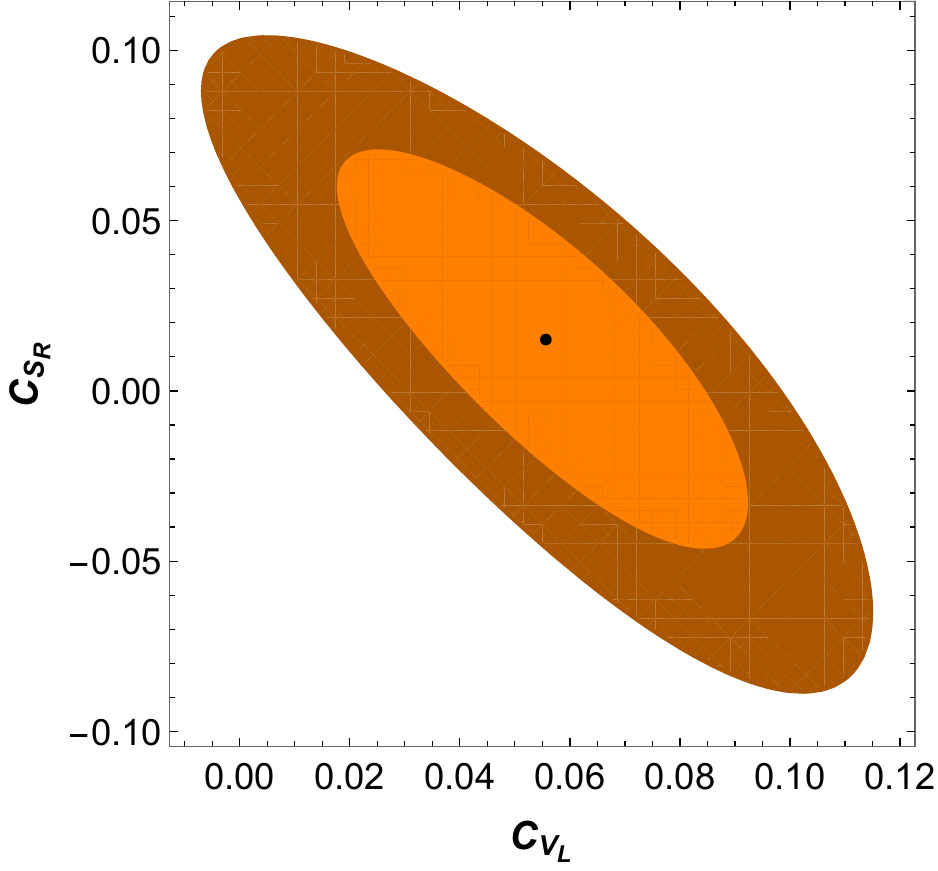}}\quad \subfloat[]{\includegraphics[width=6cm, height=5cm]{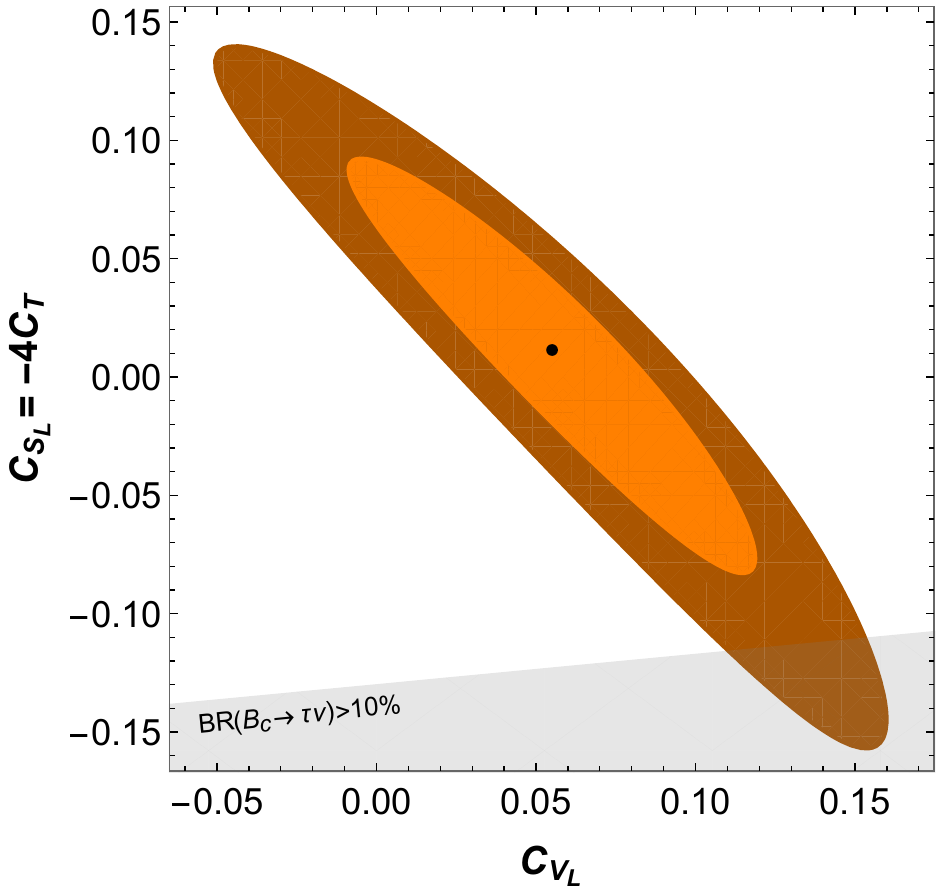}} 
\centering{} \subfloat[]{\includegraphics[width=6cm, height=5cm]{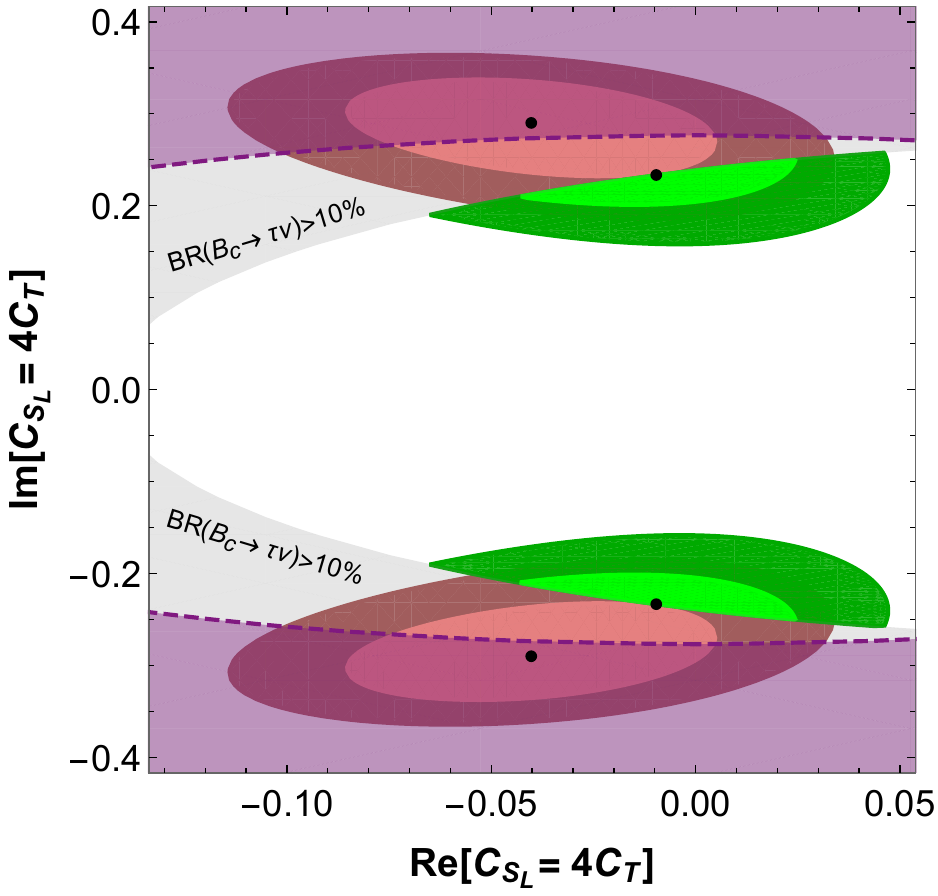}}\quad \subfloat[]{\includegraphics[width=6cm, height=5cm]{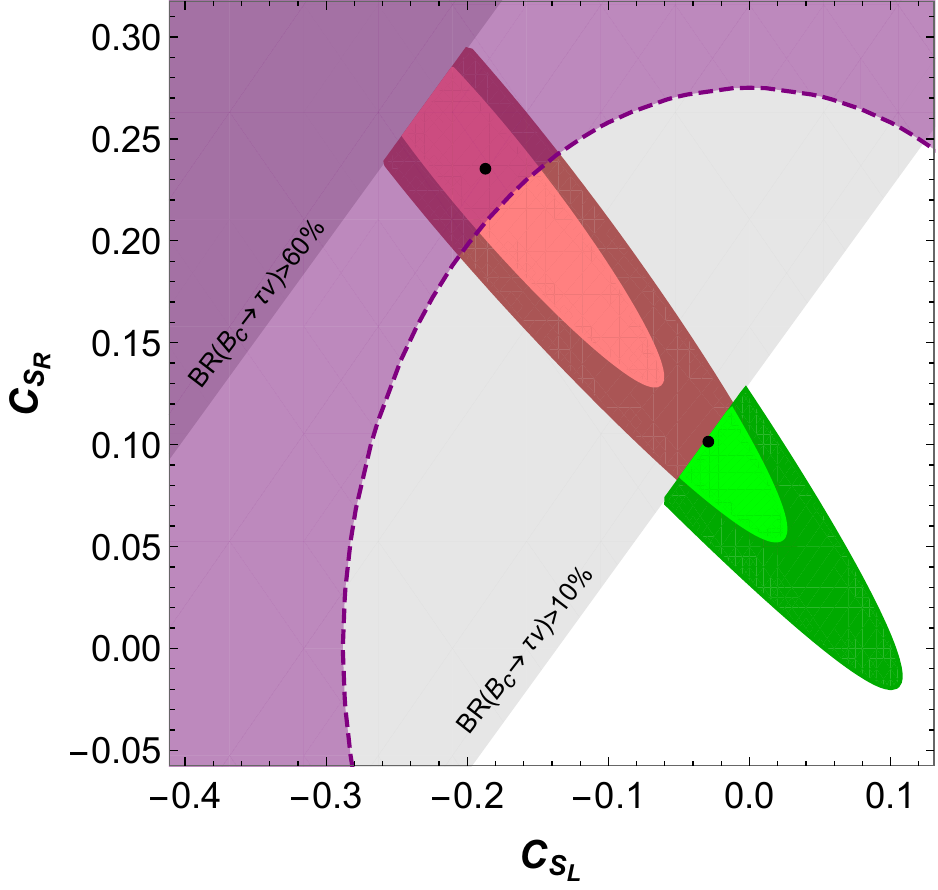}}

\caption{\label{s2d-fig}
Results of the fits for NP scenarios at the scale $\mu = 2\,\text{TeV}$. 
The light- and dark-gray shaded regions indicate the constraints 
$\mathcal{B}(B_c^- \to \tau^- \bar{\nu}_\tau) < 10\%$ and $< 60\%$, respectively. 
The light (dark) contours correspond to the $1\sigma$ ($2\sigma$) regions around the BFP, shown in black. 
In panels (a) and (b), the orange regions are unaffected by either branching-ratio constraint, while in panels (c) and (d) the red and green regions represent the $60\%$ and $10\%$ constraints, respectively. 
The purple-shaded area outside the dashed ellipse denotes the region excluded by collider bounds at an integrated luminosity of $139\,\text{fb}^{-1}$.
}
\end{figure}

For the two-dimensional fits, the corresponding BFPs, $\chi^2_{\text{min}}$, $p$-values, and SM pulls are listed in Table~\ref{2d-table-1}, while their $1\sigma$ and $2\sigma$ allowed regions are shown in Fig.~\ref{s2d-fig}. The orange contours represent parameter regions unconstrained by $\mathcal{B}(B_c^- \to \tau^- \bar{\nu}_\tau)$, whereas the red and green contours illustrate the impact of the $60\%$ and $10\%$ branching-ratio bounds, respectively. The light- and dark-gray bands denote the regions excluded by the $10\%$ and $60\%$ constraints, and any point lying within these bands is considered excluded.

Among the 2D scenarios, the $(C_{V_L},C_{S_R})$ and $(C_{V_L},C_{S_L}=-4C_T)$ solutions provide the best description of the data, with very high $p$-values of approximately $87\%$ and $85\%$, respectively, corresponding to a strong $\sim4.2\sigma$ pull from the SM. The scenario with $C_{S_L}=4C_T$ remains allowed but is clearly less favored, particularly under the stringent constraint
$\mathcal{B}(B_c^- \to \tau^- \bar{\nu}_\tau)<10\%$, where the fit quality deteriorates significantly, yielding a $p$-value of about $22\%$. The pure scalar combination $(C_{S_L},C_{S_R})$ remains viable for relaxed branching-ratio bounds ($<60\%$), but its preference decreases sharply as the constraint tightens, resulting in a $p$-value of only $\sim13\%$ for the $<10\%$ case, which
indicates an increasing tension with the data. 

\begin{table}[H]
\centering{}%
\renewcommand{\arraystretch}{1.1}
\begin{tabular}{|c|c|c|c|c|c|c|}
\toprule 
\multicolumn{7}{c}{( $\chi_{\text{SM}}^{2}=15.12$, $p-\text{value}=4.46\times10^{-3}$
)}\tabularnewline
\midrule
\midrule 
\hline\hline
WC & BFP & $\chi_{\text{min}}^{2}$ & $p-\text{value}$ $\%$ & $\text{pull}_{\text{SM}}$ & $1\sigma$-range & $2\sigma$-range\tabularnewline
\hline
\midrule
\midrule 
$C_{V_{L}}$ & $0.06$ & $0.43$ & $93.36$ & $4.19$ & $\left[0.04,0.09\right]$ & $\left[0.02,0.10\right]$\tabularnewline
 \hline
\midrule 
$C_{S_{R}}$ & $0.08$ & $5.24$ & $15.54$ & $3.58$ & $\left[0.04,0.12\right]$ & $\left[0.02,0.14\right]$\tabularnewline
 \hline
\midrule 
$C_{S_{L}}$ & $0.08$ & $9.19$ & $2.69$ & $2.97$ & $\left[0.03,0.12\right]$ & $\left[0.01,0.14\right]$\tabularnewline
 \hline
\midrule 
$C_{S_{L}}=4C_{T}$ & $0.02$ & $17.73$ & $0.05$ & $0.54$ & $\left[-0.04,0.07\right]$ & $\left[-0.08,0.10\right]$\tabularnewline
 \hline\hline
\bottomrule
\end{tabular}
\caption{\label{1d-table}
Results of the one-dimensional fits for real NP WCs. 
Shown are the BFPs, the minimum $\chi^2$, the $p$-value (in \%), the SM pull, and the corresponding $1\sigma$ and $2\sigma$ ranges of the WCs. 
The fits are performed under the constraints 
$\mathcal{B}(B_c^- \to \tau^- \bar{\nu}_\tau) < 60\%,\,30\%$, and $10$. 
We find that the results are insensitive to the choice among these three branching-ratio limits.
}
\end{table}

\begin{table}[H]
\centering{}%
\renewcommand{\arraystretch}{1.35}
\begin{tabular}{|c|c|c|c|c|c|}
\toprule 
\multicolumn{6}{c}{( $\chi_{\text{SM}}^{2}=15.12$, $p-\text{value}=4.46\times10^{-3}$
)}\tabularnewline
\hline 
WC & BR & BFP & $\chi_{min}^{2}$ & $p-value$ $\%$ & $pull_{SM}$\tabularnewline
\hline 
\hline 
$\left(C_{V_{L}},C_{S_{R}}\right)$ & - & $\left(0.06,0.01\right)$ & $0.28$ & $86.99$ & $4.21$\tabularnewline
\hline 
$\left(C_{V_{L}},C_{S_{L}}=-4C_{T}\right)$ & - & $\left(0.06,0.01\right)$ & \multirow{1}{*}{$0.32$} & \multirow{1}{*}{$85.04$} & \multirow{1}{*}{$4.21$}\tabularnewline
\hline 
\multirow{2}{*}{$\left(\Re\left[C_{S_{L}}=4C_{T}\right],\Im\left[C_{S_{L}}=4C_{T}\right]\right)$} & $\begin{array}{c}
<60\%\\
\&30\%
\end{array}$ & $\left(-0.04,-0.29\right)$ & $0.53$ & $76.63$ & $4.18$\tabularnewline
\cline{2-6} \cline{3-6} \cline{4-6} \cline{5-6} \cline{6-6} 
 & $<10\%$ & $\left(-0.01,-0.23\right)$ & $3.01$ & $22.19$ & $3.87$\tabularnewline
\hline 
\multirow{3}{*}{$\left(C_{S_{L}},C_{S_{R}}\right)$} & $<60\%$ & $\left(-0.19,0.24\right)$ & $0.54$ & $76.25$ & $4.18$\tabularnewline
\cline{2-6} \cline{3-6} \cline{4-6} \cline{5-6} \cline{6-6} 
 & $<30\%$ & $\left(-0.71,0.98\right)$ & $0.84$ & $65.62$ & $3.78$\tabularnewline
\cline{2-6} \cline{3-6} \cline{4-6} \cline{5-6} \cline{6-6} 
 & $<10\%$ & $\left(-0.03,0.10\right)$ & $4.06$ & $13.15$ & $3.74$\tabularnewline
\hline 
\end{tabular}\caption{\label{2d-table-1} The results of the two-dimensional fit for NP WCs, including BFP, $\chi_{\text{min}}^{2}$,\; $p-\text{value}$ $\%$,\; and $\text{pull}_{\text{SM}}$,\; of the corresponding WCs. These numbers are obtained by incorporating bounds on $\mathcal{B}\left(B_{c}^{-}\to\tau^{-}\bar{\nu}_{\tau}\right) <60\%,\;30\%,\;10\%$
for set of observables.}
\end{table}

We emphasize that our results are in agreement with Ref.~\cite{Blanke:2018yud} for the one-dimensional scenario, where only the $C_{V_L}$ operator provides the best fit ($p \simeq 93\%$, corresponding to an approximately $4\sigma$ pull from the SM). 
The scalar solutions are less favored, and the $C_{S_L}=4C_T$ scenario is strongly disfavored, confirming the suppression of scalar–tensor NP contributions. 

In contrast, for the two-dimensional fits, our conclusions differ from those of Ref.~\cite{Blanke:2018yud}. 
While the $(C_{S_L},C_{S_R})$ scenario was found to yield the highest $p$-value in that study, our analysis identifies it as the least favored among the two-parameter solutions. 
Instead, the $(C_{V_L},C_{S_R})$ and $(C_{V_L},C_{S_L}=-4C_T)$ scenarios provide the best fits, with $p$-values exceeding $85\%$ and an approximately $4\sigma$ pull from the SM. 
This inversion indicates that the updated dataset and the treatment of observables in our analysis strongly favor correlated vector–scalar or vector–tensor NP scenarios over purely scalar combinations.

Consequently, in the following sections, we focus our phenomenological discussion on the NP scenarios satisfying $\chi^2_{\mathrm{min}} \leq 1$.

\subsection{Impact of collider (LHC) bounds}\label{subcolb}
The current collider bounds on the NP WCs, based on the $\tau^{\pm}\nu$ search at $\mu=m_b$ with an integrated luminosity of $139\;\text{fb}^{-1}$, and the projected bounds for the HL-LHC with $1000\,(3000)\;\text{fb}^{-1}$, are \cite{Greljo:2018tzh, Faroughy:2016osc, Iguro:2018fni, Endo:2021lhi}:
\begin{equation}
\left|\widetilde{C}_{V_{L}}\right|<0.30\,(0.14),\quad
\left|\widetilde{C}_{V_{R}}\right|<0.32\,(0.15),\quad
\left|\widetilde{C}_{S_{L,R}}\right|<0.55\,(0.25),\quad
\left|\widetilde{C}_{T}\right|<0.15\,(0.07).
\end{equation}

Applying these bounds to various combinations of NP WCs, we illustrate their impact on the allowed parameter space with the purple-shaded ellipse in Fig.~\ref{s2d-fig}. 
In particular, the combinations $\left(\Re\!\left[C_{S_{L}}=4C_{T}\right], \Im\!\left[C_{S_{L}}=4C_{T}\right]\right)$ and $(C_{S_{L}}, C_{S_{R}})$ are now subject to stringent constraints, as the lower portion of the previously allowed region \cite{Hu:2020axt} is excluded. 
Furthermore, when imposing a $60\%$ branching ratio constraint, the regions corresponding to the BFPs are also excluded, resulting in a substantial reduction of the viable parameter space. 
In contrast, all other benchmark scenarios remain consistent with the current collider limits.

\section{Five-fold $\Lambda_b^0 \to \Lambda_c^+(\to \Lambda^0 \pi^+) \tau^-(\to \pi^- \nu_\tau)\bar{\nu}_\tau$ decay distribution and angular observables}
\label{secIII}

In this section, we present the analytical expressions for the angular
distribution of the decay
$\Lambda_b^0 \to \Lambda_c^+(\to \Lambda^0 \pi^+) \tau^-(\to \pi^- \nu_\tau)\bar{\nu}_\tau$.
The details of the calculation and the conventions adopted are provided
in Appendix~\ref{AppendixA}.

\subsection{Transversity amplitudes}

As an exclusive decay, the hadronic matrix elements of the vector and
axial-vector currents governing the $\Lambda_b \to \Lambda_c$ transition
can be parameterized in terms of six helicity form factors,
$F_{+}$, $F_{\perp}$, $F_{0}$, $G_{+}$, $G_{\perp}$, and $G_{0}$
\cite{Datta:2017aue}.
Using Ward identities for the $\Lambda_b \to \Lambda_c$ matrix elements,
the scalar and pseudoscalar current contributions can be expressed in
terms of $F_{0}$ and $G_{0}$, respectively. We used the numerical values of these form factors along with their different fit parameters from \cite{Detmold:2015aaa,Datta:2017aue}. 

In the absence of tensor operators, six independent transversity
amplitudes can be defined as \cite{Hu:2020axt}:
\begin{align}
\mathcal{A}_{\perp t} &=
\mathcal{A}_{\perp t}^{SP}
+ \frac{m_\tau}{\sqrt{q^{2}}}\,
\mathcal{A}_{\perp t}^{VA},
&
\mathcal{A}_{\parallel t} &=
\mathcal{A}_{\parallel t}^{SP}
+ \frac{m_\tau}{\sqrt{q^{2}}}\,
\mathcal{A}_{\parallel t}^{VA},\label{eq:11}
\\
\mathcal{A}_{\perp 1} &=
-2 F_{\perp}\sqrt{Q_{-}}
\left(1+C_{V_L}+C_{V_R}\right),
&
\mathcal{A}_{\parallel 1} &=
-2 G_{\perp}\sqrt{Q_{+}}
\left(-1-C_{V_L}+C_{V_R}\right),\label{eq:12}
\\
\mathcal{A}_{\perp 0} &=
F_{+}\sqrt{2Q_{-}}
\frac{m_1+m_2}{\sqrt{q^{2}}}
\left(1+C_{V_L}+C_{V_R}\right),
&
\mathcal{A}_{\parallel 0} &=
G_{+}\sqrt{2Q_{+}}
\frac{m_1-m_2}{\sqrt{q^{2}}}
\left(-1-C_{V_L}+C_{V_R}\right).\label{eq:13}
\end{align}
Here, $\perp$ and $\parallel$ denote the two transversity states.
The subscript $t$ corresponds to the time-like polarization of
the off-shell $\tau^-\bar{\nu}_\tau$ system, while the subscripts
$1$ and $0$ indicate the magnitude of the $z$-component of the total
angular momentum of the $\tau^-\bar{\nu}_\tau$ pair.
The kinematic factors are defined as
$Q_{\pm} = (m_1 \pm m_2)^2 - q^2$.

The time-like transversity amplitudes are given by
\begin{align}
\mathcal{A}_{\perp t}^{SP} &=
F_{0}\sqrt{2Q_{+}}
\frac{m_1-m_2}{m_b-m_c}
\left(C_{S_L}+C_{S_R}\right),\label{eq:14}
\\
\mathcal{A}_{\parallel t}^{SP} &=
- G_{0}\sqrt{2Q_{-}}
\frac{m_1+m_2}{m_b+m_c}
\left(-C_{S_L}+C_{S_R}\right),\label{eq:15}
\\
\mathcal{A}_{\perp t}^{VA} &=
F_{0}\sqrt{2Q_{+}}
\frac{m_1-m_2}{\sqrt{q^{2}}}
\left(1+C_{V_L}+C_{V_R}\right),\label{eq:16}
\\
\mathcal{A}_{\parallel t}^{VA} &=
G_{0}\sqrt{2Q_{-}}
\frac{m_1+m_2}{\sqrt{q^{2}}}
\left(-1-C_{V_L}+C_{V_R}\right).\label{eq:17}
\end{align}

When tensor operators are included, the matrix elements can be
parameterized by four additional helicity form factors,
$h_{+}$, $h_{\perp}$, $\tilde{h}_{+}$, and $\tilde{h}_{\perp}$.
This leads to four extra transversity amplitudes \cite{Hu:2020axt}:
\begin{align}
\mathcal{A}_{\perp 1}^{T} &=
4 h_{\perp}\sqrt{Q_{-}}
\frac{m_1+m_2}{\sqrt{q^{2}}} C_T,
&
\mathcal{A}_{\parallel 1}^{T} &=
4 \tilde{h}_{\perp}\sqrt{Q_{+}}
\frac{m_1-m_2}{\sqrt{q^{2}}} C_T,\label{eq:18}
\\
\mathcal{A}_{\perp 0}^{T} &=
-2 h_{+}\sqrt{2Q_{-}}\, C_T,
&
\mathcal{A}_{\parallel 0}^{T} &=
-2 \tilde{h}_{+}\sqrt{2Q_{+}}\, C_T.\label{eq:19}
\end{align}

The superscript $T$ indicates contributions arising exclusively from
tensor operators.

\subsection{Angular distribution}
The measurable angular distribution of the five-body decay
$\Lambda_{b}^{0}\to\Lambda_{c}^{+}(\to\Lambda^{0}\pi^{+})
\tau^{-}(\to\pi^{-}\nu_{\tau})\bar{\nu}_{\tau}$,
for an unpolarized $\Lambda_b^0$, is described by the invariant mass
squared of the $\tau^-\bar{\nu}_\tau$ system, $q^{2}$;
the helicity angle $\theta_{\Lambda}$ of the $\Lambda^{0}$ baryon in the
$\Lambda_{c}^{+}$ rest frame; and the energy $E_{\pi}$, polar angle
$\theta_{\pi}$, and azimuthal angle $\phi_{\pi}$ of the $\pi^{-}$ in the
$\tau^-\bar{\nu}_\tau$ center-of-mass frame.
The kinematic conventions are illustrated in Fig.~1 and detailed in
Appendix~A.

The five-fold differential decay rate can be written as
\begin{align}
\frac{d^{5}\Gamma}
{dq^{2}\, dE_{\pi}\, d\cos\theta_{\pi}\, d\phi_{\pi}\, d\cos\theta_{\Lambda}}
&=
\frac{G_{F}^{2}\,|V_{cb}|^{2}\,|q|\,(q^{2})^{3/2}\, m_{\tau}^{2}}
{2^{8}\pi^{4} m_{1}^{2}(m_{\tau}^{2}-m_{\pi}^{2})^{2}}
\nonumber\\
&\quad\times
\mathcal{B}(\Lambda_{c}\to\Lambda\pi^{+})\,
\mathcal{B}(\tau\to\pi^{-}\nu_{\tau})\,
\mathcal{K}(q^{2},E_{\pi},\cos\theta_{\Lambda},
\cos\theta_{\pi},\phi_{\pi}),
\end{align}
where $|\textbf{q}|=\sqrt{Q_{+}Q_{-}}/(2m_{1})$ is the magnitude of the
$\Lambda_{c}$ three-momentum in the $\Lambda_{b}$ rest frame, and
$m_{1}$ denotes the $\Lambda_{b}$ mass. In terms of angular coefficients, 
the angular function $\mathcal{K}$ can be expressed as \cite{Hu:2020axt}:
\begin{align}
\mathcal{K}
&=
\sum_{i=1}^{10}
\mathcal{K}_{i}(q^{2},E_{\pi})\,
\Omega_{i}(\cos\theta_{\Lambda},\cos\theta_{\pi},\phi_{\pi})
\nonumber\\
&=
\left(\mathcal{K}_{1ss}\sin^{2}\theta_{\pi}
+\mathcal{K}_{1cc}\cos^{2}\theta_{\pi}
+\mathcal{K}_{1c}\cos\theta_{\pi}\right)+
\left(\mathcal{K}_{2ss}\sin^{2}\theta_{\pi}
+\mathcal{K}_{2cc}\cos^{2}\theta_{\pi}
+\mathcal{K}_{2c}\cos\theta_{\pi}\right)\cos\theta_{\Lambda}
\nonumber\\
&\quad+
\left(\mathcal{K}_{3sc}\sin\theta_{\pi}\cos\theta_{\pi}
+\mathcal{K}_{3s}\sin\theta_{\pi}\right)
\sin\theta_{\Lambda}\sin\phi_{\pi}+
\left(\mathcal{K}_{4sc}\sin\theta_{\pi}\cos\theta_{\pi}
+\mathcal{K}_{4s}\sin\theta_{\pi}\right)
\sin\theta_{\Lambda}\cos\phi_{\pi}\;.
\end{align}

The ten angular observables $\mathcal{K}_{i}(q^{2},E_{\pi})$
can be written entirely in terms of the transversity amplitudes,
the dimensionless kinematic factors defined in
Eqs.~(\ref{eq:11})--(\ref{eq:19}).
Their explicit expressions are
\begin{align}
\mathcal{K}_{1ss} & =S_{t}\left|\mathcal{A}_{\perp t}\right|^{2}+\left(S_{1}-S_{3}\right)\left|\mathcal{A}_{\perp1}\right|^{2}+\left(S_{1}+S_{3}\right)\left|\mathcal{A}_{\perp0}\right|^{2}+\left(S_{1}^{T}-S_{3}^{T}\right)\left|\mathcal{A}_{\perp1}^{T}\right|^{2}+\left(S_{1}^{T}+S_{3}^{T}\right)\left|\mathcal{A}_{\perp0}^{T}\right|^{2}\notag\\
 & +\Re\left[\left(R_{1}-R_{3}\right)\mathcal{A}_{\perp1}\mathcal{A}_{\perp1}^{T*}+\left(R_{1}+R_{3}\right)\mathcal{A}_{\perp0}\mathcal{A}_{\perp0}^{T*}\right]+\left(\perp\leftrightarrow\parallel\right),\label{eq:22}\\
\mathcal{K}_{1cc} & =S_{t}\left|\mathcal{A}_{\perp t}\right|^{2}+\left(S_{1}+S_{3}\right)\left|\mathcal{A}_{\perp1}\right|^{2}+\left(S_{1}-3S_{3}\right)\left|\mathcal{A}_{\perp0}\right|^{2}+\left(S_{1}^{T}+S_{3}^{T}\right)\left|\mathcal{A}_{\perp1}^{T}\right|^{2}+\left(S_{1}^{T}-3S_{3}^{T}\right)\left|\mathcal{A}_{\perp0}^{T}\right|^{2}\notag\\
 & +\Re\left[\left(R_{1}+R_{3}\right)\mathcal{A}_{\perp1}\mathcal{A}_{\perp1}^{T*}+\left(R_{1}-3R_{3}\right)\mathcal{A}_{\perp0}\mathcal{A}_{\perp0}^{T*}\right]+\left(\perp\leftrightarrow\parallel\right),\label{eq:23}\\
\mathcal{K}_{1c} & =2\Re\left[S_{2}\mathcal{A}_{\perp1}\mathcal{A}_{\parallel1}^{*}+S_{2}^{T}\mathcal{A}_{\perp1}^{T}\mathcal{A}_{\parallel1}^{T*}\right]+\Re\left[R_{2}\mathcal{A}_{\perp1}\mathcal{A}_{\parallel1}^{T*}-\sqrt{2}R_{t}\mathcal{A}_{\perp t}\mathcal{A}_{\perp0}^{*}-\sqrt{2}R_{t}^{T}\mathcal{A}_{\perp t}\mathcal{A}_{\perp0}^{T*}+\left(\perp\leftrightarrow\parallel\right)\right],\label{eq:24}\\
\mathcal{K}_{2ss} & =2\alpha_{\Lambda_{c}}\Re\left[S_{t}\mathcal{A}_{\perp t}\mathcal{A}_{\parallel t}^{*}+\left(S_{1}-S_{3}\right)\mathcal{A}_{\perp1}\mathcal{A}_{\parallel1}^{*}+\left(S_{1}+S_{3}\right)\mathcal{A}_{\perp0}\mathcal{A}_{\parallel0}^{*}+\left(S_{1}^{T}-S_{3}^{T}\right)\mathcal{A}_{\perp1}^{T}\mathcal{A}_{\parallel1}^{T*}\right.\notag\\
 & \left.+\left(S_{1}^{T}+S_{3}^{T}\right)\mathcal{A}_{\perp0}^{T}\mathcal{A}_{\parallel0}^{T*}\right]+\alpha_{\Lambda_{c}}\Re\left[\left(R_{1}+R_{3}\right)\mathcal{A}_{\perp0}\mathcal{A}_{\parallel0}^{T*}+\left(R_{1}-R_{3}\right)\mathcal{A}_{\perp1}\mathcal{A}_{\parallel1}^{T*}+\left(\perp\leftrightarrow\parallel\right)\right],\label{eq:25}\\
\mathcal{K}_{2cc} & =2\alpha_{\Lambda_{c}}\Re\left[S_{t}\mathcal{A}_{\perp t}\mathcal{A}_{\parallel t}^{*}+\left(S_{1}+S_{3}\right)\mathcal{A}_{\perp1}\mathcal{A}_{\parallel1}^{*}+\left(S_{1}-3S\right)\mathcal{A}_{\perp0}\mathcal{A}_{\parallel0}^{*}+\left(S_{1}^{T}+S_{3}^{T}\right)\mathcal{A}_{\perp1}^{T}\mathcal{A}_{\parallel1}^{T*}\right.\notag\\
 & \left.+\left(S_{1}^{T}-3S_{3}^{T}\right)\mathcal{A}_{\perp0}^{T}\mathcal{A}_{\parallel0}^{T*}\right]+\alpha_{\Lambda_{c}}\Re\left[\left(R_{1}+R_{3}\right)\mathcal{A}_{\perp1}\mathcal{A}_{\parallel1}^{T*}+\left(R_{1}-3R_{3}\right)\mathcal{A}_{\perp0}\mathcal{A}_{\parallel0}^{T*}+\left(\perp\leftrightarrow\parallel\right)\right],\label{eq:26}\\
\mathcal{K}_{2c} & =\alpha_{\Lambda_{c}}\left[S_{2}\left|\mathcal{A}_{\perp1}\right|^{2}+S_{2}^{T}\left|\mathcal{A}_{\perp1}^{T}\right|^{2}\right]+\alpha_{\Lambda_{c}}\Re\left[R_{2}\mathcal{A}_{\perp1}\mathcal{A}_{\perp1}^{T*}-\sqrt{2}R_{t}\mathcal{A}_{\perp t}\mathcal{A}_{\parallel0}^{*}-\sqrt{2}R_{t}^{T}\mathcal{A}_{\perp t}\mathcal{A}_{\parallel0}^{T*}\right]+\left(\perp\leftrightarrow\parallel\right),\label{eq:27}\\
\mathcal{K}_{3sc} & =2\sqrt{2}\alpha_{\Lambda_{c}}\Im\left[2S_{3}\mathcal{A}_{\perp1}\mathcal{A}_{\perp0}^{*}-2S_{3}^{T}\mathcal{A}_{\perp1}^{T}\mathcal{A}_{\perp0}^{T*}+R_{3}\mathcal{A}_{\perp1}\mathcal{A}_{\perp0}^{T*}-R_{3}\mathcal{A}_{\perp0}\mathcal{A}_{\perp1}^{T*}\right]-\left(\perp\leftrightarrow\parallel\right),\label{eq:28}\\
\mathcal{K}_{3s} & =-\frac{\alpha_{\Lambda_{c}}}{\sqrt{2}}\Im\left[\sqrt{2}R_{t}\mathcal{A}_{\perp t}\mathcal{A}_{\perp1}^{*}+\sqrt{2}R_{t}^{T}\mathcal{A}_{\perp t}\mathcal{A}_{\perp1}^{T*}+2S_{2}\mathcal{A}_{\perp1}\mathcal{A}_{\parallel0}^{*}+2S_{2}^{T}\mathcal{A}_{\perp1}^{T}\mathcal{A}_{\parallel0}^{T*}\right.\notag\\
 & \left.+R_{2}\mathcal{A}_{\perp1}\mathcal{A}_{\parallel0}^{T*}-R_{2}\mathcal{A}_{\perp0}\mathcal{A}_{\parallel1}^{T*}-\left(\perp\leftrightarrow\parallel\right)\right],\label{eq:29}\\
\mathcal{K}_{4sc} & =2\sqrt{2}\alpha_{\Lambda_{c}}\Re\left[R_{3}\mathcal{A}_{\perp0}\mathcal{A}_{\parallel1}^{T*}-R_{3}\mathcal{A}_{\perp1}\mathcal{A}_{\parallel0}^{T*}-2S_{3}\mathcal{A}_{\perp1}\mathcal{A}_{\parallel0}^{*}-2S_{3}^{T}\mathcal{A}_{\perp1}^{T}\mathcal{A}_{\parallel0}^{T*}\right]-\left(\perp\leftrightarrow\parallel\right),\label{eq:30}\\
\mathcal{K}_{4s} & =\frac{\alpha_{\Lambda_{c}}}{\sqrt{2}}\Re\left[\sqrt{2}R_{t}\mathcal{A}_{\perp t}\mathcal{A}_{\parallel1}^{*}+\sqrt{2}R_{t}^{T}\mathcal{A}_{\perp t}\mathcal{A}_{\parallel1}^{T*}+2S_{2}\mathcal{A}_{\perp1}\mathcal{A}_{\perp0}^{*}+2S_{2}^{T}\mathcal{A}_{\perp1}^{T}\mathcal{A}_{\perp0}^{T*}\right.\notag\\
 & \left.+R_{2}\mathcal{A}_{\perp1}\mathcal{A}_{\perp0}^{T*}+R_{2}\mathcal{A}_{\perp0}\mathcal{A}_{\perp1}^{T*}-\left(\perp\leftrightarrow\parallel\right)\right],\label{eq:31}
\end{align}
The decay asymmetry parameter $\alpha_{\Lambda_c}$ is defined via the
angular distribution of the $\Lambda^0$ baryon in the $\Lambda_c^+$ rest frame,
\begin{equation}
\frac{1}{\Gamma_{\Lambda_c}}
\frac{d\Gamma_{\Lambda_c}}{d\cos\theta_{\Lambda}}
= \frac{1}{2}
\left(1+\alpha_{\Lambda_c}\cos\theta_{\Lambda}\right),
\end{equation}
where $\theta_{\Lambda}$ is the angle between the $\Lambda^0$ momentum
and the polarization direction of the $\Lambda_c^+$.

\section{Numerical Analysis }\label{Num-anlaysis}
In this section, we analyze the above-mentioned angular observables within the SM and in the presence of NP, incorporating the constraints on the allowed parameter space of various NP scenarios derived from the latest HFLAV results discussed in Sec.~\ref{subsec3}, together with the collider bounds presented in Sec.~\ref{subcolb}.

\begin{figure}[H]
\centering 
\begin{subfigure}[b]{0.32\textwidth}
\centering 
\includegraphics[width=5.6cm, height=3.5cm]{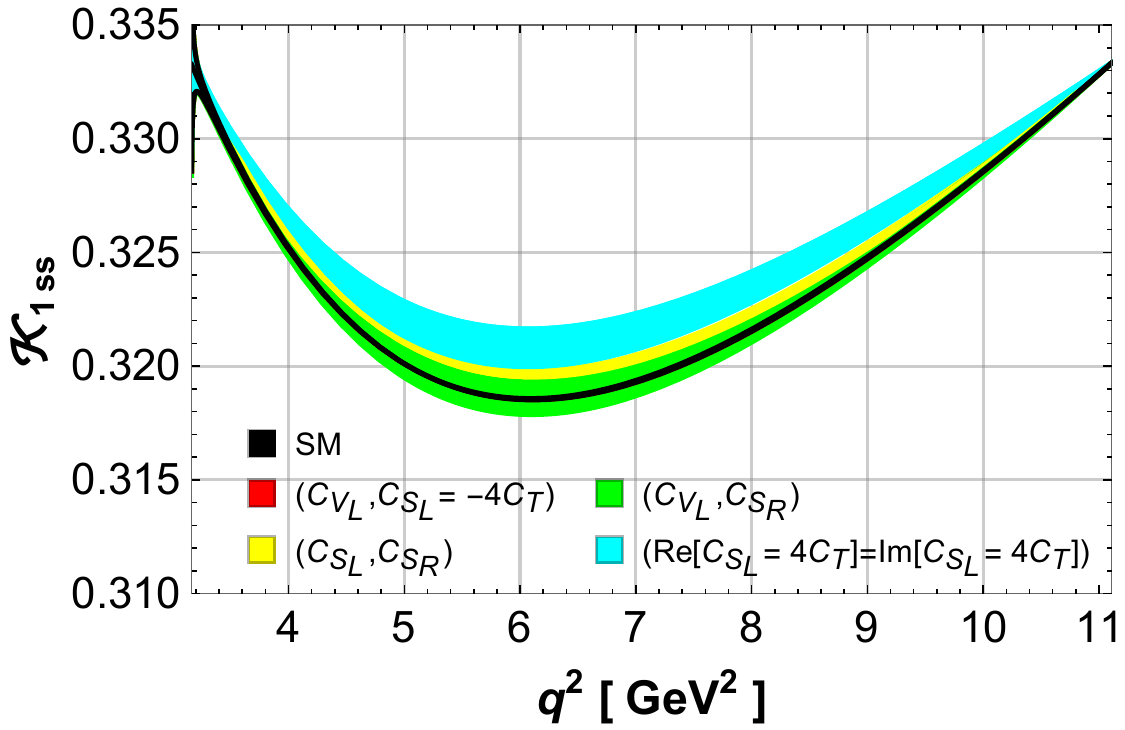}
\caption{}
\end{subfigure}
\begin{subfigure}[b]{0.32\textwidth}
\centering 
\includegraphics[width=5.6cm, height=3.5cm]{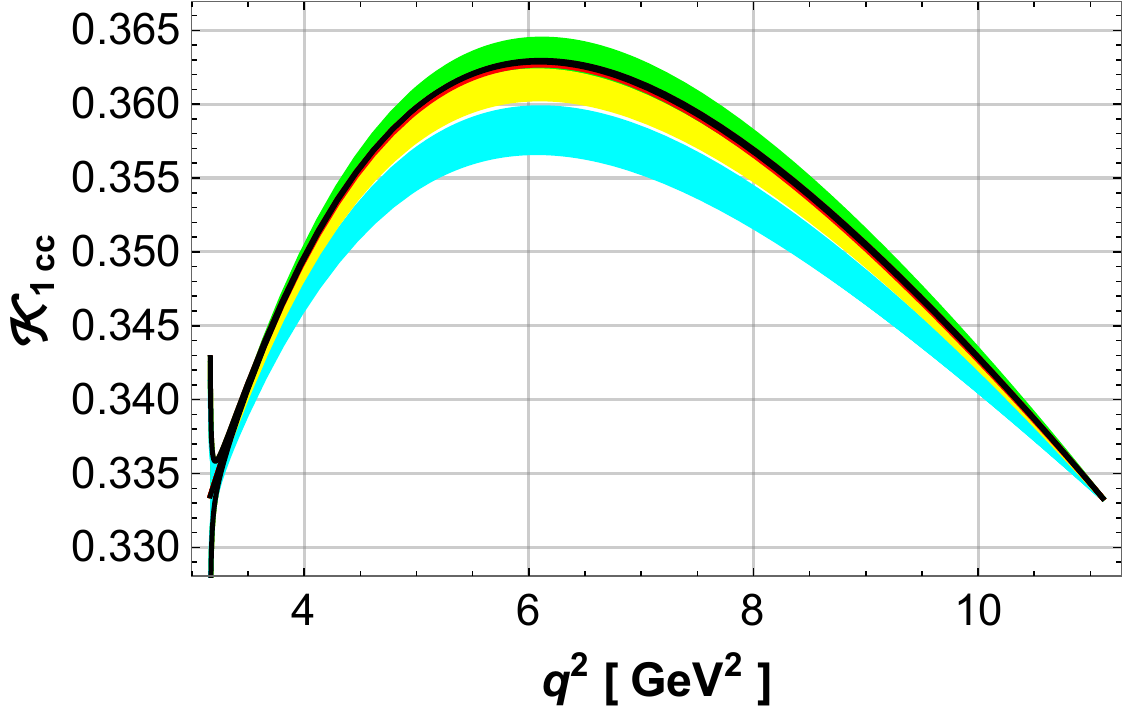} 
\caption{}
\end{subfigure}
\begin{subfigure}[b]{0.32\textwidth}
\centering 
\includegraphics[width=5.6cm, height=3.5cm]{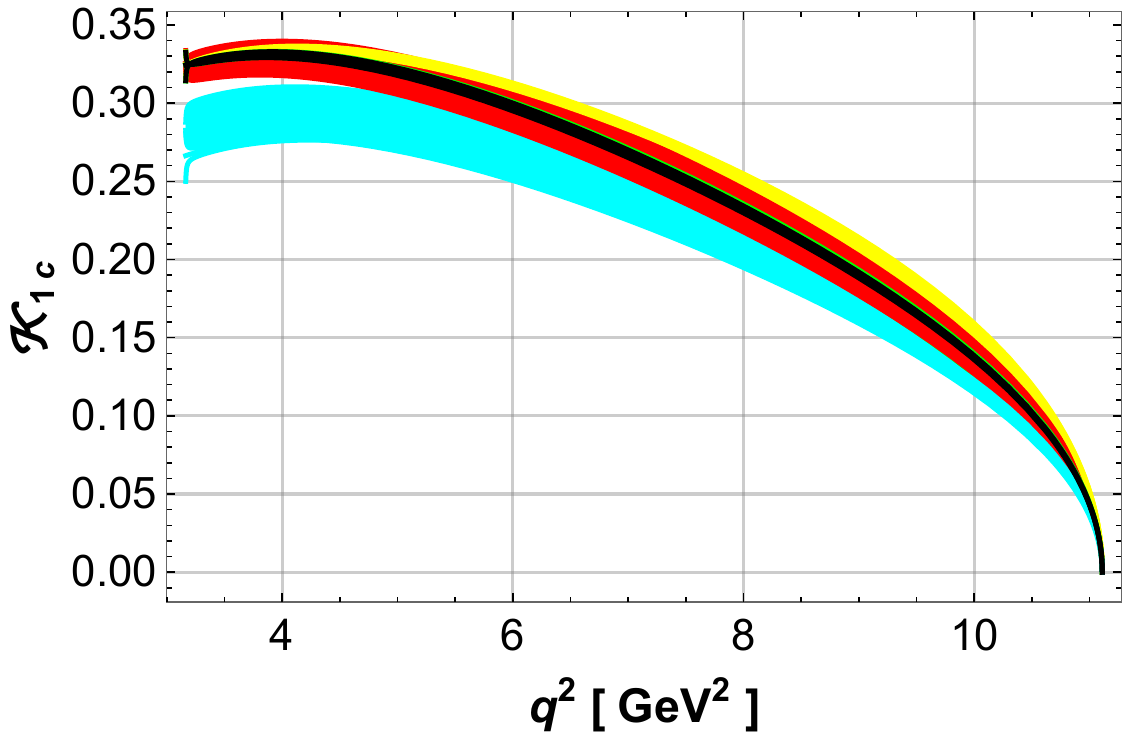}
\caption{}
\end{subfigure}
\begin{subfigure}[b]{0.32\textwidth}
\centering 
\includegraphics[width=5.6cm, height=3.5cm]{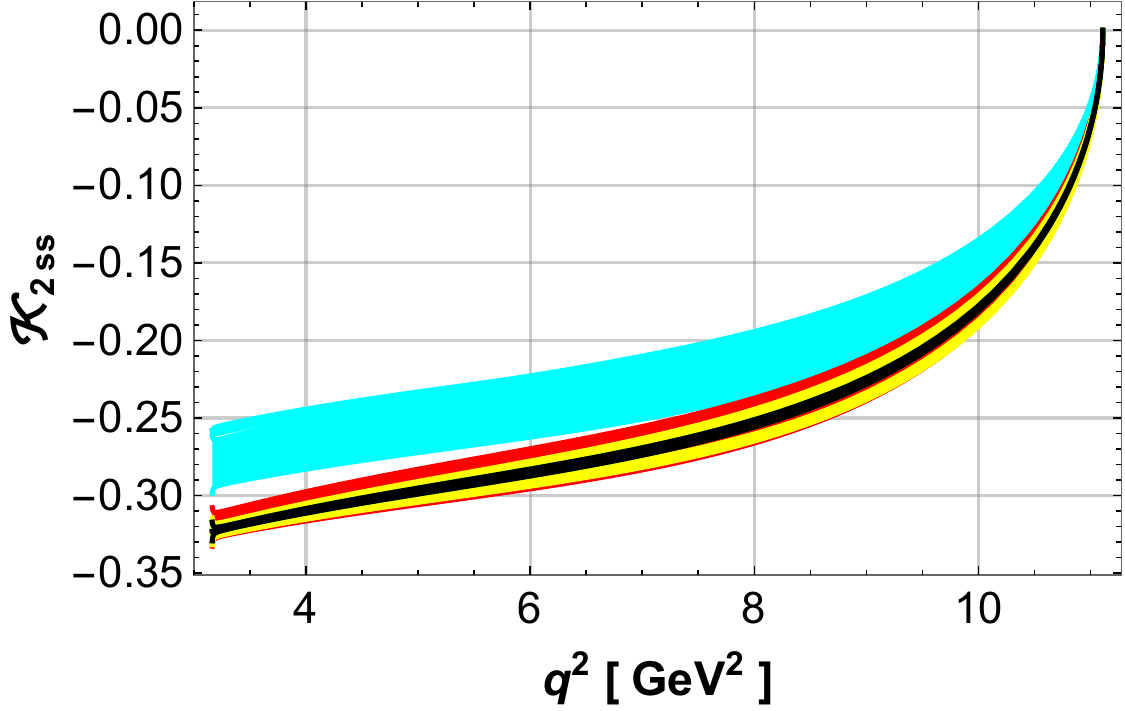}
\caption{}
\end{subfigure}
\begin{subfigure}[b]{0.32\textwidth}
\centering 
\includegraphics[width=5.6cm, height=3.5cm]{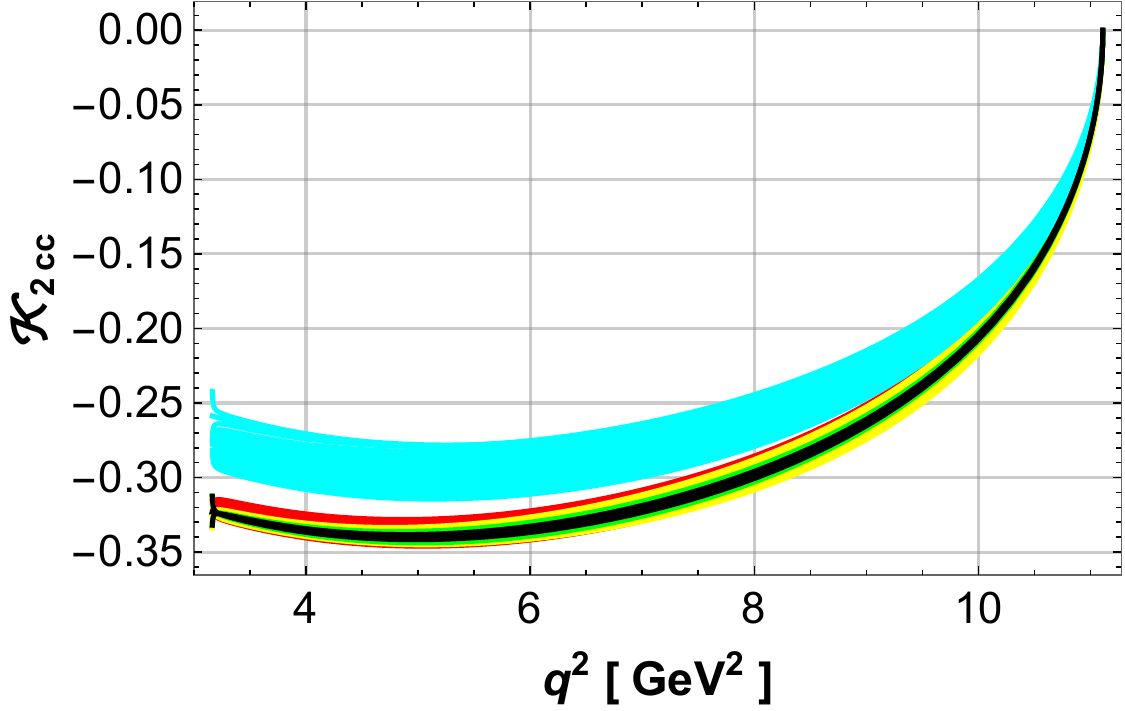} 
\caption{}
\end{subfigure}
\begin{subfigure}[b]{0.32\textwidth}
\centering 
\includegraphics[width=5.6cm, height=3.5cm]{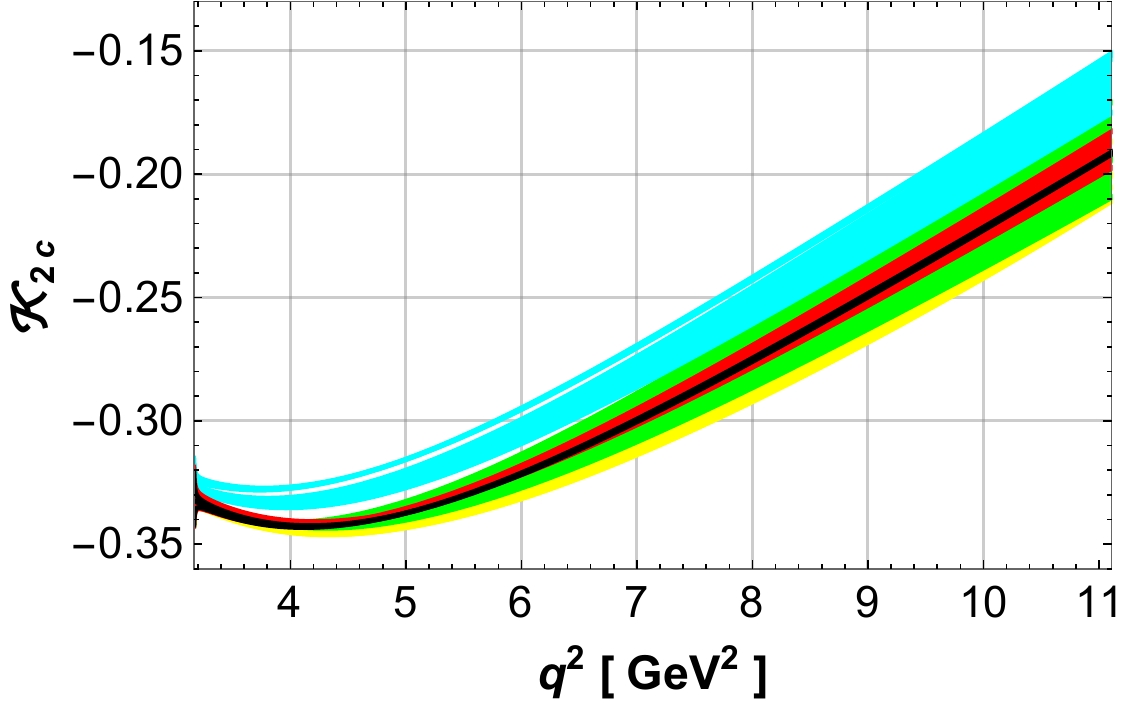}
\caption{}
\end{subfigure}
\begin{subfigure}[b]{0.32\textwidth}
\centering 
\includegraphics[width=5.6cm, height=3.5cm]{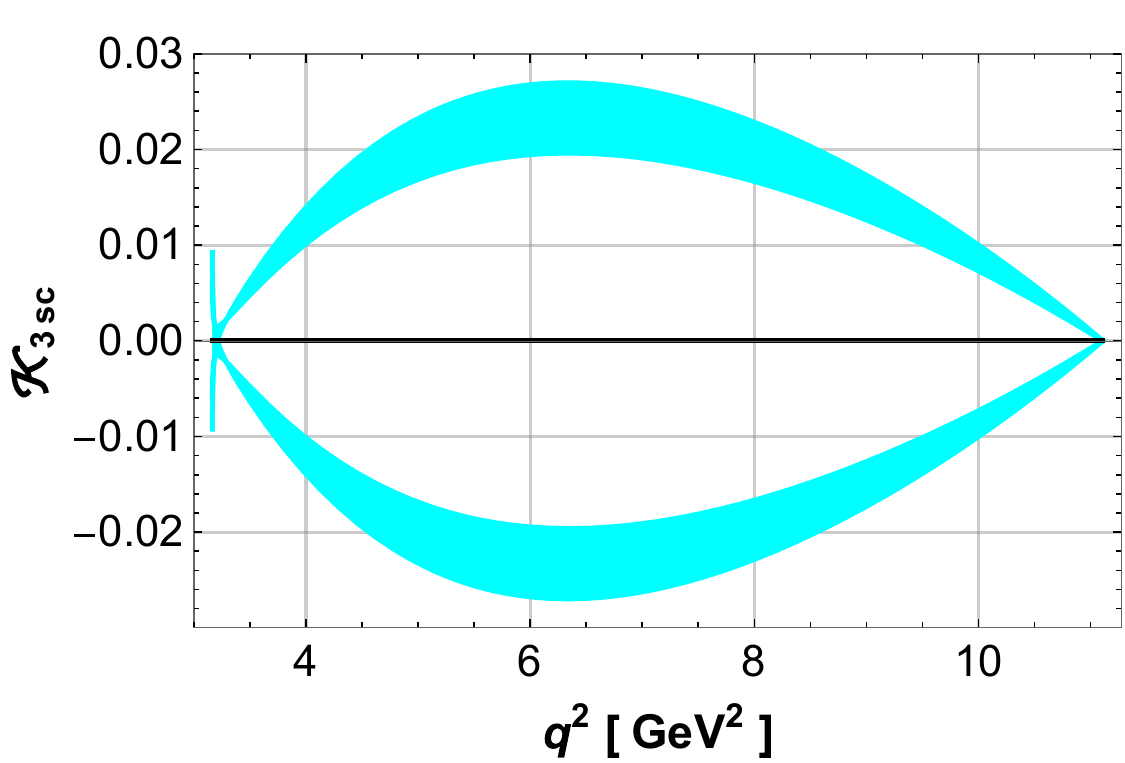}
\caption{}
\end{subfigure}
\begin{subfigure}[b]{0.32\textwidth}
\centering 
\includegraphics[width=5.6cm, height=3.5cm]{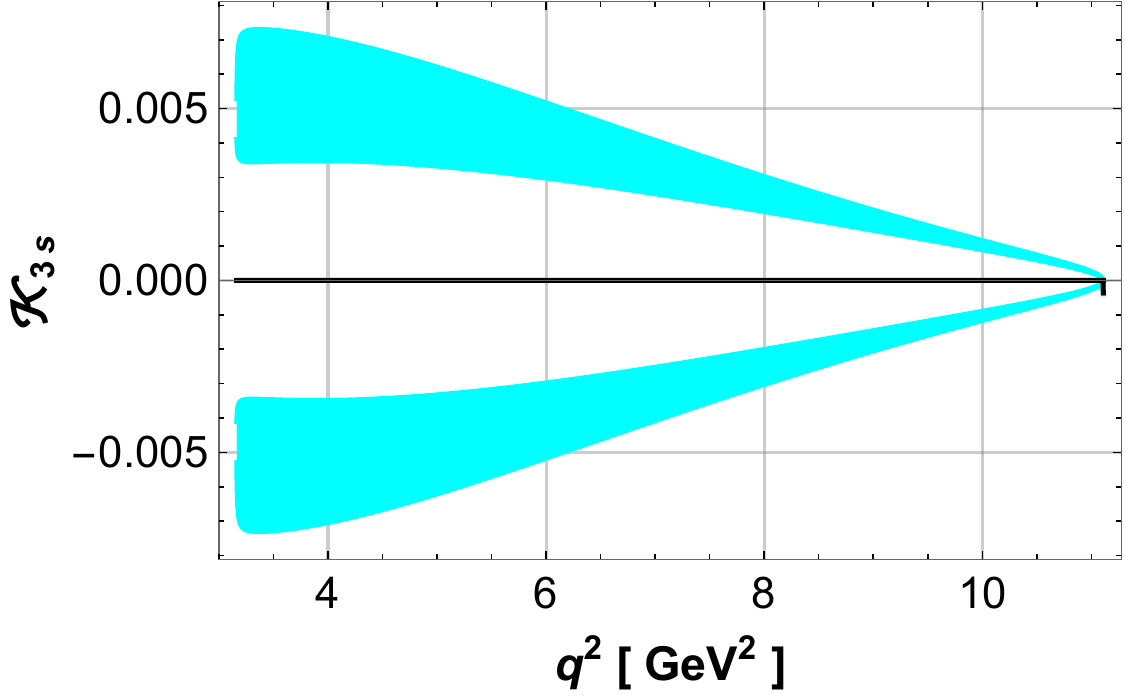} 
\caption{}
\end{subfigure}
\begin{subfigure}[b]{0.32\textwidth}
\centering 
\includegraphics[width=5.6cm, height=3.5cm]{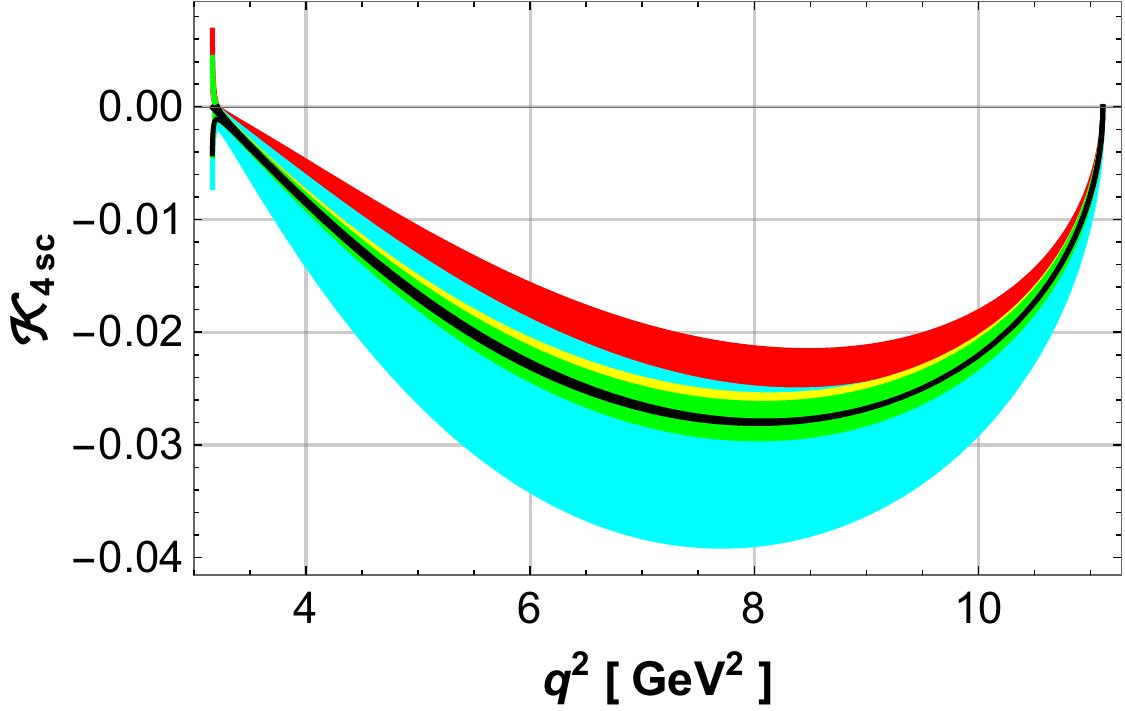}
\caption{}
\end{subfigure}
\begin{subfigure}[b]{0.32\textwidth}
\centering 
\includegraphics[width=5.6cm, height=3.5cm]{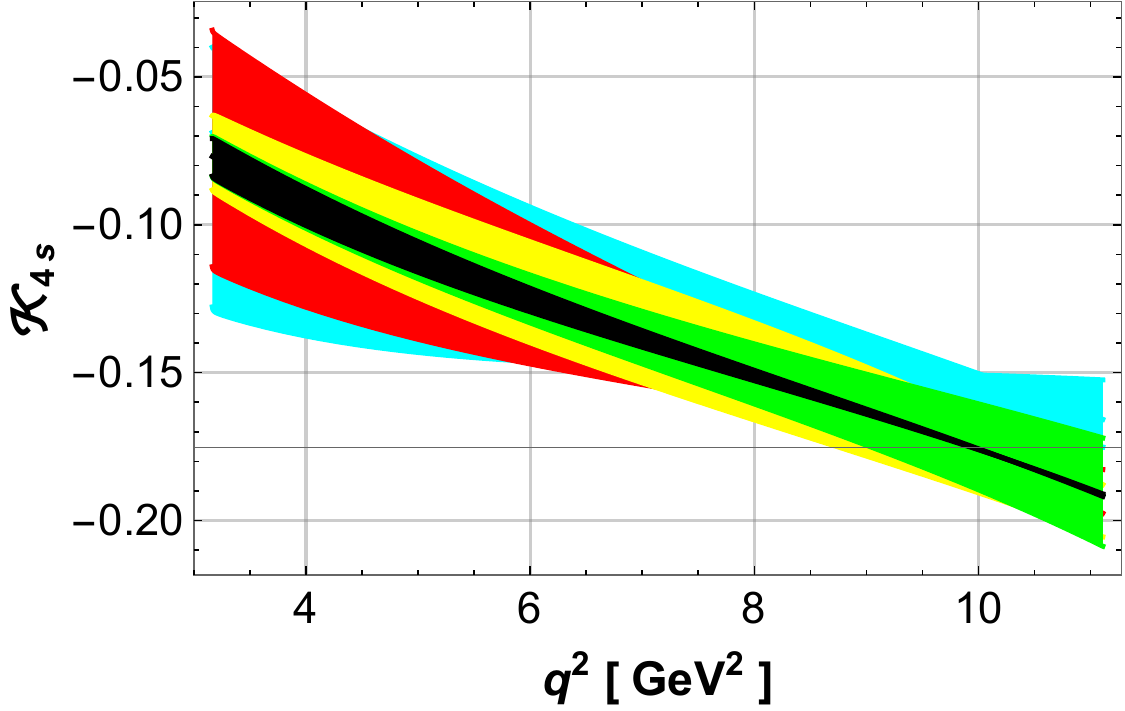}
\caption{}
\end{subfigure}

\caption{\label{phen-2d}The angular observables $\mathcal{K}_{i}(q^{2})$ are shown as functions of $q^{2}$ both within the SM and for various NP scenarios.
The bands of the curves reflect the theoretical uncertainties arising from the hadronic form factors and other input parameters.
The SM predictions are represented by the black bands, while the NP scenarios are indicated by colored bands.
}
\end{figure}

Figure~\ref{phen-2d} illustrates the $q^{2}$ dependence of the angular observables $\mathcal{K}_{i}$, with 
$i = 1ss,\,1cc,\,1c,\,2ss,\,2cc,\,2c,\,3sc,\,3s,\,4sc,\,4s$, 
evaluated in the SM and for different two-dimensional NP benchmark scenarios listed in Table~\ref{2d-table-1}. 
In the presence of NP, deviations from the SM predictions are shown as colored bands, each corresponding to a specific combination of Wilson coefficients:
$(C_{V_L}, C_{S_R})$, 
$(C_{V_L}, C_{S_L}=-4C_T)$, 
$\left(\Re[C_{S_L}=4C_T],\,\Im[C_{S_L}=4C_T]\right)$, 
and $(C_{S_L}, C_{S_R})$.
With reference to the SM predictions, the impact of NP on these observables can be summarized as follows:
\begin{itemize}
\item \textbf{$\bm{\mathcal{K}_{1ss}}$:}  
NP effects are most pronounced in the intermediate $q^{2}$ region for all scenarios. 
The $\left(\Re[C_{S_L}=4C_T],\,\Im[C_{S_L}=4C_T]\right)$ scenario yields the largest upward shift, while the remaining scenarios induce comparatively mild deviations. 
Consequently, precise measurements of $\mathcal{K}_{1ss}$ can effectively constrain NP scenarios and, in particular, discriminate $\left(\Re[C_{S_L}=4C_T],\,\Im[C_{S_L}=4C_T]\right)$ interactions from other possibilities.

\item \textbf{$\bm{\mathcal{K}_{1cc}}$:}  
A noticeable deviation is observed only in the 
$\left(\Re[C_{S_L}=4C_T],\,\Im[C_{S_L}=4C_T]\right)$ scenario, which induces a downward shift of approximately $0.005$ with an uncertainty of $0.002$ (corresponding to about $2.5\sigma$). 
The $(C_{V_L},C_{S_R})$ scenario leads to a minor upward shift. 
Overall, $\mathcal{K}_{1cc}$ remains relatively stable and does not serve as a strong discriminator among $(C_{S_L}, C_{S_R})$ NP fits, although it shows moderate sensitivity to $\left(\Re[C_{S_L}=4C_T],\,\Im[C_{S_L}=4C_T]\right)$ contributions.

\item \textbf{$\bm{\mathcal{K}_{1c}}$:}  
This observable exhibits the strongest sensitivity to NP effects. 
The $\left(\Re[C_{S_L}=4C_T],\,\Im[C_{S_L}=4C_T]\right)$ scenario produces a significant downward shift of about $-0.05$, while the $(C_{S_L},C_{S_R})$ scenario leads to a clear upward shift of approximately $+0.032$. 
In contrast, the $(C_{V_L},C_{S_R})$ and $(C_{V_L},C_{S_L}=-4C_T)$ scenarios remain consistent with the SM within uncertainties. 
Thus, $\mathcal{K}_{1c}$ serves as a powerful discriminator between different NP scenarios, particularly between $\left(\Re[C_{S_L}=4C_T],\,\Im[C_{S_L}=4C_T]\right)$ and $(C_{S_L}, C_{S_R})$ interactions.

\item \textbf{$\bm{\mathcal{K}_{2ss}}$:}  
This observable is largely insensitive to $(C_{V_L}, C_{S_R})$ NP scenarios but exhibits strong sensitivity to the 
$\left(\Re[C_{S_L}=4C_T],\,\Im[C_{S_L}=4C_T]\right)$ contributions, which produce a large and highly significant upward shift across the entire $q^{2}$ region. 
The $(C_{S_L},C_{S_R})$ scenario induces only a mild reduction.

\item \textbf{$\bm{\mathcal{K}_{2cc}}$:}  
The observable remains close to the SM prediction for most NP scenarios. 
The largest deviation, corresponding to approximately $1.7\sigma$, occurs for the 
$\left(\Re[C_{S_L}=4C_T],\,\Im[C_{S_L}=4C_T]\right)$ case, where the distribution is shifted to values around $0.25$--$0.30$ over the full $q^{2}$ range.

\item \textbf{$\bm{\mathcal{K}_{2c}}$:}  
A moderate sensitivity to NP effects is observed. 
The $\left(\Re[C_{S_L}=4C_T],\,\Im[C_{S_L}=4C_T]\right)$ scenario produces the largest upward shift (about $1.8\sigma$), reducing the magnitude of the negative values across all $q^{2}$. 
The maximum separation from the SM occurs in the intermediate region, $q^{2}\approx6$--$11~\mathrm{GeV}^{2}$. 
Other scenarios yield smaller negative shifts. 
Hence, $\mathcal{K}_{2c}$ provides a useful probe of $\left(\Re[C_{S_L}=4C_T],\,\Im[C_{S_L}=4C_T]\right)$ interference effects.

\item \textbf{$\bm{\mathcal{K}_{3sc}}$:}  
This observable remains consistent with zero within uncertainties for all scenarios. 
The largest apparent effect, of order $0.03$, appears in the 
$\left(\Re[C_{S_L}=4C_T],\,\Im[C_{S_L}=4C_T]\right)$ scenario.

\item \textbf{$\bm{\mathcal{K}_{3s}}$:}  
The observable shows negligible sensitivity to NP and exhibits no significant shape distortion, making it less effective for NP discrimination.

\item \textbf{$\bm{\mathcal{K}_{4sc}}$:}  
A pronounced negative dip is observed around $q^{2}\approx6$--$10~\mathrm{GeV}^{2}$, reaching values close to $-0.04$ in the 
$\left(\Re[C_{S_L}=4C_T],\,\Im[C_{S_L}=4C_T]\right)$ scenario. 
The $(C_{V_L},C_{S_L}=-4C_T)$ and $(C_{S_L},C_{S_R})$ scenarios show moderate upward shifts, although the observable remains negative throughout the entire $q^{2}$ range. 
Thus, $\mathcal{K}_{4sc}$ provides a sensitive probe of $\left(\Re[C_{S_L}=4C_T],\,\Im[C_{S_L}=4C_T]\right)$ effects.

\item \textbf{$\bm{\mathcal{K}_{4s}}$:}  
This observable exhibits a smooth negative distribution over the full $q^{2}$ region,  reaching values near $-0.14$. 
While some NP scenarios induce mild shifts, the overall shape remains largely unchanged. 
Notably, the 
$\left(\Re[C_{S_L}=4C_T],\,\Im[C_{S_L}=4C_T]\right)$ scenario in the intermediate $q^{2}$ region and the $(C_{V_L},C_{S_L}=-4C_T)$ scenario at low $q^{2}$ show significant deviations from the SM.
\end{itemize}

The averaged values of the angular observables $\mathcal{K}_{i}$,
with $i=1ss,\,1cc,\,1c,\,2ss,\,2cc,\,2c,\,3sc,\,3s,\,4sc,\,4s$,
evaluated at the BFPs for the two-dimensional NP scenarios,
are listed in Table~\ref{2d-obs-table}.
We observe that, except for the total decay rate, the uncertainties arising from the input parameters do not mimic the effects induced by NP.
Consequently, these angular observables provide clean and robust probes for establishing NP in these FCCC decays.

\begin{table}[H]
\centering{}%
\renewcommand{\arraystretch}{1.5}
\begin{tabular}{|c|c|c|c|c|c|}
\hline 
WCs & SM & $\left(C_{V_{L}},C_{S_{R}}\right)$ & $\left(C_{V_{L}},C_{S_{L}}=-4C_{T}\right)$ & $\left(\Re\left[C_{S_{L}}=4C_{T}\right],\Im\left[C_{S_{L}}=4C_{T}\right]\right)$ & $\left(C_{S_{L}},C_{S_{R}}\right)$ \tabularnewline
\hline 
\hline 
BFP & $C_{i}=0$ & $\left(0.06,0.01\right)$ & $\left(0.06,0.01\right)$ & $\left(-0.04,-0.29\right)$ & $\left(-0.19,0.24\right)$
\tabularnewline
\hline 
$\mathcal{K}_{1ss} $ & $0.323\pm0$ & $0.323\pm0.001$ & $0.323\pm0.001$ & $0.325\pm0.001$ & $0.324\pm0.001$
\tabularnewline
\hline 
$\mathcal{K}_{1cc} $ & $0.354\pm0$ & $0.353\pm0.002$ & $0.353\pm0.002$ & $0.349\pm0.002$ & $0.351\pm0.001$
\tabularnewline\hline 
$\mathcal{K}_{1c} $ & $0.224\pm0.004$ & $0.226\pm0.025$ & $0.222\pm0.008$ & $0.191\pm0.025$ & $0.256\pm0.043$
\tabularnewline\hline 
$\mathcal{K}_{2ss} $ & $-0.240\pm0.003$ & $-0.240\pm0.003$ & $-0.236\pm0.017$ & $-0.171\pm0.041$ & $-0.256\pm0.033$
\tabularnewline\hline 
$\mathcal{K}_{2cc} $ & $-0.279\pm0.003$ & $-0.278\pm0.009$ & $-0.275\pm0.0017$ & $-0.211\pm0.039$ & $-0.290\pm0.031$
\tabularnewline\hline 
$\mathcal{K}_{2c} $ & $-0.274\pm0.002$ & $-0.278\pm0.033$ & $-0.276\pm0.0014$ & $-0.215\pm0.033$ & $-0.294\pm0.013$
\tabularnewline\hline 
$\mathcal{K}_{3sc} $ & $0\pm0$ & $0\pm0$ & $0\pm0$ & $\pm0.017\pm0.034$ & $0\pm0.009$
\tabularnewline\hline 
$\mathcal{K}_{3s} $ & $0\pm0$ & $0\pm0$ & $0\pm0$ & $0\pm0.001$ & $0\pm0$
\tabularnewline\hline 
$\mathcal{K}_{4sc} $ & $-0.024\pm0$ & $-0.023\pm0.003$ & $-0.024\pm0.005$ & $-0.026\pm0.007$ & $-0.021\pm0.002$
\tabularnewline\hline 
$\mathcal{K}_{4s} $ & $-0.149\pm0.003$ & $-0.152\pm0.025$ & $-0.156\pm0.023$ & $-0.111\pm0.021$ & $-0.136\pm0.016$
\tabularnewline
\hline 
\end{tabular}\caption{\label{2d-obs-table} BFPs for the two-dimensional NP scenarios obtained under the constraint
$\mathcal{B}\!\left(B_{c}^{-}\to\tau^{-}\bar{\nu}_{\tau}\right)<60\%$, together with the corresponding
predictions for the full set of angular observables $\mathcal{K}_{i}$ with
$i=1ss,\,1cc,\,1c,\,2ss,\,2cc,\,2c,\,3sc,\,3s,\,4sc,\,4s$.
The quoted uncertainties include contributions from hadronic form factors,
other input parameters, and the uncertainties in the BFP determination.
}
\end{table}

\section{Correlating different Physical observables}\label{correlation}

\begin{figure}[H]
\centering 
\begin{subfigure}[b]{0.32\textwidth}
\centering 
\includegraphics[width=5.6cm, height=3.5cm]{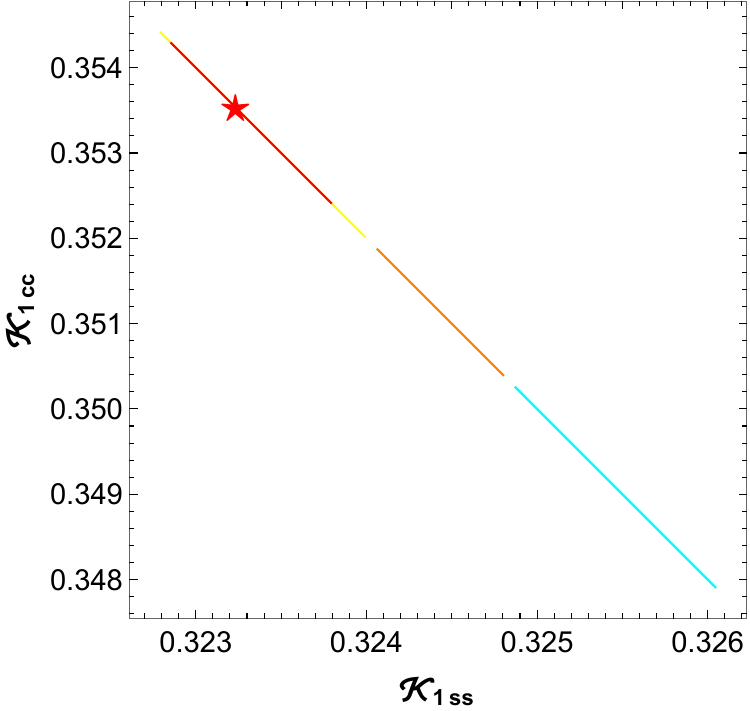}
\caption{}
\end{subfigure}
\begin{subfigure}[b]{0.32\textwidth}
\centering 
\includegraphics[width=5.6cm, height=3.5cm]{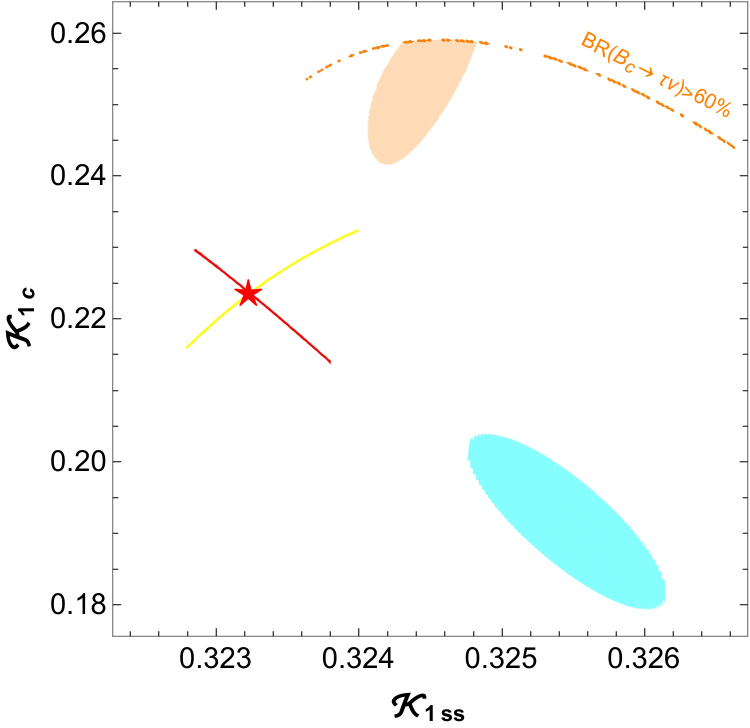} 
\caption{}
\end{subfigure}
\begin{subfigure}[b]{0.32\textwidth}
\centering 
\includegraphics[width=5.6cm, height=3.5cm]{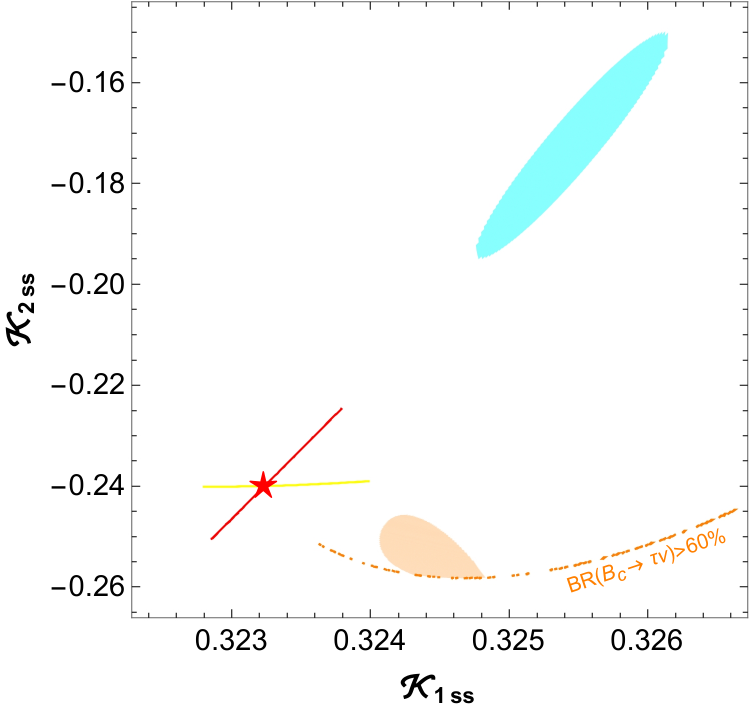}
\caption{}
\end{subfigure}
\begin{subfigure}[b]{0.32\textwidth}
\centering 
\includegraphics[width=5.6cm, height=3.5cm]{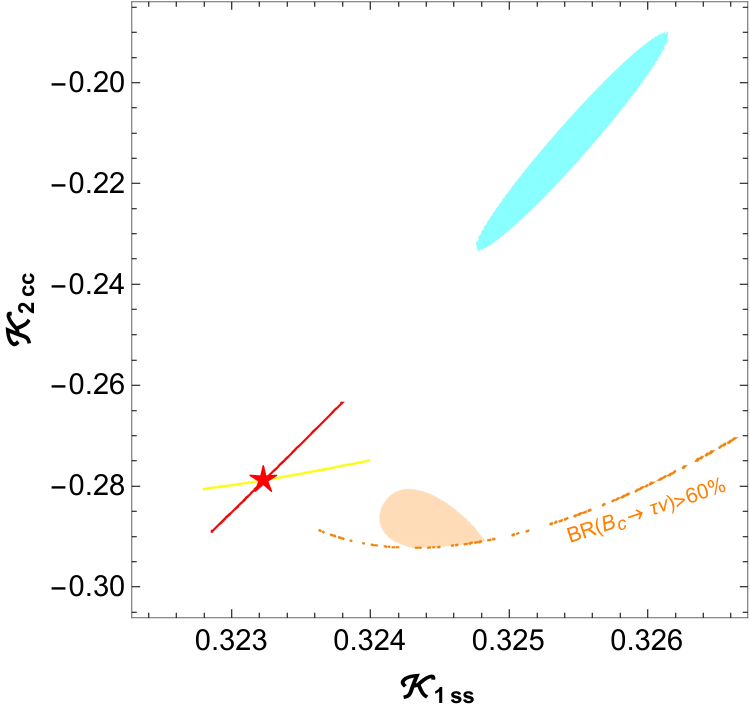}
\caption{}
\end{subfigure}
\begin{subfigure}[b]{0.32\textwidth}
\centering 
\includegraphics[width=5.6cm, height=3.5cm]{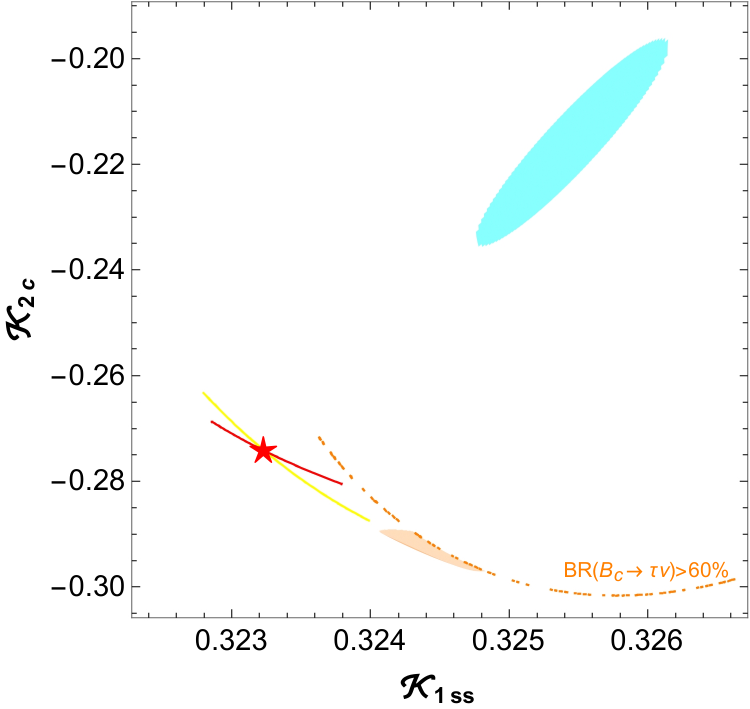} 
\caption{}
\end{subfigure}
\begin{subfigure}[b]{0.32\textwidth}
\centering 
\includegraphics[width=5.6cm, height=3.5cm]{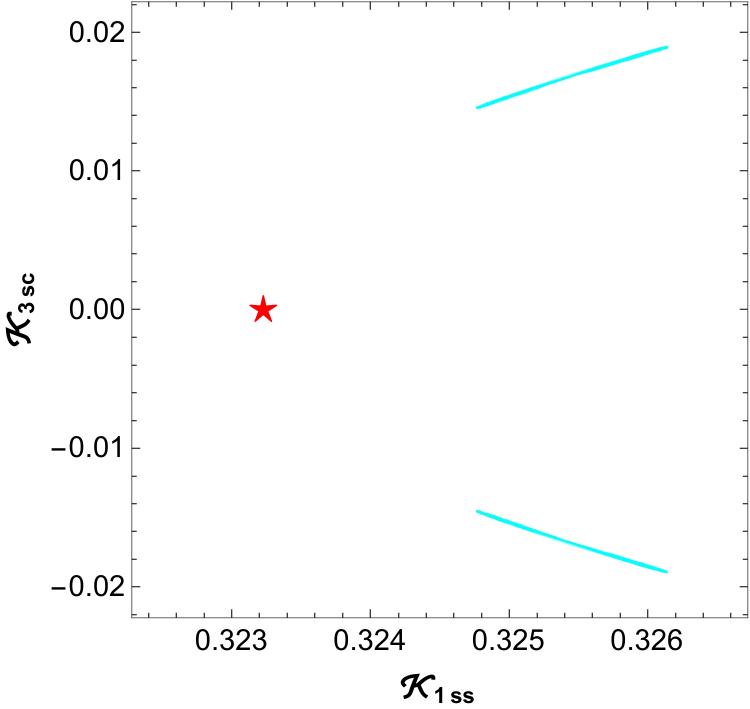}
\caption{}
\end{subfigure}
\begin{subfigure}[b]{0.32\textwidth}
\centering 
\includegraphics[width=5.6cm, height=3.5cm]{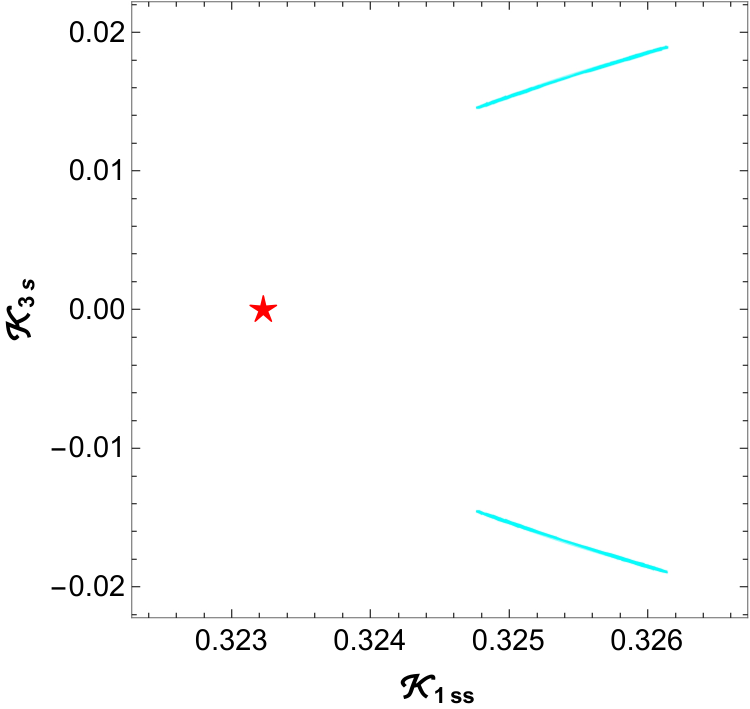}
\caption{}
\end{subfigure}
\begin{subfigure}[b]{0.32\textwidth}
\centering 
\includegraphics[width=5.6cm, height=3.5cm]{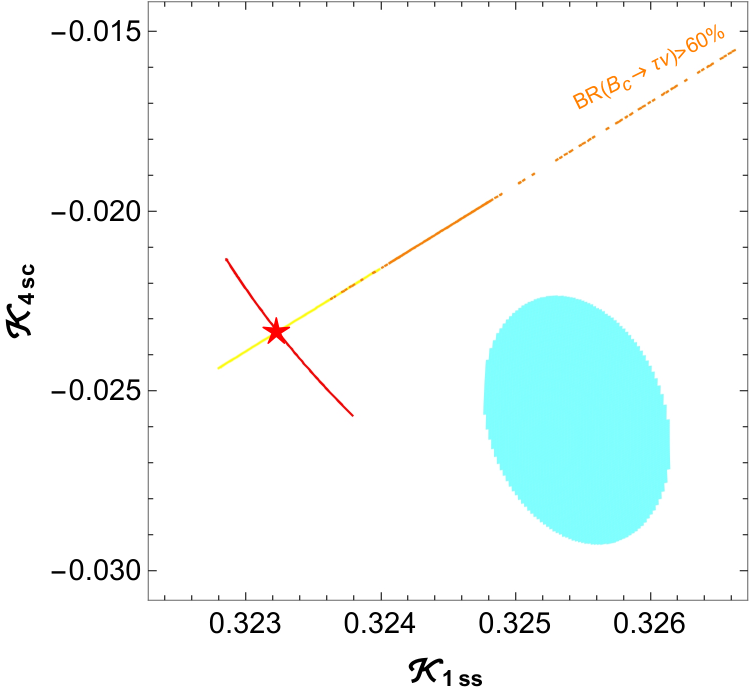} 
\caption{}
\end{subfigure}
\begin{subfigure}[b]{0.32\textwidth}
\centering 
\includegraphics[width=5.6cm, height=3.5cm]{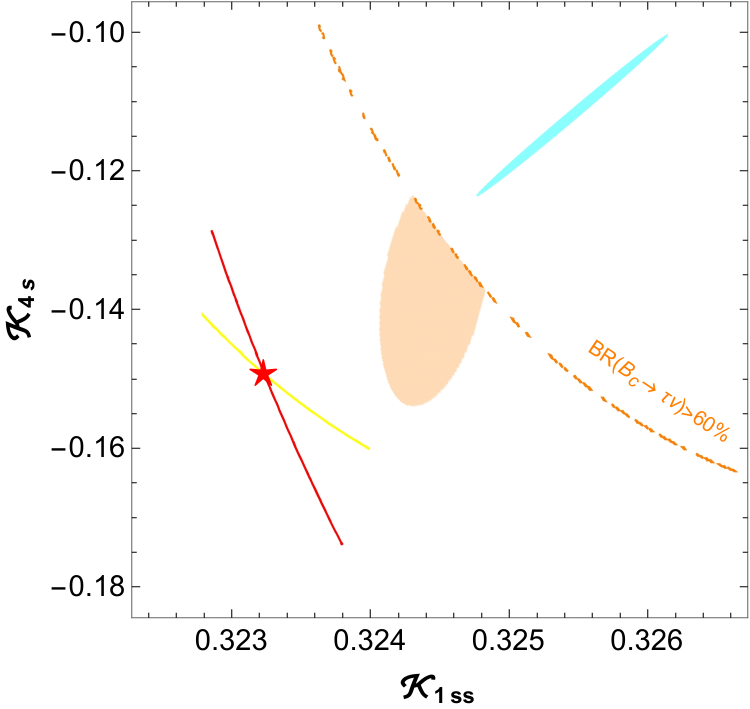}
\caption{}
\end{subfigure}
\begin{subfigure}[b]{0.32\textwidth}
\centering 
\includegraphics[width=5.6cm, height=3.5cm]{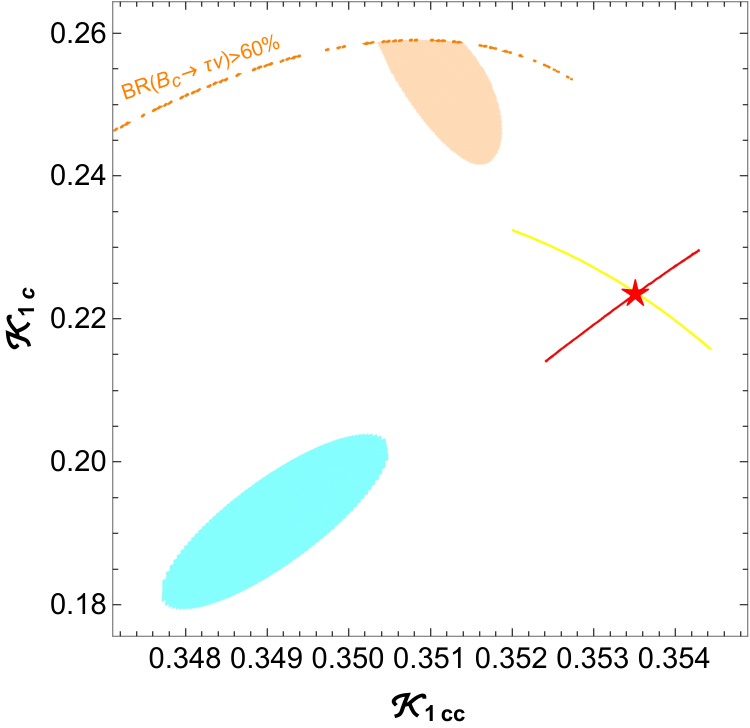}
\caption{}
\end{subfigure}
\begin{subfigure}[b]{0.32\textwidth}
\centering 
\includegraphics[width=5.6cm, height=3.5cm]{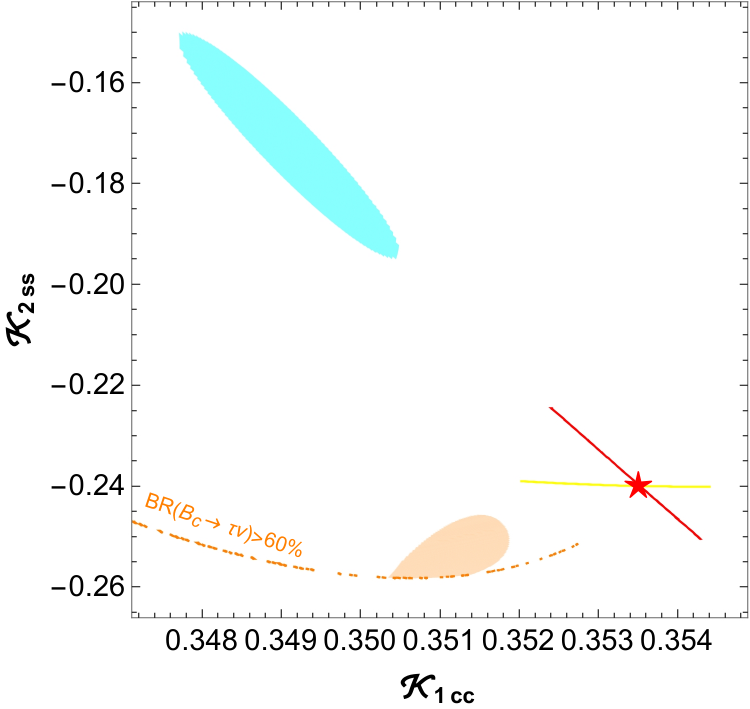}
\caption{}
\end{subfigure}
\begin{subfigure}[b]{0.32\textwidth}
\centering 
\includegraphics[width=5.6cm, height=3.5cm]{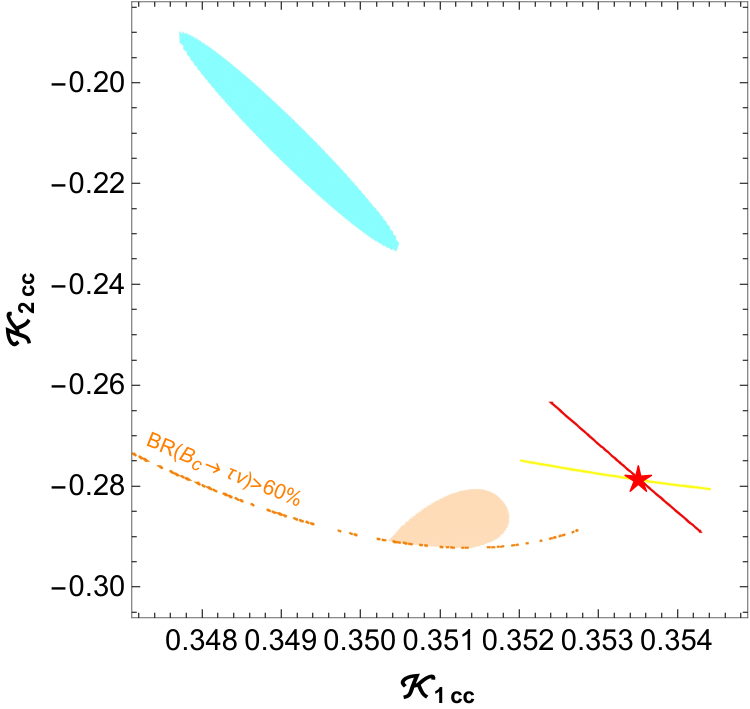}
\caption{}
\end{subfigure}
\begin{subfigure}[b]{0.32\textwidth}
\centering 
\includegraphics[width=5.6cm, height=3.5cm]{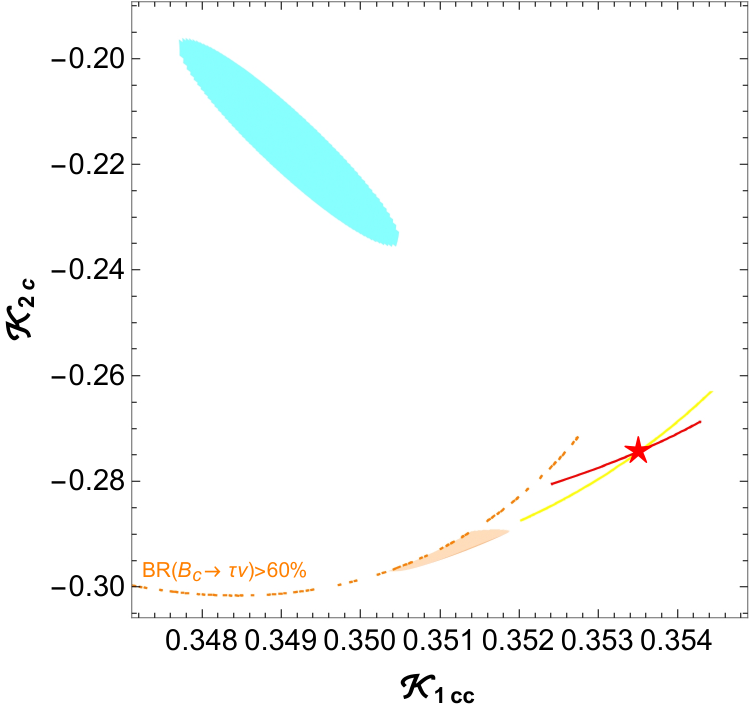}
\caption{}
\end{subfigure}
\begin{subfigure}[b]{0.32\textwidth}
\centering 
\includegraphics[width=5.6cm, height=3.5cm]{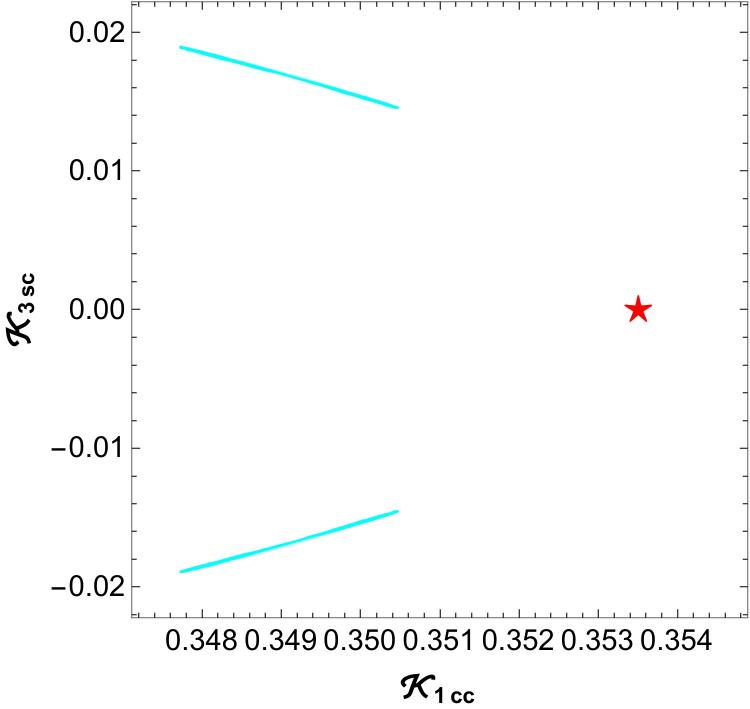}
\caption{}
\end{subfigure}
\begin{subfigure}[b]{0.32\textwidth}
\centering 
\includegraphics[width=5.6cm, height=3.5cm]{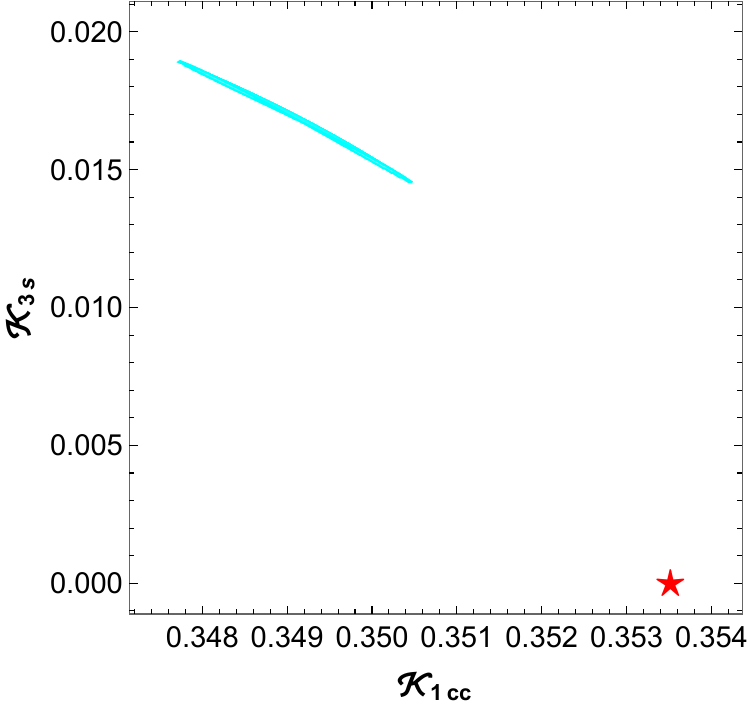}
\caption{}
\end{subfigure}
\begin{subfigure}[b]{0.32\textwidth}
\centering 
\includegraphics[width=5.6cm, height=3.5cm]{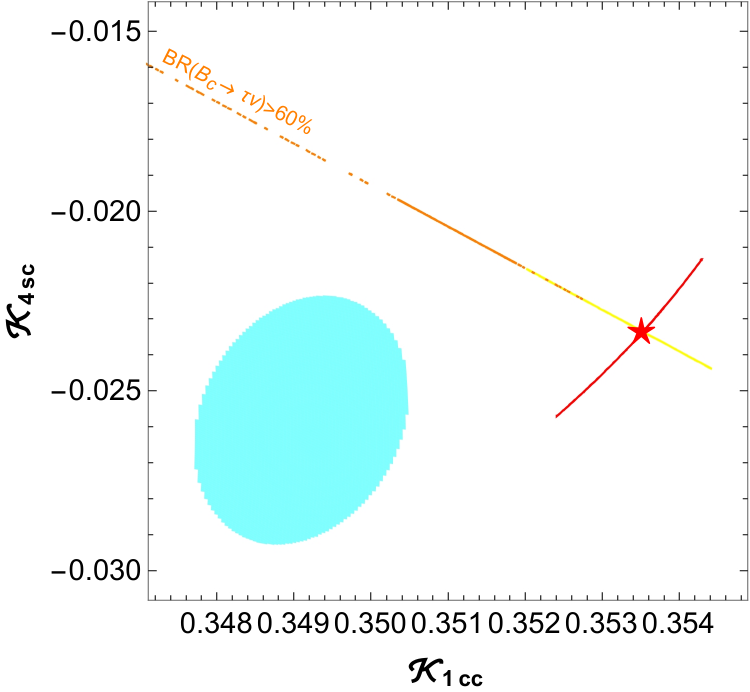}
\caption{}
\end{subfigure}
\begin{subfigure}[b]{0.32\textwidth}
\centering 
\includegraphics[width=5.6cm, height=3.5cm]{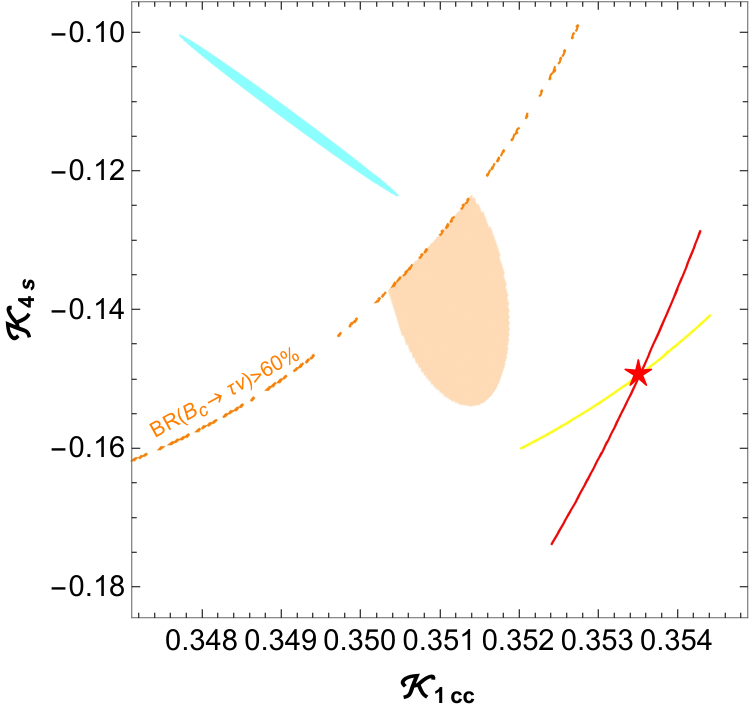}
\caption{}
\end{subfigure}
\begin{subfigure}[b]{0.32\textwidth}
\centering 
\includegraphics[width=5.6cm, height=3.5cm]{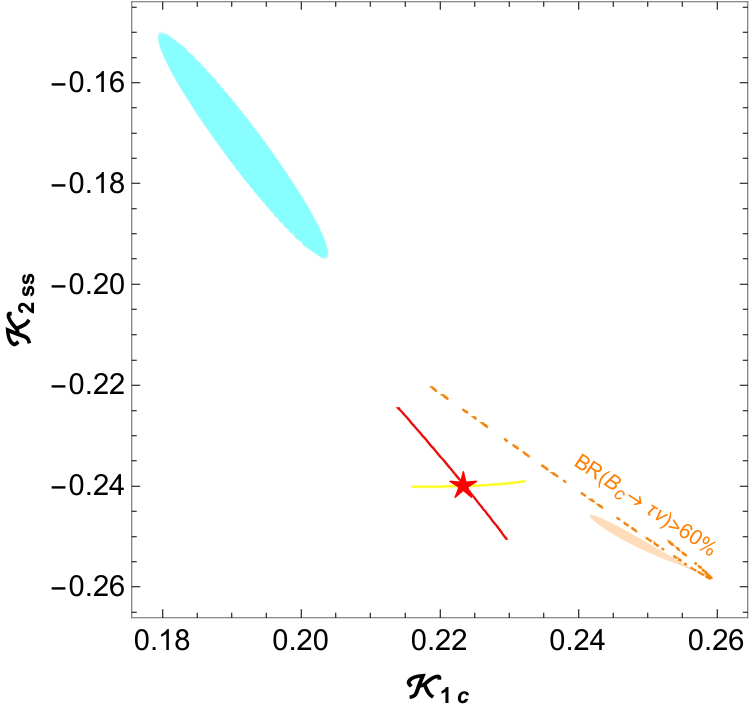}
\caption{}
\end{subfigure}
\caption{\label{corr-2d}
Preferred $1\sigma$ regions for the four two-WC scenarios (listed in panel (a)),
obtained from correlation plots of the angular observables
$\mathcal{K}_{i}(q^{2})$, subject to the constraint
$\mathcal{B}(B_{c}^{-}\to\tau^{-}\bar{\nu}_{\tau})<60\%$.
The orange dashed line indicates the
$\mathcal{B}(B_{c}^{-}\to\tau^{-}\bar{\nu}_{\tau})<60\%$ bound,
while the red stars denote the SM predictions.
The legend and color scheme follow those shown in Fig.~\ref{corr-2d}(a).
}
\end{figure}

\begin{figure}[H]
\centering 
\begin{subfigure}[b]{0.32\textwidth}
\centering 
\includegraphics[width=5.6cm, height=3.5cm]{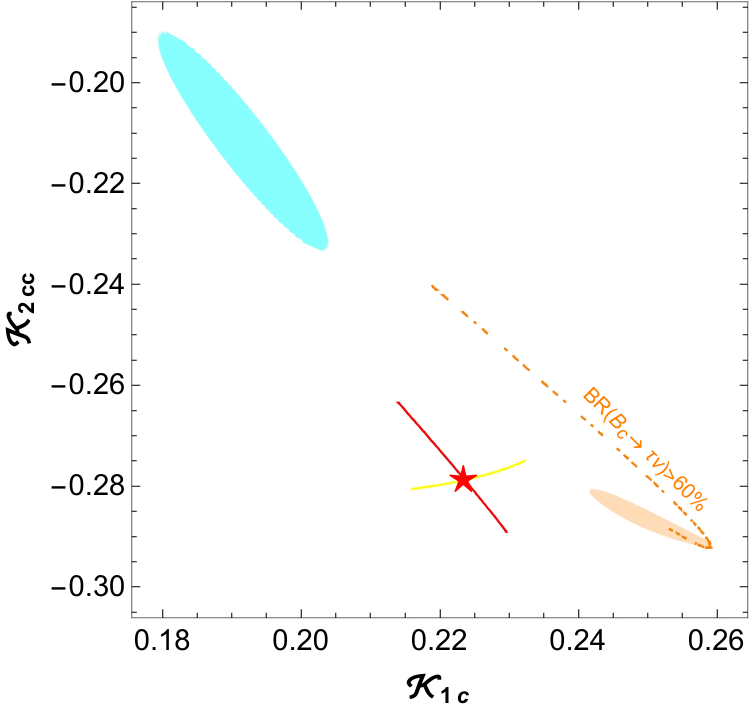}
\caption{}
\end{subfigure}
\begin{subfigure}[b]{0.32\textwidth}
\centering 
\includegraphics[width=5.6cm, height=3.5cm]{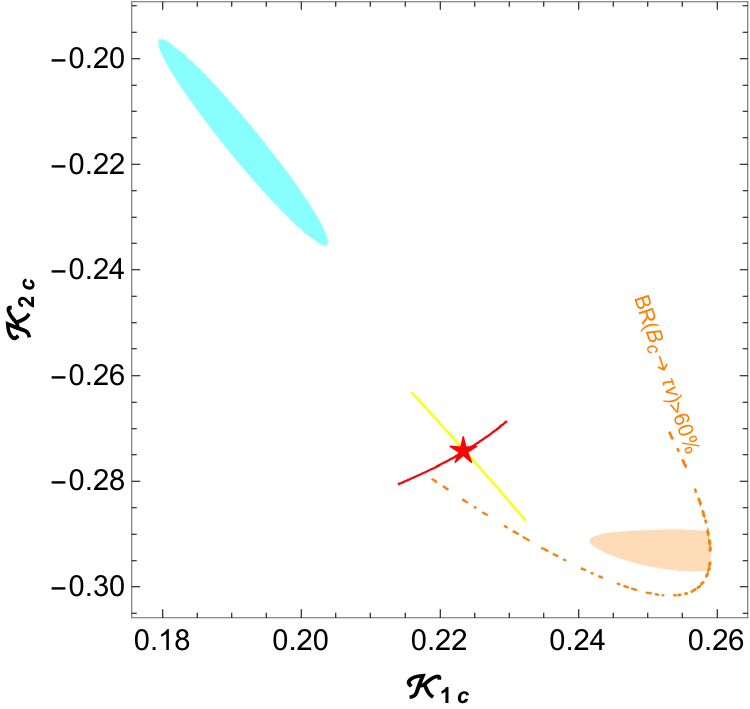} 
\caption{}
\end{subfigure}
\begin{subfigure}[b]{0.32\textwidth}
\centering 
\includegraphics[width=5.6cm, height=3.5cm]{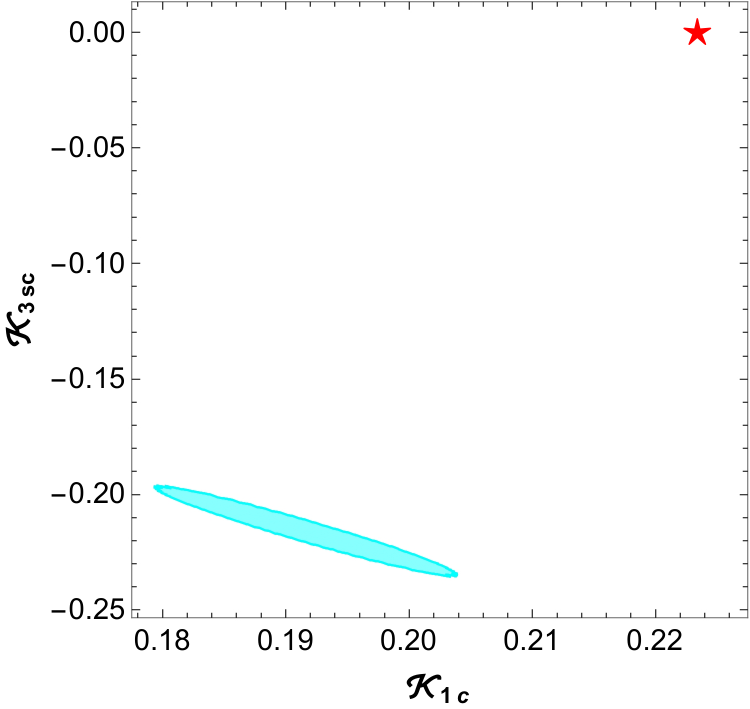}
\caption{}
\end{subfigure}
\begin{subfigure}[b]{0.32\textwidth}
\centering 
\includegraphics[width=5.6cm, height=3.5cm]{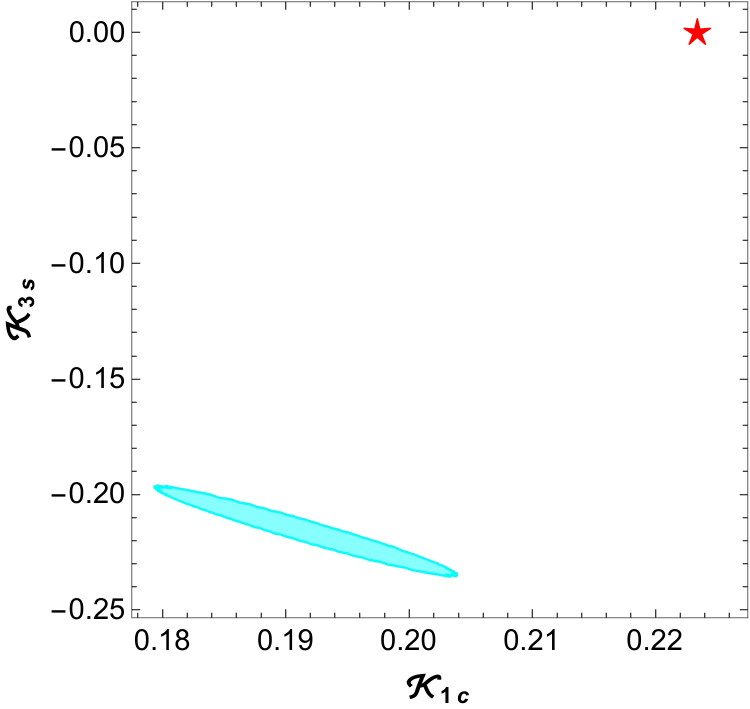}
\caption{}
\end{subfigure}
\begin{subfigure}[b]{0.32\textwidth}
\centering 
\includegraphics[width=5.6cm, height=3.5cm]{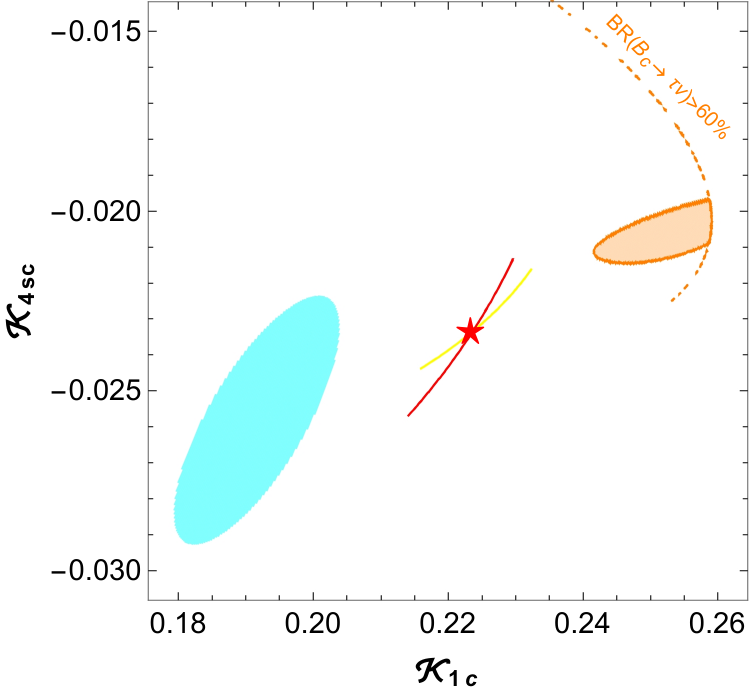} 
\caption{}
\end{subfigure}
\begin{subfigure}[b]{0.32\textwidth}
\centering 
\includegraphics[width=5.6cm, height=3.5cm]{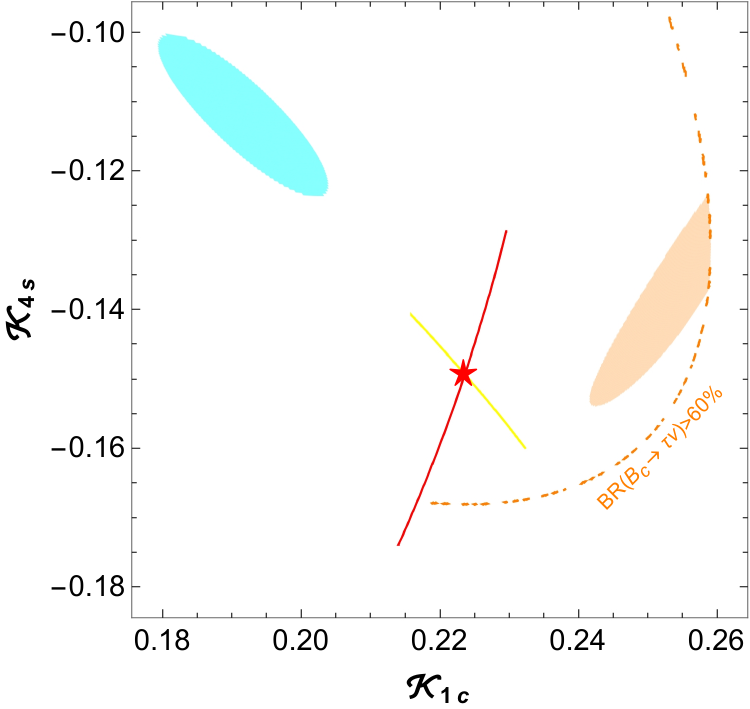}
\caption{}
\end{subfigure}
\begin{subfigure}[b]{0.32\textwidth}
\centering 
\includegraphics[width=5.6cm, height=3.5cm]{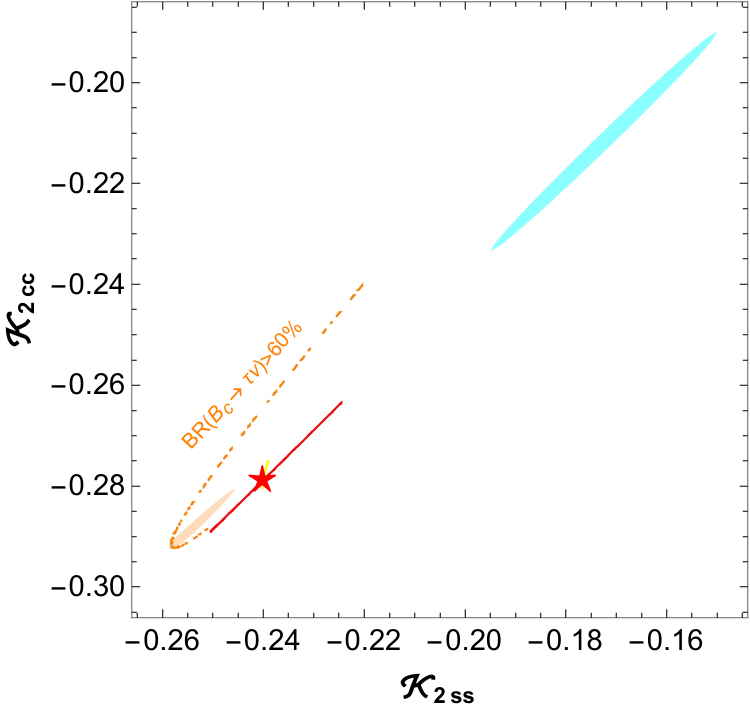}
\caption{}
\end{subfigure}
\begin{subfigure}[b]{0.32\textwidth}
\centering 
\includegraphics[width=5.6cm, height=3.5cm]{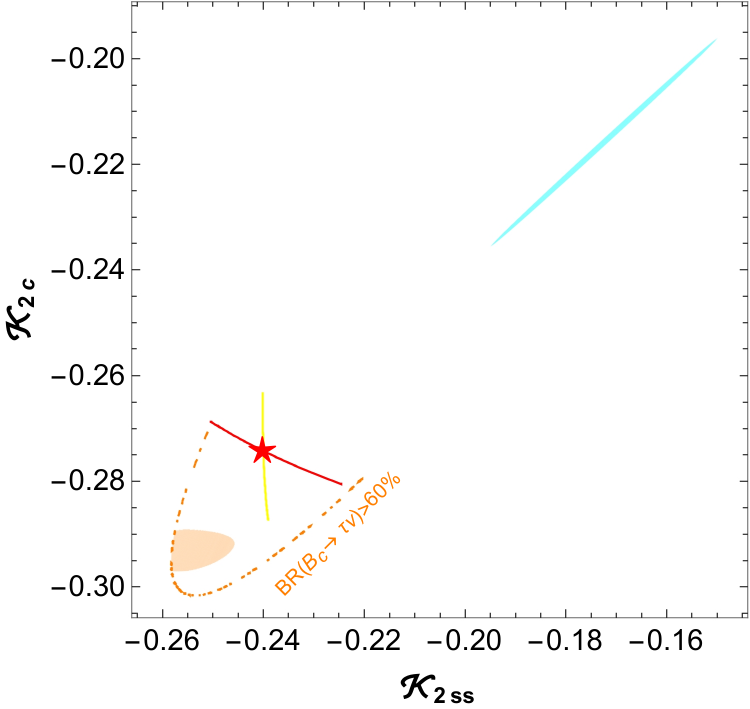} 
\caption{}
\end{subfigure}
\begin{subfigure}[b]{0.32\textwidth}
\centering 
\includegraphics[width=5.6cm, height=3.5cm]{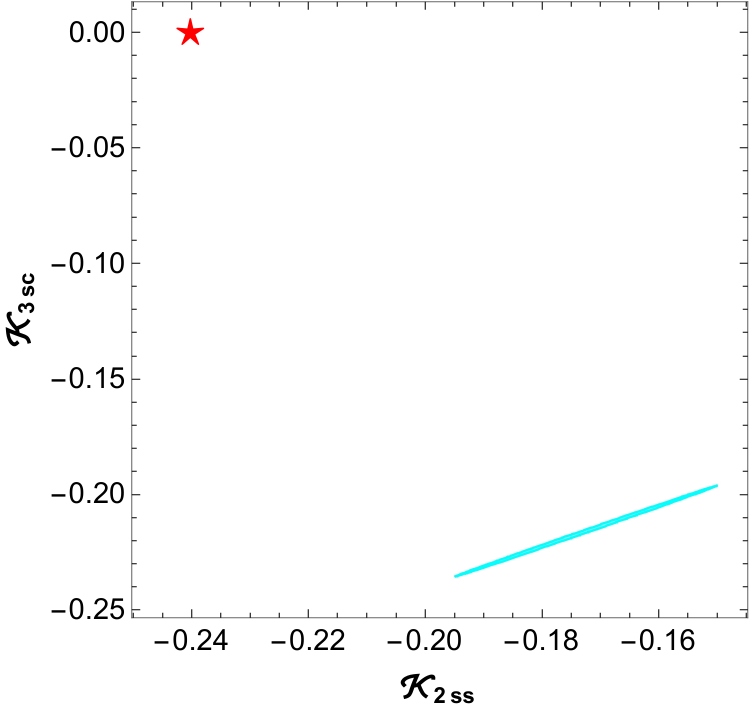}
\caption{}
\end{subfigure}
\begin{subfigure}[b]{0.32\textwidth}
\centering 
\includegraphics[width=5.6cm, height=3.5cm]{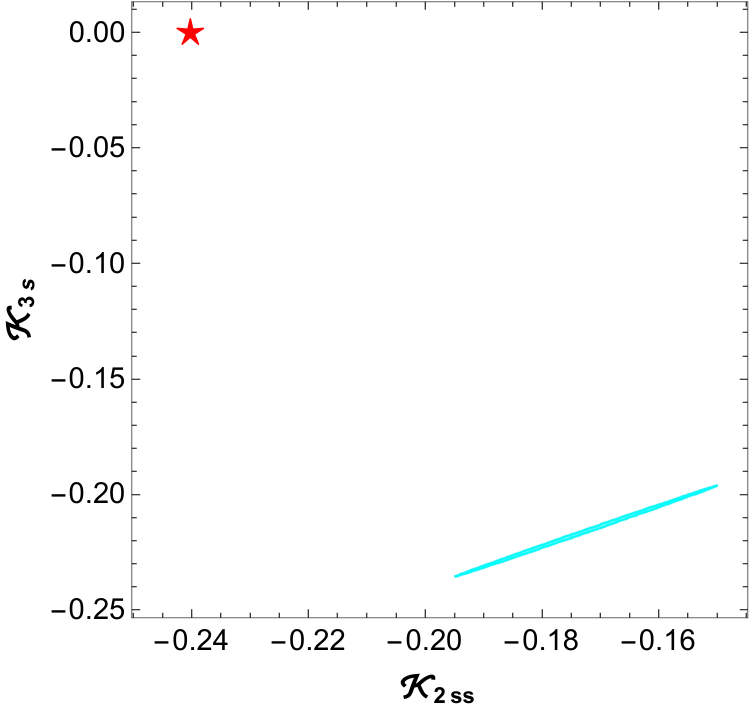}
\caption{}
\end{subfigure}
\begin{subfigure}[b]{0.32\textwidth}
\centering 
\includegraphics[width=5.6cm, height=3.5cm]{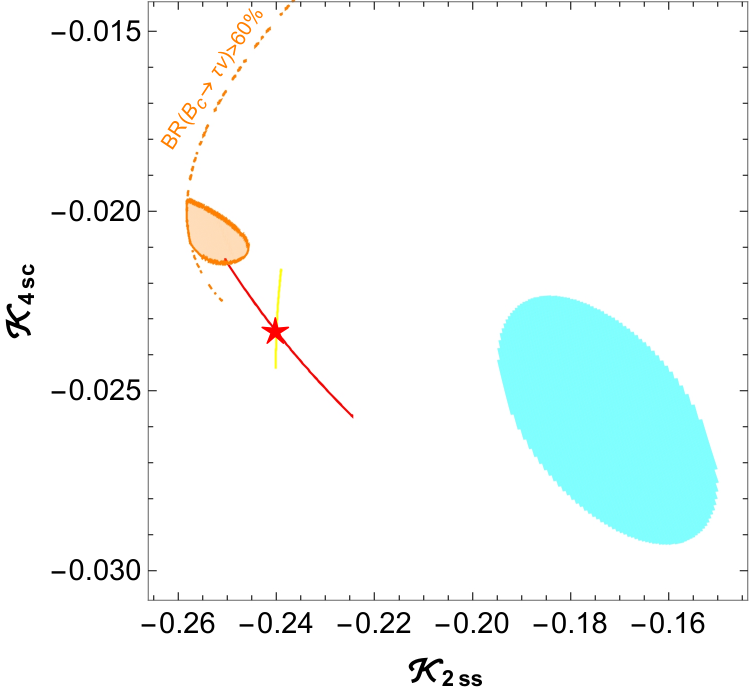}
\caption{}
\end{subfigure}
\begin{subfigure}[b]{0.32\textwidth}
\centering 
\includegraphics[width=5.6cm, height=3.5cm]{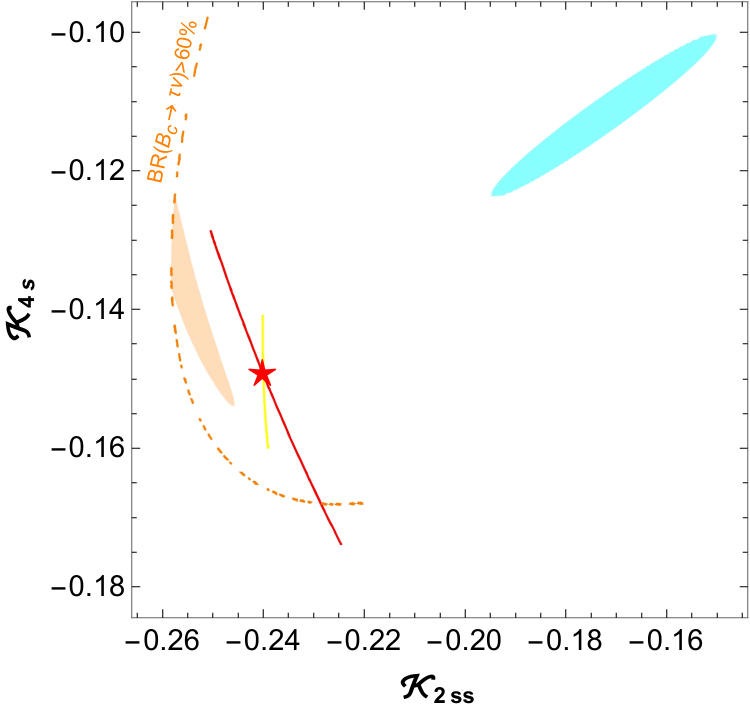}
\caption{}
\end{subfigure}
\begin{subfigure}[b]{0.32\textwidth}
\centering 
\includegraphics[width=5.6cm, height=3.5cm]{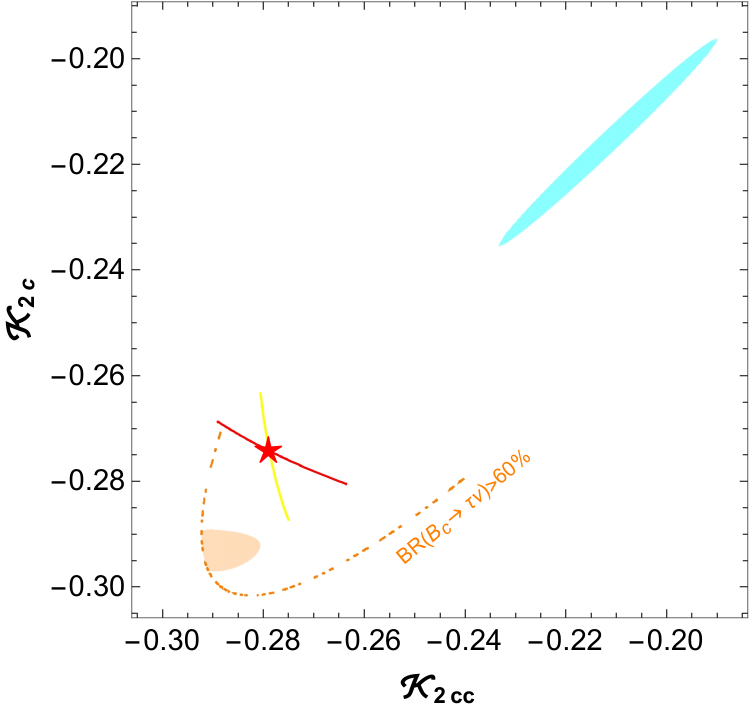}
\caption{}
\end{subfigure}
\begin{subfigure}[b]{0.32\textwidth}
\centering 
\includegraphics[width=5.6cm, height=3.5cm]{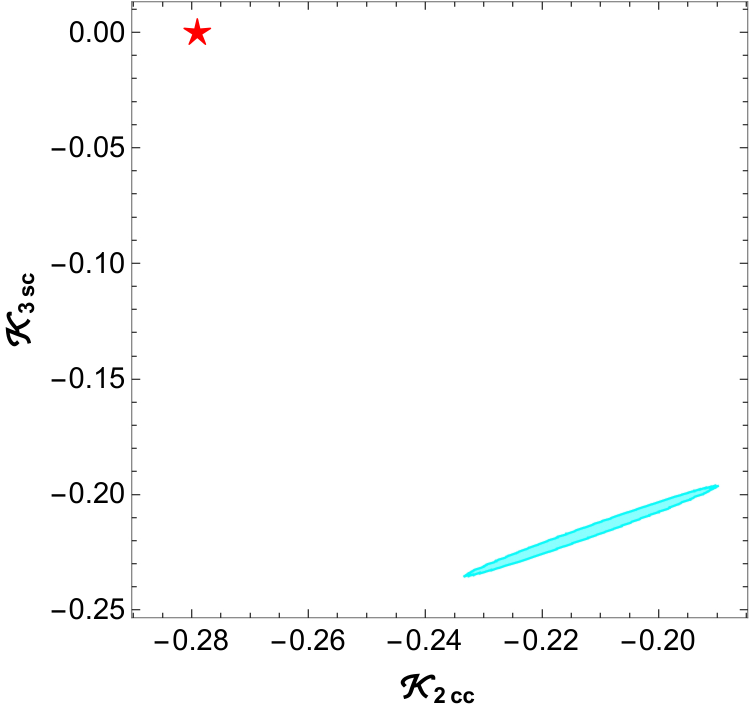}
\caption{}
\end{subfigure}
\begin{subfigure}[b]{0.32\textwidth}
\centering 
\includegraphics[width=5.6cm, height=3.5cm]{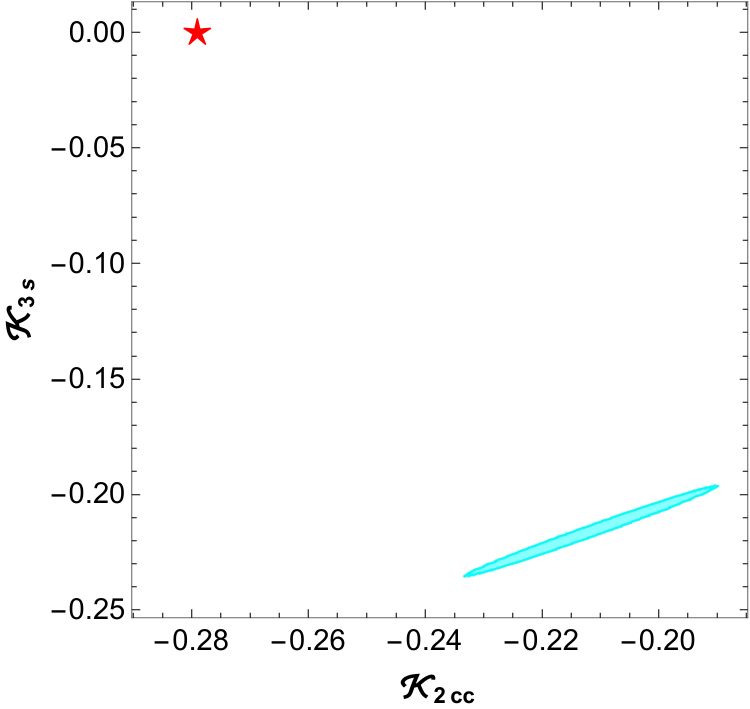}
\caption{}
\end{subfigure}
\begin{subfigure}[b]{0.32\textwidth}
\centering 
\includegraphics[width=5.6cm, height=3.5cm]{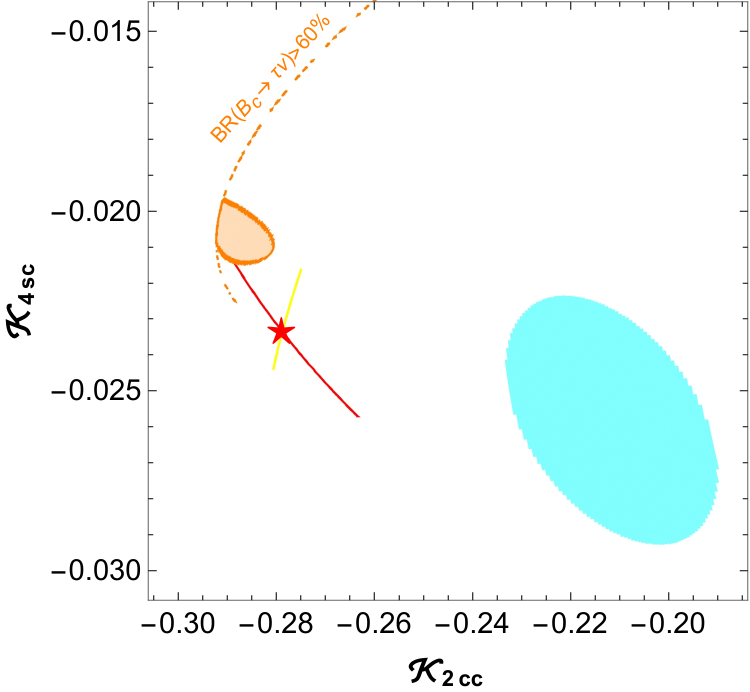}
\caption{}
\end{subfigure}
\begin{subfigure}[b]{0.32\textwidth}
\centering 
\includegraphics[width=5.6cm, height=3.5cm]{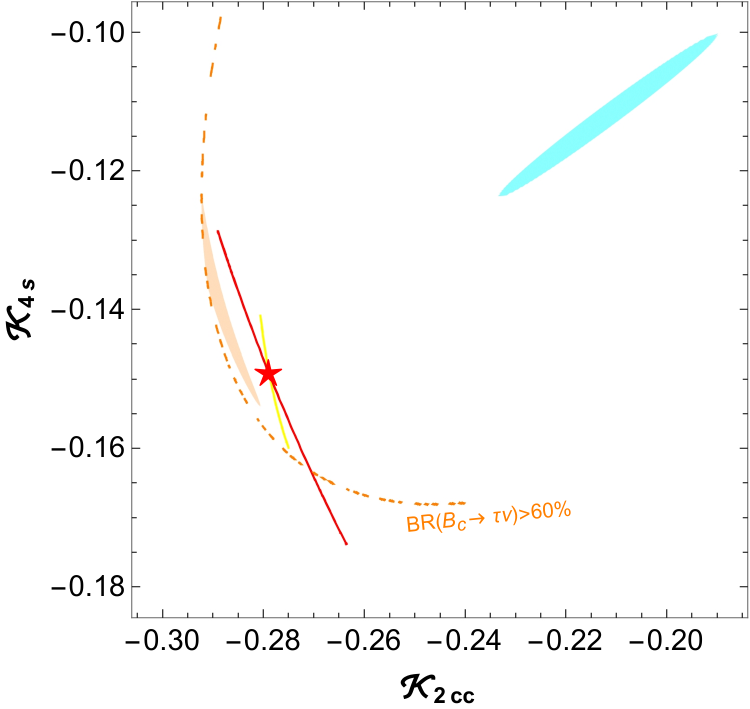}
\caption{}
\end{subfigure}
\begin{subfigure}[b]{0.32\textwidth}
\centering 
\includegraphics[width=5.6cm, height=3.5cm]{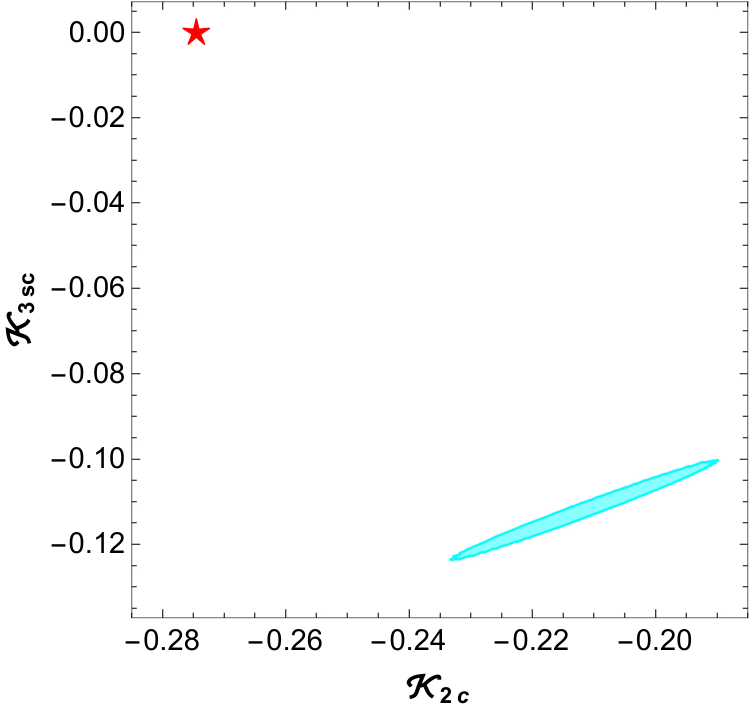}
\caption{}
\end{subfigure}
\caption{\label{corr-2d-1}
 Correlation plots of the angular observables $\mathcal{K}_{i}(q^{2})$.
The legend, color coding, and conventions are the same as in
Fig.~\ref{corr-2d}.
}
\end{figure}

\begin{figure}[H]
\centering 
\begin{subfigure}[b]{0.32\textwidth}
\centering 
\includegraphics[width=5.6cm, height=3.5cm]{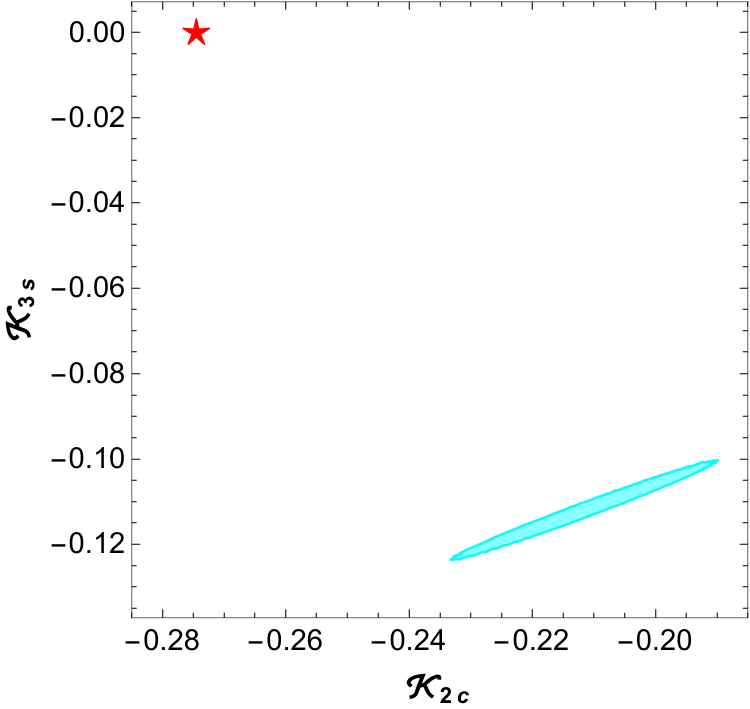}
\caption{}
\end{subfigure}
\begin{subfigure}[b]{0.32\textwidth}
\centering 
\includegraphics[width=5.6cm, height=3.5cm]{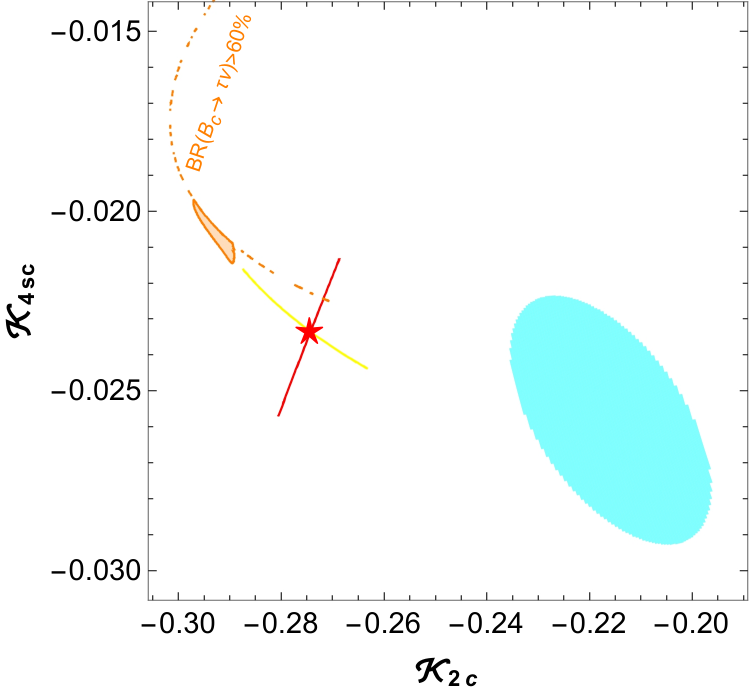} 
\caption{}
\end{subfigure}
\begin{subfigure}[b]{0.32\textwidth}
\centering 
\includegraphics[width=5.6cm, height=3.5cm]{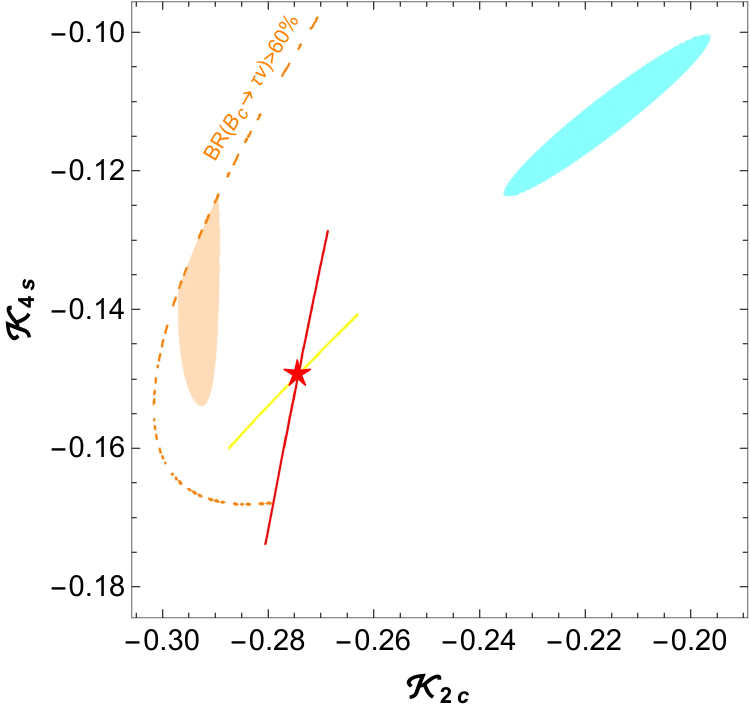}
\caption{}
\end{subfigure}
\begin{subfigure}[b]{0.32\textwidth}
\centering 
\includegraphics[width=5.6cm, height=3.5cm]{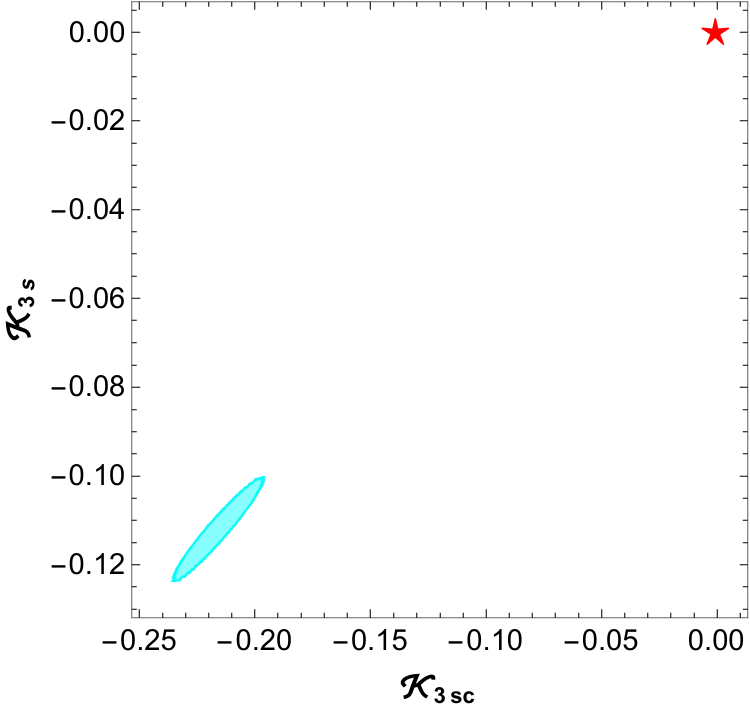}
\caption{}
\end{subfigure}
\begin{subfigure}[b]{0.32\textwidth}
\centering 
\includegraphics[width=5.6cm, height=3.5cm]{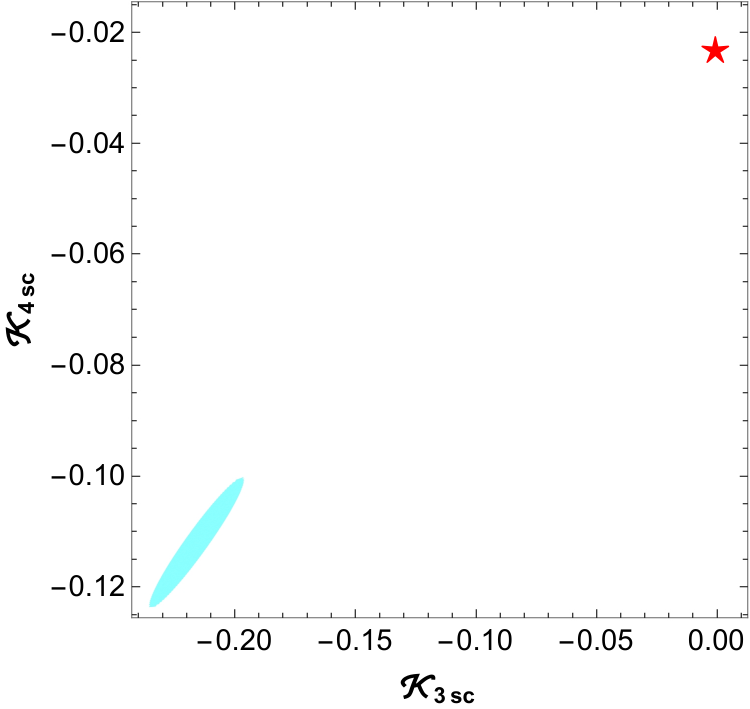} 
\caption{}
\end{subfigure}
\begin{subfigure}[b]{0.32\textwidth}
\centering 
\includegraphics[width=5.6cm, height=3.5cm]{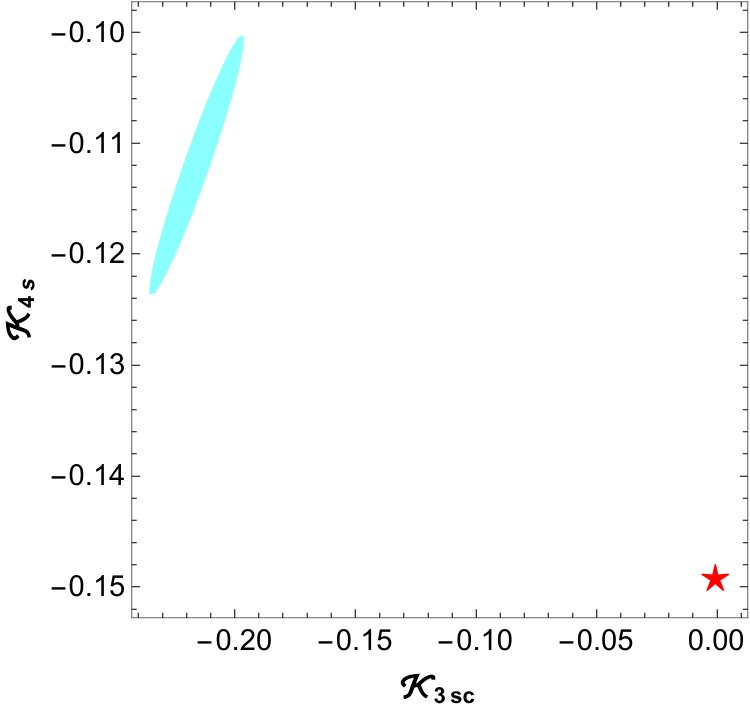}
\caption{}
\end{subfigure}
\begin{subfigure}[b]{0.32\textwidth}
\centering 
\includegraphics[width=5.6cm, height=3.5cm]{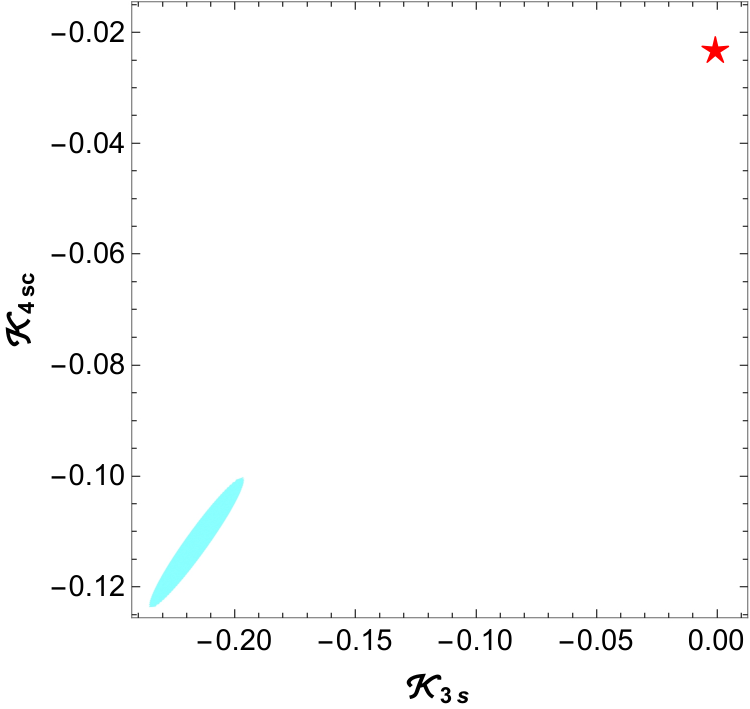}
\caption{}
\end{subfigure}
\begin{subfigure}[b]{0.32\textwidth}
\centering 
\includegraphics[width=5.6cm, height=3.5cm]{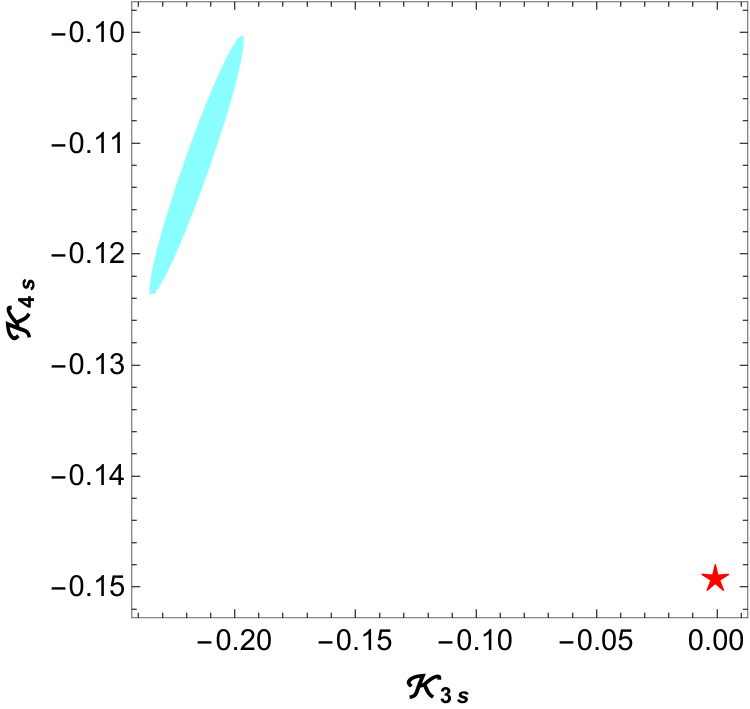} 
\caption{}
\end{subfigure}
\begin{subfigure}[b]{0.32\textwidth}
\centering 
\includegraphics[width=5.6cm, height=3.5cm]{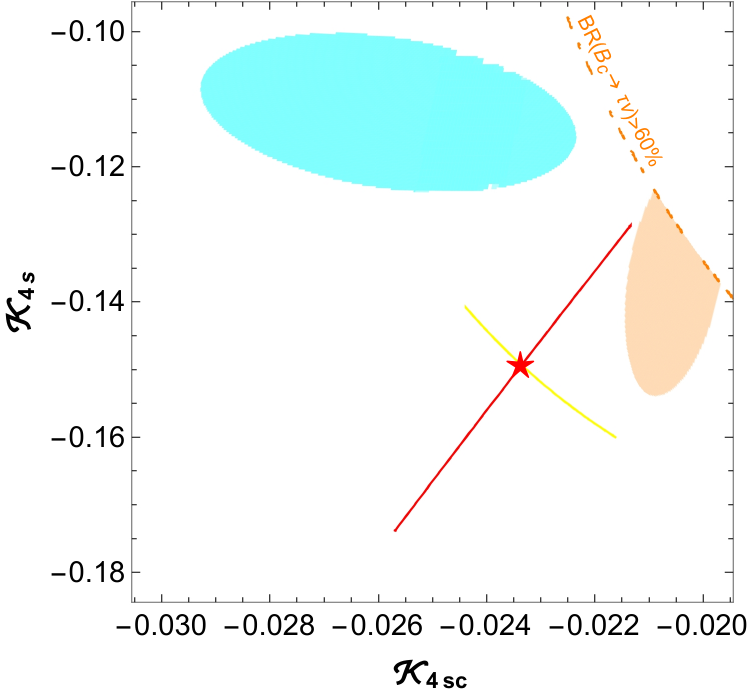}
\caption{}
\end{subfigure}

\caption{\label{corr-2d-2}
 Correlation plots of the angular observables $\mathcal{K}_{i}(q^{2})$.
The legend, color coding, and conventions are the same as in
Fig.~\ref{corr-2d}.
}
\end{figure}

In this section, we investigate the correlations among the angular observables
$\mathcal{K}_{i}$ with
$i=1ss,\,1cc,\,1c,\,2ss,\,2cc,\,2c,\,3sc,\,3s,\,4sc,\,4s$
using the $1\sigma$ allowed parameter space of the two-dimensional NP scenarios.
Such correlation studies provide complementary and incisive information beyond
individual observable analyses.
The explicit expressions employed in this analysis are given in
Appendix~\ref{obsRLc}, and the resulting correlation plots are shown in
Fig. \ref{corr-2d} - \ref{corr-2d-2}.

We find that the correlations exhibit clear NP-dependent patterns.
For the scenario
$\left(\Re\!\left[C_{S_{L}}=4C_{T}\right],
\Im\!\left[C_{S_{L}}=4C_{T}\right]\right)$ (cyan),
an inverse correlation is observed for
$\mathcal{K}_{1ss}$--$\mathcal{K}_{1cc}$ and
$\mathcal{K}_{1cc}$--$\mathcal{K}_{4s}$,
whereas direct correlations appear in
$\mathcal{K}_{1ss}$--$\mathcal{K}_{4s}$,
$\mathcal{K}_{2ss}$--$\mathcal{K}_{2c}$,
$\mathcal{K}_{2cc}$--$\mathcal{K}_{2c}$, and
$\mathcal{K}_{2ss}$--$\mathcal{K}_{4s}$.
In contrast, for the scenario
$\left(C_{S_{L}},C_{S_{R}}\right)$ (orange),
$\mathcal{K}_{1ss}$--$\mathcal{K}_{1cc}$ and
$\mathcal{K}_{1c}$--$\mathcal{K}_{2ss}$ show inverse correlations,
while $\mathcal{K}_{1ss}$--$\mathcal{K}_{4sc}$ and
$\mathcal{K}_{2cc}$--$\mathcal{K}_{4s}$ exhibit direct correlations,
highlighting their distinct dynamical origins.

We now examine the correlations among the observables $\mathcal{K}_{1c}$, $\mathcal{K}_{2ss}$,
$\mathcal{K}_{2cc}$, and $\mathcal{K}_{4s}$, which display the strongest NP sensitivity,
as already indicated in Sec.~\ref{Num-anlaysis}.
The pair $\mathcal{K}_{1c}$--$\mathcal{K}_{2ss}$ shows an inverse correlation for
$\left(C_{S_{L}},C_{S_{R}}\right)$ and a pronounced negative correlation for
$\left(\Re\!\left[C_{S_{L}}=4C_{T}\right],
\Im\!\left[C_{S_{L}}=4C_{T}\right]\right)$.
A similar behavior is observed for $\mathcal{K}_{1c}$--$\mathcal{K}_{2cc}$,
with a negative correlation in the $\left(\Re\!\left[C_{S_{L}}=4C_{T}\right],
\Im\!\left[C_{S_{L}}=4C_{T}\right]\right)$ case.
Conversely, $\mathcal{K}_{1c}$--$\mathcal{K}_{4s}$ exhibits a strong negative
correlation for the $\left(\Re\!\left[C_{S_{L}}=4C_{T}\right],
\Im\!\left[C_{S_{L}}=4C_{T}\right]\right)$ scenario, but a strong positive
correlation for the $\left(C_{S_{L}},C_{S_{R}}\right)$ case, signaling distinct interference
patterns.

For $\mathcal{K}_{2ss}$--$\mathcal{K}_{2cc}$, both NP scenarios $\left(\Re\!\left[C_{S_{L}}=4C_{T}\right],
\Im\!\left[C_{S_{L}}=4C_{T}\right]\right)$ and $\left(C_{S_{L}},C_{S_{R}}\right)$ lead to a direct
correlation, while $\mathcal{K}_{2ss}$--$\mathcal{K}_{4s}$ displays a strong
positive correlation in the $\left(\Re\!\left[C_{S_{L}}=4C_{T}\right],
\Im\!\left[C_{S_{L}}=4C_{T}\right]\right)$ and a pronounced negative
correlation in the $\left(C_{S_{L}},C_{S_{R}}\right)$ scenario.
Similarly, $\mathcal{K}_{2cc}$--$\mathcal{K}_{4s}$ remains directly correlated
for the $\left(\Re\!\left[C_{S_{L}}=4C_{T}\right],
\Im\!\left[C_{S_{L}}=4C_{T}\right]\right)$ case but becomes inversely correlated for $\left(C_{S_{L}},C_{S_{R}}\right)$.
These correlation flips shows contrasting behaviors underscore the sensitivity of angular correlations
to the phase structure of NP couplings.

The inverse-correlation patterns observed in the $\left(\Re\!\left[C_{S_{L}}=4C_{T}\right],
\Im\!\left[C_{S_{L}}=4C_{T}\right]\right)$ scenario point to destructive helicity interference and provide
indirect sensitivity to possible CP-violating phases in the NP sector.
In contrast, the $\left(C_{S_{L}},C_{S_{R}}\right)$ scenario predominantly yields constructive
correlations, consistent with CP-conserving interactions.
Therefore, correlation observables constitute a powerful diagnostic tool for
disentangling $\left(\Re\!\left[C_{S_{L}}=4C_{T}\right],
\Im\!\left[C_{S_{L}}=4C_{T}\right]\right)$ NP from $\left(C_{S_{L}},C_{S_{R}}\right)$ scenarios.

Finally, among the remaining combinations, $\mathcal{K}_{1ss}$ shows strong
positive correlations with $\mathcal{K}_{2ss}$, $\mathcal{K}_{2cc}$, and
$\mathcal{K}_{2c}$ in the $\left(\Re\!\left[C_{S_{L}}=4C_{T}\right],
\Im\!\left[C_{S_{L}}=4C_{T}\right]\right)$ case, while exhibiting an
opposite trend in the $\left(C_{S_{L}},C_{S_{R}}\right)$ scenario.
Conversely, $\mathcal{K}_{1cc}$ correlates positively with
$\mathcal{K}_{2ss}$, $\mathcal{K}_{2cc}$, and $\mathcal{K}_{2c}$ for
$\left(C_{S_{L}},C_{S_{R}}\right)$ but negatively for
$\left(\Re\!\left[C_{S_{L}}=4C_{T}\right],
\Im\!\left[C_{S_{L}}=4C_{T}\right]\right)$.
Additionally, $\mathcal{K}_{1c}$--$\mathcal{K}_{2c}$ exhibits a strong negative
correlation in the $\left(\Re\!\left[C_{S_{L}}=4C_{T}\right],
\Im\!\left[C_{S_{L}}=4C_{T}\right]\right)$ scenario, whereas
$\mathcal{K}_{2c}$--$\mathcal{K}_{4s}$ shows a strong positive correlation,
further emphasizing the distinct interference patterns and phase dependence of
the two NP scenarios.

\section{Conclusion}\label{sec6}

The experimental results for $R_{\tau/(\mu,e)}(D^{(*)})$ from BaBar, Belle, and LHCb show a $3.8\sigma$ deviation from SM predictions. Using the latest HFLAV results and imposing branching ratio constraints of $60\%$, $30\%$, and $10\%$ from the $B_c$ meson lifetime, we determine the WCs for various NP models. Our $\chi^2$ analysis indicates that the $(C_{V_L},C_{S_R})$ scenario achieves the highest $p$-value, $87\%$, and a maximum pull of $4.2\sigma$ from the SM predictions, demonstrating a significant improvement in fitting the data. Additionally, one-dimensional NP scenarios, $C_{V_L}$ also achieve a $p$-value of $93\%$, highlighting multiple pathways for NP contributions. NP scenarios are particularly sensitive to $B_c \to \tau \nu$ branching ratio constraints, underscoring the importance of precise measurements.

The baryonic decay $\Lambda_b \to \Lambda_c \tau \bar{\nu}_\tau$ offers a complementary avenue to explore the $R_{\tau/(\mu,e)}(D^{(*)})$ anomaly. Unlike mesonic decays, where form factors are well studied, the $\Lambda_b \to \Lambda_c$ form factors are not yet experimentally determined, requiring reliance on lattice QCD computations. After constraining the NP WCs, we investigate their impact on ten angular observables in the full five-folded cascade decay $\Lambda_b^0 \to \Lambda_c^+ (\to \Lambda^0 \pi^+) \tau^- (\to \pi^- \nu_\tau) \bar{\nu}_\tau$. Among these, $\mathcal{K}_{1c}$, $\mathcal{K}_{2ss}$, $\mathcal{K}_{2cc}$, and $\mathcal{K}_{4s}$ show the largest deviations from SM predictions, with the $(\Re[C_{S_L}=4C_T], \Im[C_{S_L}=4C_T])$ scenario producing the most pronounced shifts across the $q^2$ spectrum, while $(C_{S_L},C_{S_R})$ exhibits comparatively smaller effects.

Correlation analysis of these observables reveals distinct NP signatures. In the $(\Re[C_{S_L}=4C_T], \Im[C_{S_L}=4C_T])$ scenario, $\mathcal{K}_{1c}-\mathcal{K}_{2ss}$, $\mathcal{K}_{1c}-\mathcal{K}_{2cc}$, and $\mathcal{K}_{1c}-\mathcal{K}_{4s}$ show inverse correlations, whereas $\mathcal{K}_{2ss}-\mathcal{K}_{2cc}$ and $\mathcal{K}_{2ss}-\mathcal{K}_{4s}$ remain directly correlated, signaling destructive helicity interference and potential CP-violating phases. In contrast, the $(C_{S_L},C_{S_R})$ scenario exhibits constructive correlations consistent with CP-conserving interactions. These patterns underscore the power of correlation observables in distinguishing NP scenarios.

In summary, the baryonic decay $\Lambda_b^0 \to \Lambda_c^+ (\to \Lambda^0 \pi^+) \tau^- (\to \pi^- \nu_\tau) \bar{\nu}_\tau$ provides a unique and sensitive probe for NP, complementing mesonic $R_{\tau/(\mu,e)}(D^{(*)})$ anomalies. The angular observables $\mathcal{K}_{1c}$, $\mathcal{K}_{2ss}$, $\mathcal{K}_{2cc}$, and $\mathcal{K}_{4s}$ are particularly sensitive to $(\Re[C_{S_L}=4C_T], \Im[C_{S_L}=4C_T])$ and  $(C_{S_L},C_{S_R})$ NP effects, with distinct correlation patterns revealing possible CP-violating phases. These results establish baryonic modes as a promising laboratory for future experimental tests at LHCb Upgrade II and next-generation $b$-decay facilities.

\section*{Data Availability Statement}

Data sharing not applicable to this article as no datasets were generated or analysed during the current study.

\section*{Code Availability Statement}

Code/Software sharing not applicable to this article as no code/software was generated or analysed during the current study.

\appendix
\numberwithin{equation}{section}

\section{Expressions of Physical observables in terms of NP WCs}\label{AppendixA}
The expressions of the physical observables used to fit the data are given below \cite{Arslan:2023wgk}:
\begin{eqnarray}
R_{\tau/{\mu,e}}(D) & = & R_{\tau/{\mu,e}}^{\text{SM}}(D)\left\{ \left|1+\widetilde{C}_{V_{L}}+\widetilde{C}_{V_{R}}\right|^{2}+1.01\left|\widetilde{C}_{S_{L}}+\widetilde{C}_{S_{R}}\right|^{2}+0.84\left|\widetilde{C}_{T}\right|^{2}\right.\nonumber \\
 &  & \left.+1.49\Re\left[\left(1+\widetilde{C}_{V_{L}}+\widetilde{C}_{V_{R}}\right)\left(\widetilde{C}_{S_{L}}+\widetilde{C}_{S_{R}}\right)^{*}\right]+1.08\Re\left[\left(1+\widetilde{C}_{V_{L}}+\widetilde{C}_{V_{R}}\right)\left(\widetilde{C}_{T}\right)^{*}\right]\right\} ,\label{eqn1}\\
R_{\tau/{\mu,e}}(D^{*}) & = & R_{\tau/{\mu,e}}^{\text{SM}}(D^{*})\left\{ \left|1+\widetilde{C}_{V_{L}}\right|^{2}+\left|\widetilde{C}_{V_{R}}\right|^{2}+0.04\left|\widetilde{C}_{S_{L}}-\widetilde{C}_{S_{R}}\right|^{2}+16\left|\widetilde{C}_{T}\right|\right.\nonumber \\
 &  & \left.-1.83\Re\left[\left(1+\widetilde{C}_{V_{L}}\right)\left(\widetilde{C}_{V_{R}}\right)^{*}\right]-0.11\Re\left[\left(1+\widetilde{C}_{V_{L}}-\widetilde{C}_{V_{R}}\right)\left(\widetilde{C}_{S_{L}}-\widetilde{C}_{S_{R}}\right)^{*}\right]\right.\nonumber \\
 &  & \left.-5.17\Re\left[\left(1+\widetilde{C}_{V_{L}}\right)\left(\widetilde{C}_{T}\right)^{*}\right]+6.60\Re\left[\widetilde{C}_{V_{R}}\left(\widetilde{C}_{T}\right)^{*}\right]\right\} ,\label{eqn2}\\
 P_{\tau}\left(D\right) & = & P_{\tau}^{SM}\left(D\right)\left(\frac{R_{\tau/{\mu,e}}(D)}{R_{\tau/{\mu,e}}^{\text{SM}}(D)}\right)^{-1}\left\{ \left|1+\widetilde{C}_{V_{L}}\widetilde{C}_{V_{R}}\right|^{2}+3.04\left|\widetilde{C}_{S_{L}}+\widetilde{C}_{S_{R}}\right|^{2}+0.17\left|\widetilde{C}_{T}\right|^{2}\right.\nonumber \\
 &  & \left.+4.50\Re\left[\left(1+\widetilde{C}_{V_{L}}+\widetilde{C}_{V_{R}}\right)\left(\widetilde{C}_{S_{L}}+\widetilde{C}_{S_{R}}\right)^{*}\right]-1.09\Re\left[\left(1+\widetilde{C}_{V_{L}}+\widetilde{C}_{V_{R}}\right)\left(\widetilde{C}_{T}\right)^{*}\right]\right.\label{eqn8}\\
P_{\tau}\left(D^{*}\right) & = & P_{\tau}^{SM}\left(D^{*}\right)\left(\frac{R_{\tau/{\mu,e}}(D^{*})}{R_{\tau/{\mu,e}}^{\text{SM}}(D^{*})}\right)^{-1}\left\{ \left|1+\widetilde{C}_{V_{L}}\right|^{2}+\left|\widetilde{C}_{V_{R}}\right|^{2}-0.07\left|\widetilde{C}_{S_{L}}-\widetilde{C}_{S_{R}}\right|^{2}-1.85\left|\widetilde{C}_{T}\right|^{2}\right.\nonumber \\
 &  & \left.-1.79\Re\left[\left(1+\widetilde{C}_{V_{L}}\right)\left(\widetilde{C}_{V_{R}}\right)^{*}\right]+0.23\Re\left[\left(1+\widetilde{C}_{V_{L}}-\widetilde{C}_{V_{R}}\right)\left(\widetilde{C}_{S_{L}}-\widetilde{C}_{S_{R}}\right)^{*}\right]\right.\nonumber \\
 &  & \left.\left.-3.47\Re\left[\left(1+\widetilde{C}_{V_{L}}\right)\left(\widetilde{C}_{T}\right)^{*}\right]+4.41\Re\left[\widetilde{C}_{V_{R}}\left(\widetilde{C}_{T}\right)^{*}\right]\right\} \right.\label{eqn3}\\
F_{L}\left(D^{*}\right) & = & F_{L}^{SM}\left(D^{*}\right)\left(\frac{R_{\tau/{\mu,e}}(D^{*})}{R_{\tau/{\mu,e}}^{\text{SM}}(D^{*})}\right)^{-1}\left\{ \left|1+\widetilde{C}_{V_{L}}-\widetilde{C}_{V_{R}}\right|^{2}+0.08\left|\widetilde{C}_{S_{L}}-\widetilde{C}_{S_{R}}\right|^{2}+6.9\left|\widetilde{C}_{T}\right|^{2}\right.\nonumber \\
 &  & \left.\left.-0.25\Re\left[\left(1+\widetilde{C}_{V_{L}}-\widetilde{C}_{V_{R}}\right)\left(\widetilde{C}_{S_{L}}-\widetilde{C}_{S_{R}}\right)^{*}\right]-4.3\Re\left[\left(1+\widetilde{C}_{V_{L}}-\widetilde{C}_{V_{R}}\right)\left(\widetilde{C}_{T}\right)^{*}\right]\right\} \right.\label{eqn4}\\
\end{eqnarray}
Similarly, the expression of branching ratio of the $B_{c}\to\tau\bar{\nu}_{\tau}$
decay read as 
\begin{eqnarray*}
\mathcal{B}\left(B_{c}^{-}\rightarrow\tau^{-}\bar{\nu}_{\tau}\right) & = & \mathcal{B}\left(B_{c}^{-}\rightarrow\tau^{-}\bar{\nu}_{\tau}\right)^{\text{SM}}\left\{ \left|1+\widetilde{C}_{V_{L}}-\widetilde{C}_{V_{R}}-4.35\left(\widetilde{C}_{S_{L}}-\widetilde{C}_{S_{R}}\right)\right|^{2}\right\} ,
\end{eqnarray*}
where in the SM $\mathcal{B}\left(B_{c}^{-}\rightarrow\tau^{-}\bar{\nu}_{\tau}\right)^{\text{SM}}\approx 0.022$  \cite{Iguro:2022yzr}.

\section{The detailed calculation of the measurable angular distribution}\label{obsRLc}

The differential decay rate of the unpolarized $\Lambda_{b}^{0}\left(p_1,m_1\right)\rightarrow\Lambda_{c}^{+}\left(p_2,m_2\right)\left[\rightarrow\Lambda^{0}\pi^{+}\right]\tau^{-}\left(\rightarrow\pi^{-}\nu_{\tau}\right)\bar{\nu}_{\tau}$
decay can be written as:
\begin{equation}
d\Gamma=\frac{1}{2m_{1}}\left|\mathcal{M}\right|^{2}d\Pi_{5}\left(p_{1};p_{2},p_{\pi^{+}},p_{\pi^{-}},p_{\nu},p_{\bar{\nu}}\right)\label{B1}\;,
\end{equation}
where the prefactor $1/2m_{1}$ comes from relativistic flux normalization,
$\left|\mathcal{M}\right|^{2}$ encodes all the dynamics (including
spins, helicities), the matrix elements $\Lambda_b \to \Lambda_c$ transitions for various four-quark operators and their corresponding WCs. It square reads as:
\begin{align}
\left|\mathcal{M}\right|^{2} & =\sum_{\lambda_{3}}\frac{1}{2}\sum_{\lambda_{1}}\left|\mathcal{M}_{\lambda_{1}}^{\lambda_{3}}\right|^{2},\notag\\
\mathcal{M}_{\lambda_{1}}^{\lambda_{3}} & =\sum_{\lambda_{2},\lambda_{\tau}}\frac{\mathcal{M}_{\lambda_{1}}^{\lambda_{2},\lambda_{\tau}}\left(\Lambda_{b}\rightarrow\Lambda_{c}\tau\bar{\nu}_{\tau}\right)\mathcal{M}_{\lambda_{2}}^{\lambda_{3}}\left(\Lambda_{c}\rightarrow\Lambda\pi^{+}\right)\mathcal{M}_{\lambda_{\tau}}\left(\tau\rightarrow\pi^{-}\nu_{\tau}\right)}{\left(p_{2}^{2}-m_{2}^{2}+im_{2}\Gamma_{2}\right)\left(p_{\tau}^{2}-m_{\tau}^{2}+im_{\tau}\Gamma_{\tau}\right)}\;\label{B2}.
\end{align}
Here,  $\lambda_1$, $\lambda_2$ and $\lambda_\tau$ represent the helicities of $\Lambda_b$, $\Lambda_c$ and $\tau$, respectively.  We drop
the helicity indices $\lambda_{\pi^{\pm}}$ as they are scalars and hence they do not carry and spin. In above equation we assumed that due to the unstable nature of $\Lambda_{c}$ and $\tau$, they will decay further and act as resonances, i.e., 
\begin{equation}
\Lambda_{c}\rightarrow\Lambda\pi^{+},\quad\tau^{-}\rightarrow\pi^{-}\nu_{\tau}\;.
\end{equation}
When a particle decays via an intermediate unstable state like $\Lambda_{c}$ or $\tau$, its amplitude contains a propagator, which accounts
for the possibility that the particle goes slightly off-shell. Each
unstable intermediate state has a Breit-Wigner propagator:
\begin{equation}
\frac{1}{p^{2}-m^{2}+im\Gamma}\;.
\end{equation}
In our case, $\Gamma_2$ and $\Gamma_\tau$ correspond to the total decay widths of $\Lambda_c$ and $\tau$, respectively. 

The five-body phase space describes the allowed kinematics
for the $5-$body final state, and this can be chopped as:
\begin{equation}
d\Pi_{5}\left(p_{1};p_{2},p_{\pi^{+}},p_{\pi^{-}},p_{\nu},p_{\bar{\nu}}\right)=\frac{dq^{2}dp_{\tau}^{2}dp_{2}^{2}}{\left(2\pi\right)^{3}}d\Pi_{2}\left(p_{1};q,p_{2}\right)d\Pi_{2}\left(p_{2};p_{3},p_{\pi^{+}}\right)d\Pi_{2}\left(q;p_{\tau},p_{\bar{\nu}}\right)d\Pi_{2}\left(p_{\tau};p_{\pi^{-}},p_{\nu}\right)\;,
\end{equation}
where, $d\Pi_{5}$ is $5-$body Lorentz-invariant phase space, integrated over the invariant mass of intermediate states, i.e., $dq^{2}$, and $dp_{\tau}^{2}$. The 
$\lambda_{x}$ stands for the helicity of the particle $x$.  As neutrinos (anti-neutrinos) are left-(right) handed in the SM, therefore, their helicities are fixed: $\lambda_{\bar{\nu}_{\tau}}\left(\lambda_{\nu_{\tau}}\right)$
to $\frac{1}{2}\left(-\frac{1}{2}\right)$. These are not summed over because
they are dictated by the structure of the $V-A$ in the weak-interactions.

For a given WEH, one can express the helicity
amplitude of $\Lambda_{b}\rightarrow\Lambda_{c}\tau\bar{\nu}_{\tau}$
decay as:
\begin{equation}
\mathcal{M}_{\lambda_{1}}^{\lambda_{2},\lambda_{\tau}}\left(\Lambda_{b}\rightarrow\Lambda_{c}\tau\bar{\nu}_{\tau}\right)=\sqrt{2}G_{F}V_{cb}\left(H_{\lambda_{1}}^{\lambda_{2}}L^{\lambda_{\tau}}+\sum\eta_{\lambda}H_{\lambda_{1}}^{\lambda_{2},\lambda}L_{\lambda}^{\lambda_{\tau}}+\sum\eta_{\lambda}\eta_{\lambda^{'}}H_{\lambda_{1}}^{\lambda_{2},\lambda,\lambda^{'}}L_{\lambda,\lambda^{'}}^{\lambda_{\tau}}\right).
\end{equation}
Here, $\lambda^{\left(\prime\right)}=t,\pm1,0$ indicates the
helicity of the virtual vector boson $W^{*}$. The number of helicity
indices depends on the Lorentz structure of the effective operator.
The factor $\eta$ that appears here is due to the use of the completeness
relation of the polarization vectors of the virtual vector boson.
The hadronic and leptonic helicity amplitudes are, respectively, defined
as
\begin{eqnarray}
H_{\lambda_{1}}^{\lambda_{2}} & =&\left\langle \Lambda_{c}\left(\lambda_{2}\right)\left|g_{S}\left(\bar{c}b\right)+g_{P}\left(\bar{c}\gamma_{5}b\right)\right|\Lambda_{b}\left(\lambda_{1}\right)\right\rangle ,\notag\\
H_{\lambda_{1}}^{\lambda_{2},\lambda} & =&\epsilon^{\mu*}\left(\lambda\right)\left\langle \Lambda_{c}\left(\lambda_{2}\right)\left|g_{V}\left(\bar{c}\gamma_{\mu}b\right)+g_{A}\left(\bar{c}\gamma_{\mu}\gamma_{5}b\right)\right|\Lambda_{b}\left(\lambda_{1}\right)\right\rangle ,\notag\\
H_{\lambda_{1}}^{\lambda_{2},\lambda,\lambda^{\prime}} & =&g_{T}\epsilon^{\mu*}\left(\lambda\right)\epsilon^{\nu*}\left(\lambda^{\prime}\right)\left\langle \Lambda_{c}\left(\lambda_{2}\right)\left|\bar{c}i\sigma_{\mu\nu}\gamma_{\mu}\left(1-\gamma_{5}\right)b\right|\Lambda_{b}\left(\lambda_{1}\right)\right\rangle ,
\end{eqnarray}
and
\begin{eqnarray}
L^{\lambda_{\tau}} & =&\left\langle \tau^{-}\left(\lambda_{\tau}\right)\bar{\nu}\left|\tau P_{L}\nu\right|0\right\rangle ,\notag\\
L_{\lambda}^{\lambda_{\tau}} & =&\epsilon^{\mu}\left(\lambda\right)\left\langle \tau^{-}\left(\lambda_{\tau}\right)\bar{\nu}\left|\tau\gamma_{\mu}P_{L}\nu\right|0\right\rangle ,\notag\\
L_{\lambda,\lambda^{\prime}}^{\lambda_{\tau}} & =&\left(-i\right)\epsilon^{\mu}\left(\lambda\right)\epsilon^{\nu}\left(\lambda^{\prime}\right)\left\langle \tau^{-}\left(\lambda_{\tau}\right)\bar{\nu}\left|\tau\sigma_{\mu\nu}P_{L}\nu\right|0\right\rangle ,
\end{eqnarray}
where $\epsilon^{\mu}\left(\lambda\right)$ is the polarization vector
of the virtual vector boson with helicity $\lambda$ and $P_L = \frac{1}{2}\left(1-\gamma_5\right)$ and $P_L = \frac{1}{2}\left(1+\gamma_5\right)$.

Using the narrow width $\left(\Gamma_{a}\ll m_{a}\right)$ approximation
\begin{equation}
\frac{1}{\left(p_{a}^{2}-m_{a}^{2}\right)^{2}+im_{a}^{2}\Gamma_{a}^{2}}=\frac{\pi}{m_{a}\Gamma_{a}}\delta\left(p_{a}^{2}-m_{a}^{2}\right),\quad\left(a=\Lambda_{c},\tau\right)
\end{equation}
and integrating over $dp_{2}^{2}dp_{\tau}^{2}$, one can obtain
two on-shell relations $p_{2}^{2}=m_{2}^{2}$ and $p_{\tau}^{2}=m_{\tau}^{2}$,
as well as 
\begin{eqnarray}
d\Gamma & =&\frac{dq^{2}}{2^{5}\pi m_{1}m_{2}\Gamma_{2}m_{\tau}\Gamma_{\tau}}d\Pi_{2}\left(p_{1};q,p_{2}\right)d\Pi_{2}\left(p_{2};p_{3},p_{\pi^{+}}\right)d\Pi_{2}\left(q;p_{\tau},p_{\bar{\nu}}\right)d\Pi_{2}\left(p_{\tau};p_{\pi^{-}},p_{\nu}\right)\notag\\
 & \times&\sum_{\lambda_{1},\lambda_{3}}\left|\sum_{\lambda_{2},\lambda_{\tau}}\mathcal{M}_{\lambda_{1}}^{\lambda_{2},\lambda_{\tau}}\left(\Lambda_{b}\rightarrow\Lambda_{c}\tau\bar{\nu}_{\tau}\right)\mathcal{M}_{\lambda_{2}}^{\lambda_{3}}\left(\Lambda_{c}\rightarrow\Lambda\pi^{+}\right)\mathcal{M}_{\lambda_{\tau}}\left(\tau\rightarrow\pi^{-}\nu_{\tau}\right)\right|^{2}.
\end{eqnarray}
Since each individual two-body phase space or helicity amplitude is
Lorentz invariant, one can calculate each part of $d\Gamma$ in different
reference frames. In this work, we should consider three measurable
reference frames -- the $\Lambda_{b}$ rest frame, the $\Lambda_{c}$
rest frame and the $\tau^{-}\bar{\nu}_{\tau}$ center-of-mass frame. 

\subsection{$\Lambda_{b}$ rest frame}

In the $\Lambda_{b}$ rest frame, we calculate the hadronic helicity amplitudes
$H$ and the two-body phase space $d\Pi_{2}\left(p_{1};q,p_{2}\right)$.
The momenta of $\Lambda_{b}$, $\Lambda_{c}$ and $W^{*}$ are respectively
as
\begin{equation}
p_{1}^{\mu}=\left(m_{1},0,0,0\right),\quad p_{2}^{\mu}=\left(E_{2},0,0,\left|\bf{q}\right|\right),\quad q^{\mu}=\left(q_{0},0,0,-\left|\bf{q}\right|\right)
\end{equation}
By the virtue of the Lorentz condition $q_{\mu}\epsilon^{\mu}=0$,  the polarization
vector of virtual boson becomes
\begin{eqnarray}
\epsilon^{\mu}\left(t\right) & =&\frac{1}{\sqrt{q^{2}}}\left(q^{0},0,0,q^{3}\right)=\frac{1}{\sqrt{q^{2}}}\left(q_{0},0,0,-\left|\bf{q}\right|\right),\notag\\
\epsilon^{\mu}\left(0\right) & =&\frac{1}{\sqrt{q^{2}}}\left(-q^{3},0,0,q^{0}\right)=\frac{1}{\sqrt{q^{2}}}\left(\left|\bf{q}\right|,0,0,-q_{0}\right),\notag\\
\epsilon^{\mu}\left(\pm1\right) & =&\frac{1}{\sqrt{2}}\left(0,\pm1,i,0\right).
\end{eqnarray}

The spinors of $\Lambda_{b}$ and $\Lambda_{c}$ are described by
the Dirac spinors $u\left(p,s\right)$, which satisfies the Dirac equation. 
\begin{equation}
\left(\slashed{p}-m\right)u\left(p,s\right)=0,
\end{equation}
where $p^{\mu}$, $m$ are the four-momentum and mass of the baryon, $u\left(p,s\right)$
is a four-component spinor corresponding to a specific helicity $s$.

In the rest frame of $\Lambda_{b}$, its spinor reads as 
\begin{equation}
u_{1}\left(p_{1},s\right)=\sqrt{2m_{1}}\left(\begin{array}{c}
\chi_{s}\\
0
\end{array}\right). 
\end{equation}
The corresponding
spinors of $\Lambda_{c}$ is then
\[
u_{2}\left(p_{2},s\right)=\left(\begin{array}{c}
\beta_{2}^{+}\chi_{s}\\
\pm\beta_{2}^{-}\chi_{s}
\end{array}\right),\beta_{x}^{\pm}=\sqrt{E_{x}\pm m_{x}}\;.
\]
Also
\[
\bar{u}_{2}\left(p_{2},s\right)=u_{2}^{\dagger}\gamma^{0}=\left(\begin{array}{cc}
\beta_{2}^{+}\chi_{s} & \pm\beta_{2}^{-}\chi_{s}\end{array}\right)\left(\begin{array}{cc}
I & 0\\
0 & -I
\end{array}\right)=\left(\begin{array}{cc}
\beta_{2}^{+}\chi_{s}^{\dagger} & \mp\beta_{2}^{-}\chi_{s}^{\dagger}\end{array}\right)\;,
\]
and 
\[
\left|q\right|=\frac{\lambda^{1/2}\left(m_{1}^{2},m_{2}^{2},q^{2}\right)}{2m_{1}},\quad E_{2}=\frac{m_{1}^{2}+m_{2}^{2}-q^{2}}{2m_{1}},
\]
where K$\ddot{a}$llen function $\lambda\left(a,b,c\right)\equiv a^{2}+b^{2}+c^{2}-2ab-2bc-2ca$,
implies
\[
\lambda^{1/2}\left(m_{1}^{2},m_{2}^{2},q^{2}\right)=Q_{+}Q_{-},\quad Q_{\pm}=\left(m_{1}\pm m_{2}\right)^{2}-q^{2}.
\]

Most generally, the $n-$body phase space is defined as 
\begin{equation}
d\Pi_{n}\left(P,p_{i}\right)=\left(\Pi_{i}\frac{d^{3}p_{i}}{\left(2\pi\right)^{3}2E_{i}}\right)\left(2\pi\right)^{4}\delta^{\left(4\right)}\left(P-\sum p_{i}\right)\label{B14},
\end{equation}
and the corresponding two-body phase space  $d\Pi_{2}\left(p_{1};q,p_{2}\right)$ has the form \cite{ParticleDataGroup:2024cfk}

\[
d\Pi_{2}\left(p_{1};q,p_{2}\right)=\frac{1}{4\pi}\frac{\left|\textbf{q}\right|}{m_{1}},
\]
with $\left|\textbf{q}\right|=\frac{1}{2m_{1}}\sqrt{\lambda\left(m_{1}^{2},m_{2}^{2},q^{2}\right)}$,
$E_{2}=\frac{1}{2m_{1}}\left(m_{1}^{2}+m_{2}^{2}-q^{2}\right)$, $q_{0}=\frac{1}{2m_{1}}\left(m_{1}^{2}-m_{2}^{2}+q^{2}\right)$
and $\lambda\left(m_{1}^{2},m_{2}^{2},q^{2}\right)=Q_{+}Q_{-}.$

\subsection{$\Lambda_{c}$ rest frame}

In the $\Lambda_{c}$ rest frame, we calculate helicity amplitude
$\mathcal{M}\left(\Lambda_{c}\rightarrow\Lambda\pi^{+}\right)$ and
two-body phase space $d\Pi_{2}\left(p_{2};p_{3},p_{\pi^{+}}\right)$.
The momenta of $\Lambda_{c}$ and $\Lambda$ are respectively as
\[
\tilde{p}_{2}^{\mu}=\left(m_{2},0,0,0\right),\quad p_{3}^{\mu}=\left(E_{3},\left|\textbf{p}_{3}\right|\sin\theta_{3},0,\left|\textbf{p}_{3}\right|\cos\theta_{3}\right).
\]
The tilde over the momenta are only used to distinguish the representations
of the same kinematic quantity in different reference frames. The
spinors of $\Lambda_{c}$ and $\Lambda$ are given by:
\[
u\left(p,s\right)=\sqrt{E+m}\left(\begin{array}{c}
\chi_{s}\\
\frac{\bf{\sigma}\cdot \bf{p}}{E+m}\chi_{s}
\end{array}\right),
\]

\begin{align*}
u_{3}\left(p_{3},\frac{1}{2}\right) & =\left(\begin{array}{cccc}
\beta_{3}^{+}\cos\left(\frac{\theta_{3}}{2}\right), & \beta_{3}^{+}\sin\left(\frac{\theta_{3}}{2}\right), & \beta_{3}^{-}\cos\left(\frac{\theta_{3}}{2}\right), & \beta_{3}^{-}\sin\left(\frac{\theta_{3}}{2}\right)\end{array}\right)^{T},\\
u_{3}\left(p_{3},-\frac{1}{2}\right) & =\left(\begin{array}{cccc}
-\beta_{3}^{+}\sin\left(\frac{\theta_{3}}{2}\right), & \beta_{3}^{+}\cos\left(\frac{\theta_{3}}{2}\right), & \beta_{3}^{-}\sin\left(\frac{\theta_{3}}{2}\right), & -\beta_{3}^{-}\cos\left(\frac{\theta_{3}}{2}\right)\end{array}\right)^{T}.
\end{align*}
The corresponding two-body phase space is then
\begin{equation}
d\Pi_{2}\left(p_{2};p_{3},p_{\pi^{+}}\right)=\frac{1}{8\pi}\frac{\left|\textbf{p}_{3}\right|}{E_{3}+E_{\pi^{+}}}d\cos\theta_{3}=\frac{1}{8\pi}\frac{\left|\textbf{p}_{3}\right|}{m_{2}}d\cos\theta_{3},
\end{equation}
with
\begin{align*}
\left|p_{3}\right| & =\frac{1}{2m_{2}}\lambda^{1/2}\left(m_{2}^{2},m_{3}^{2},m_{\pi^{+}}^{2}\right)=p^{*},\\
E_{3} & =\frac{1}{2m_{2}}\left(m_{2}^{2}+m_{3}^{2}-m_{\pi^{+}}^{2}\right).
\end{align*}

Now, as the pion is a pseudoscalar (spin-0, odd parity),  the only allowed Lorentz scalars (respecting parity) are
$\bar{u}u$ and $\bar{u}\gamma^{5}u$. So, the helicity amplitude becomes
\begin{equation}
\mathcal{M}_{\lambda_{2}}^{\lambda_{3}}\left(\Lambda_{c}\rightarrow\Lambda\pi^{+}\right)=i\bar{u}_{3}\left(\lambda_{3}\right)\left(A+B\gamma_{5}\right)u_{2}\left(\lambda_{2}\right)\;.
\end{equation}
This describes a scalar--pseudoscalar structure, where $A$: scalar
coupling (parity conserving), $B$: pseudoscalar coupling (parity violating). This is a valid low-energy hadronic parameterization of the weak decay
$\Lambda_{c}\rightarrow\Lambda\pi^{+}$

\begin{equation}
\mathcal{M}_{\lambda_{2}}^{\lambda_{3}}\left(\Lambda_{c}\rightarrow\Lambda\pi^{+}\right)=i\chi_{s}^{\dagger}\left(\lambda_{3}\right)\left(S+P\bf{\sigma}.\hat{p}_{3}\right)\chi_{s}\left(\lambda_{2}\right)\;,
\end{equation}
where $S=\sqrt{2m_{2}}\beta_{3}^{+}$ and $P=-\sqrt{2m_{2}}\beta_{3}^{-}$. $\bf{\sigma}=\left(\begin{array}{ccc}
\sigma^{1}, & \sigma^{2}, & \sigma^{3}\end{array}\right)$ is a vector composed of Pauli matrices. $\hat{p}_{3}$ is the unit
vector along the direction of $\Lambda$ baryon. The four helicity
amplitudes are:
\begin{align*}
\mathcal{M}_{1/2}^{1/2}\left(\Lambda_{c}\rightarrow\Lambda\pi^{+}\right)  =i\left(S+P\right)\cos\left(\frac{\theta_{3}}{2}\right),\quad\mathcal{M}_{-1/2}^{1/2}\left(\Lambda_{c}\rightarrow\Lambda\pi^{+}\right) =i\left(S+P\right)\sin\left(\frac{\theta_{3}}{2}\right),
\end{align*}
\begin{align}
\mathcal{M}_{1/2}^{-1/2}\left(\Lambda_{c}\rightarrow\Lambda\pi^{+}\right) =-i\left(S-P\right)\sin\left(\frac{\theta_{3}}{2}\right),\quad\mathcal{M}_{-1/2}^{1/2}\left(\Lambda_{c}\rightarrow\Lambda\pi^{+}\right) =i\left(S-P\right)\cos\left(\frac{\theta_{3}}{2}\right)\;.
\end{align}
For a two body decay $\Lambda_{c}\rightarrow\Lambda\pi^{+}$
\begin{equation}
\Gamma^{\lambda_{3}=1/2}\left(\Lambda_{c}\rightarrow\Lambda\pi^{+}\right)=\frac{\left|\textbf{p}_{3}\right|}{8\pi m_{3}^{2}}\frac{1}{2}\left|\mathcal{M}^{1/2}\right|^{2}\;.
\end{equation}
The factor of $\frac{1}{2}$ comes from averaging over the initial $\Lambda_{c}$
spin. Therefore
\begin{equation}
\Gamma^{\lambda_{3}=1/2}\left(\Lambda_{c}\rightarrow\Lambda\pi^{+}\right)=\frac{\left|\textbf{p}_{3}\right|}{16\pi m_{3}^{2}}\left|S+P\right|^{2}\;.
\end{equation}
In the same way, one can obtain the decay rate for helicity $\lambda_{3}=-1/2$:
\begin{equation}
\Gamma^{\lambda_{3}=-1/2}\left(\Lambda_{c}\rightarrow\Lambda\pi^{+}\right)=\frac{\left|\textbf{p}_{3}\right|}{16\pi m_{3}^{2}}\left|S-P\right|^{2}\;.
\end{equation}
By using the two helicity amplitudes, the angular asymmetry parameter has the form
$\alpha_{\Lambda_{c}}$ has the form
\begin{equation}
\alpha_{\Lambda_{c}}=\frac{2\Re\left(S^{*}P\right)}{\left|S\right|^{2}+\left|P\right|^{2}},
\end{equation}
and one can immediately get the relations
\begin{align*}
\frac{\Gamma^{\lambda_{3}=1/2}}{\Gamma^{\lambda_{3}=1/2}+\Gamma^{\lambda_{3}=-1/2}} & =\frac{\left|S+P\right|^{2}}{\left|S+P\right|^{2}+\left|S-P\right|^{2}}\\
 & =\frac{1}{2}\left(1+\alpha_{\Lambda_{c}}\right)\;.
\end{align*}
Similarly
\[
\frac{\Gamma^{\lambda_{3}=-1/2}}{\Gamma^{\lambda_{3}=1/2}+\Gamma^{\lambda_{3}=-1/2}}=\frac{1}{2}\left(1-\alpha_{\Lambda_{c}}\right).
\]

\subsection{$\tau^{-}\bar{\nu}_{\tau}-$ center of mass frame}
In this frame, we calculate the leptonic helicity amplitudes $L$
and the helicity amplitudes $\mathcal{M}_{\lambda_{\tau}}\left(\tau^-\to\pi^{-}\nu_{\tau}\right)$,
as well as the two-body phase spaces $d\Pi_{2}\left(q;p_{\tau},p_{\bar{\nu}}\right)$
and $d\Pi_{2}\left(p_{\tau};p_{\pi^{-}},p_{\nu}\right)$. The momenta
of $\pi^{-}$ is defined as $p_{\pi}=\left(E_{\pi},\left|\textbf{p}_{\pi}\right|\hat{p}_{\pi}\right)$
with
\[
\hat{p}_{\pi}=\left(\sin\theta_{\pi}\cos\phi_{\pi},\sin\theta_{\pi}\sin\phi_{\pi},\cos\theta_{\pi}\right),
\]
is the unit vector along the direction of $\pi^{-}$. In this frame,
the polarization vectors of the virtual vector boson $W^{*}$ are
changed to 
\[
\tilde{\epsilon}^{\mu}\left(t\right)=\left(1,0,0,0\right),\quad\tilde{\epsilon}^{\mu}\left(\pm1\right)=\frac{1}{\sqrt{2}}\left(0,\pm1,-i,0\right),\quad\tilde{\epsilon}^{\mu}\left(0\right)=\left(0,0,0,-1\right)\;.
\]
Because $W^{*}$ is at rest in this frame, $4-$momentum of $W^{*}$ is
$q^{\mu}=\left(q^{0},\textbf{0}\right)=\left(\sqrt{q^{2}},0,0,0\right)$. Also, 
a massive vector boson (virtual $W^{*}$) has $4-$polarization states,
$1$ time-like/unphysical $\lambda=t$ and $3$ physical $\lambda=\pm1,0$.
For time like polarization, we can write
\[
\epsilon^{\mu}\left(t\right)=\frac{1}{\sqrt{q^{2}}}\left(q^{0},\textbf{0}\right)=\frac{1}{\sqrt{q^{2}}}\left(\sqrt{q^{2}},\textbf{0}\right)=\left(1,0,0,0\right)\;.
\]
This component is not physical for a real gauge bosons, but appears in
virtual processes like off-shell $W^{*}\rightarrow\tau^{-}\bar{\nu}_{\tau}$,
especially in the leptonic tensor decomposition. For physical polarizations,
the Lorentz condition holds: $q_{\mu}\epsilon^{\mu}\left(\lambda\right)=0$.
For $\lambda=t$, this gives 
\[
q_{\mu}\epsilon^{\mu}\left(t\right)=\sqrt{q^{2}}\neq0.
\]
Transverse polarization: $\tilde{\epsilon}^{\mu}\left(\pm1\right)=\frac{1}{\sqrt{2}}\left(0,\pm1,-i,0\right),$
represent circularly polarized states in the $xy$-plane, and longitudinal
polarizaion: $\tilde{\epsilon}^{\mu}\left(0\right)=\left(0,0,0,-1\right),$ this
points along the $-z$ axis. For transverse case, we can write
\[
q_{\mu}\epsilon^{\mu}\left(\pm1,0\right)=0.
\]

The helicity amplitudes $\mathcal{M}_{\lambda_{\tau}}\left(\tau^-\to\pi^{-}\nu_{\tau}\right)$
can be written as 
\begin{equation}
\mathcal{M}_{\lambda_{\tau}}\left(\tau^-\to\pi^{-}\nu_{\tau}\right)=i\sqrt{2}G_{F}V_{ud}^{*}f_{\pi}u_{\bar{\nu}_{\tau}}\cancel{p}_{\pi}P_{L}u_{\tau}\left(\lambda_{\tau}\right)\;.
\end{equation}
In the case of low-energy effective Lagrangian
\begin{equation}
\mathcal{L}_{eff}=\frac{G_{F}}{\sqrt{2}}V_{ud}\left[\bar{u}\gamma^{\mu}\left(1-\gamma_{5}\right)d\right]\left[\bar{\nu}_{\tau}\gamma_{\mu}\left(1-\gamma_{5}\right)\tau\right],
\end{equation}
using hadronization of the quark current, the hadronic matrix element
$\left\langle 0\left|\bar{u}\gamma^{\mu}\left(1-\gamma_{5}\right)d\right|\pi^{-}\left(p_{\pi}\right)\right\rangle$ becomes
\begin{equation}
\left\langle 0\left|\bar{u}\gamma^{\mu}\gamma_{5}d\right|\pi^{-}\left(p_{\pi}\right)\right\rangle =if_{\pi}p_{\pi}^{\mu}\;,
\end{equation}
where $f_{\pi}$ is the $\pi$-decay constant.
Using the effective Lagrangian and the above hadronic matrix element,
the full amplitude is
\begin{align}
\mathcal{M}_{\lambda_{\tau}} & =\left(-\frac{G_{F}}{\sqrt{2}}V_{ud}\right)\left(if_{\pi}p_{\pi}^{\mu}\right)\left[\bar{u}_{\nu_{\tau}}\gamma_{\mu}\left(1-\gamma_{5}\right)u_{\tau}\right]=-i\frac{G_{F}}{\sqrt{2}}V_{ud}f_{\pi}\left[\bar{u}_{\nu_{\tau}}\left(\gamma_{\mu}p_{\pi}^{\mu}\right)\left(2P_{L}\right)u_{\tau}\right],\notag\\
 & =i\sqrt{2}G_{F}V_{ud}^{*}f_{\pi}\left[\bar{u}_{\nu_{\tau}}\cancel{p}_{\pi}P_{L}u_{\tau}\left(\lambda_{\tau}\right)\right]\;.
\end{align}
The corresponding decay rate is then
\begin{equation}
\Gamma\left(\tau^-\to\pi^{-}\nu_{\tau}\right)=\frac{G_{F}^{2}\left|V_{ud}\right|^{2}f_{\pi}^{2}\left(m_{\tau}^{2}-m_{\pi}^{2}\right)^{2}}{16\pi m_{\tau}}\;.
\end{equation}

\end{document}